*To be submitted*

**From Ferroionic Instability to Domain Patterns at an Exposed Surfaces of Multiaxial Ferroelectrics**

Sergei V. Kalinin

Department of Materials Science and Engineering, University of Tennessee, Knoxville, Tennessee 37996, USA

## Abstract

We analyze the formation of domain structures in multiaxial ferroelectrics with an open, electrochemically compensated surface. The model combines vector polarization, surface electrochemistry, electrostatics, compatible strain, and competitive adsorption on common sites. We construct homogeneous phase maps and analyze polarization instabilities, first assuming perturbations that are uniform through the film thickness and then allowing depth-dependent polarization profiles. Nonlinear continuation follows the growth of these modulations into developed domains. Optimization of their period and tests of their stability establish how the initial instability wavelength relates to the preferred spacing of an established domain structure. We also consider surface ordering above a stable, weakly anisotropic bulk as a constitutive limit relevant to ferroelectric relaxors, providing insight into possible origins of their periodic surface modulations. Surface energies with the same quadratic terms can produce the same initial instability but favor different developed states, including periodic domains and uniform in-plane polarization. This analysis provides a procedure for connecting the initial development of ferroelectric modulations to amplitude selection, wavelength adjustment, and the subsequent evolution of the resulting domain structures.

## 1 Introduction

Ferroelectric domains arise from the competition between a local preference for polar order and the electrical and mechanical constraints imposed on a finite body [1]. A domain pattern can reduce depolarization fields, accommodate spontaneous strain, or satisfy both requirements, while paying the energetic cost of polarization gradients and domain walls [2–3]. Film orientation, substrate constraint, electrical contacts, and accessible polarization variants determine which arrangement is favorable [4–5]. At an exposed surface, the compensating charge and interfacial chemistry provide additional thermodynamic controls [6]. Accordingly, the pattern appearing at the onset of ordering may differ in terms of amplitude, wall profile, or spacing with that after relaxation.

The distinction between polarization reversal and ferroelastic variant formation is central to thin films. Restricting polarization to the surface normal admits oppositely polarized domains but excludes rotation into in-plane variants. Pompe, Gong, Suo, and Speck established how domain formation releases elastic energy in strained epitaxy [4]. Speck and Pompe subsequently analyzed competition between different mechanisms of misfit accommodation [7]. Streiffer and co-workers developed the geometric description of rhombohedral film patterns [8], and Romanov and co-workers analyzed their interfacial defects and energetics [9]. Roytburd formulated the thermodynamics of polydomain heterostructures in terms of macroscopic and microscopic stresses [10–11]. These studies establish the importance of the admissible variant mixture and the compatibility of its strain field.

The phase-diagram approach of Pertsev, Zembilgotov, and Tagantsev showed that epitaxial boundary conditions change the stability of homogeneous polarization states [5]. Extensions to dense domain structures and in-plane polarization instabilities clarified why a single-domain phase diagram cannot be used directly as a domain-phase diagram [12–14]. Phase-field calculations by Li, Hu, Liu, and Chen provided a route to spatially resolved polarization and compatible elastic relaxation [15–16]. Experiments resolved nanoscale reversal stripes [17], very narrow ferroelastic domains [18], and controlled domain arrays [19]; studies of layered and incompletely screened films further broadened the accessible configurations [20–21]. Higher-order Devonshire theory demonstrated that low-symmetry polarization states require an adequate angular free-energy landscape [22].

Descriptions of ferroelectric domains have commonly adopted limiting electrical boundary conditions. Ideal metallic compensation suppresses the macroscopic depolarization field, whereas an uncompensated surface with no mobile screening charge makes domain formation a principal route to reducing its electrical energy [3,21,23]. Real interfaces can lie between these limits, and their compensation can change with the electrochemical environment [6]. Fong and co-workers established the emergence and evolution of polar order in ultrathin films [24–25] and the stabilization of monodomain polarization under appropriate interfacial conditions [26–27]. Atomic-resolution studies resolved polarization variations near charged and uncharged walls [28], continuous rotation and flux-closure structures [29–30], and polar vortices and skyrmions in suitable heterostructures [31–32]. These observations motivate retaining vector polarization and more than one lateral coordinate. The functional properties of domain walls provide a further reason to resolve the actual structure [33].

The local-potential studies of Kalinin and Bonnell distinguished polarization bound charge from compensating surface charge and its response to changing conditions [34–35]. Chemical switching experiments demonstrated that the environment can change polarization without changing electrode voltage [36]. First-principles studies connected polarization to surface bonding and chemical reactivity [37–39]. Local probes and surface diffraction established effects of polar distortion, adsorption, and water exposure [40–42], while subsequent work resolved charged surface species and electrochemical behavior [43]. Ievlev and co-workers showed that ionic redistribution affects local switching and memory [44], that humidity changes switching pathways [45], and that repeated local excitation can produce complex domain sequences [46]. Surface compensation must therefore be allowed to respond to the local potential and reservoir conditions.

Morozovska, Eliseev, Kalinin, and collaborators developed coupled descriptions of ferroionic states and polarization dynamics [47–50]. Related analyses included semiconductor screening [51–52], defect thermodynamics in strained films [53], screening-limited wall motion [54], and surface modifications of wall profiles [55]. Cao and Kalinin implemented chemical control of polarization within phase-field theory [56]. Chemical control of multiaxial homogeneous phase diagrams and polarization rotation has also been established [57]. The present work asks whether a spatial instability of a chemically selected reference state leads to a periodic structure that remains energetically competitive and stable after it forms. Addressing this question requires separate calculations of the uniform state, its fluctuation spectrum, its nonlinear descendants, and their stability.

We follow this sequence for tetragonal, orthorhombic, rhombohedral, and weak-anisotropy polarization landscapes. The first three use temperature-dependent $BaTiO_3$ free energies, supplemented by PZT and $BiFeO_3$ transfer examples. The weak-anisotropy case separates small angular polarization stiffness from weak electrostriction. We then examine surface ordering above a stable, nearly isotropic bulk with an explicit preference for normal surface polarization. This constitutive limit addresses a possible origin of periodic surface modulations in relaxors without replacing their disorder and distributed dynamics by a complete microscopic theory. Throughout, the onset wavelength, fastest kinetic wavelength, primitive repeat, and energy-minimizing period remain distinct quantities.

## 2 Thermodynamic setting and material classes

We consider a coherently clamped ferroelectric film occupying $0 < z < h$, with an ideal bottom electrode at zero potential. The upper surface at $z = h$ carries adsorbed ionic species and adjoins an unbounded dielectric exterior of permittivity $\epsilon_e$. There is no upper electrode or finite dielectric gap. Lateral coordinates are unbounded in the linear analysis and periodic in the nonlinear calculations. Polarization is the vector field $\mathbf{P} = (P_x, P_y, P_z)$. Mechanical displacement relaxes subject to fixed substrate displacement and zero traction at the upper surface. The imposed equal in-plane misfit is $u_m$. The geometry and representative angular landscapes are shown in Figure 1. The exterior electric field of a nonzero lateral Fourier component decays away from the surface; the uniform exterior field is zero. Its constant potential is determined by the film and is not independently fixed at infinity.

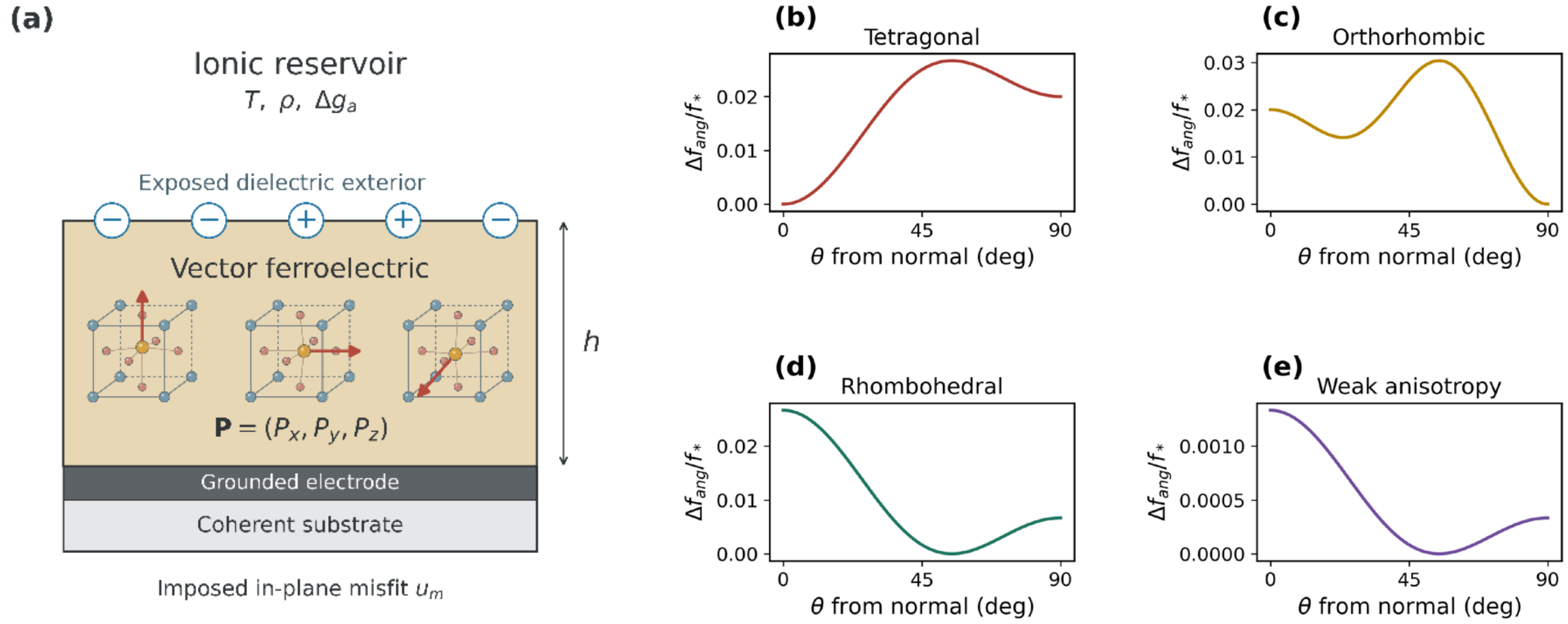

**Figure 1.** Exposed-surface geometry and parent polarization landscapes. (a) A vector ferroelectric film is coherently attached to a grounded bottom electrode and exchanges surface ions with a reservoir. The upper dielectric is unbounded; there is no top electrode or gap capacitor. (b–e) Illustrative tetragonal, orthorhombic, rhombohedral, and weak-anisotropy angular sections along $\mathbf{n} = \left(\sin\theta/\sqrt{2}, \sin\theta/\sqrt{2}, \cos\theta\right)$. The angular energies classify the model landscapes and are not measured material barriers.

At fixed bottom-electrode potential, the appropriate electrical functional is given by Eq. (1). Using absolute permittivities $\epsilon_f = \epsilon_0\epsilon_b$ and $\epsilon_e = \epsilon_0\epsilon_{e,r}$, we write

$$\begin{aligned} \mathcal{G} = \int_f \left[ f_L(\mathbf{P}) + \tfrac{1}{2} G_{ijkl}\, \partial_j P_i\, \partial_l P_k + f_{el} + \mathbf{P}\cdot\nabla\phi - \tfrac{\epsilon_f}{2}|\nabla\phi|^2 \right] dV \\ - \int_e \tfrac{\epsilon_e}{2}|\nabla\phi|^2 dV + \int_s \left[f_{ads} + \sigma\psi + f_s\right] dA, \qquad \psi = \phi(h). \end{aligned} \tag{1}$$

The subscripts denote film, exterior, and surface. The potential is eliminated by stationarity, whereas the material and ionic variables are minimized. This distinction incorporates the work of the voltage source and prevents a sign error in the reduced electrostatic energy. The resulting grand potential is denoted by $\Omega$. The same $\Omega$, with the same chemical reservoir and mechanical boundary conditions, is used for the homogeneous, linear, and nonlinear comparisons. Changing the charge constraint between those calculations would change the ensemble and invalidate an energy comparison.

The elastic energy and electrostrictive eigenstrain are

$$f_{el} = \tfrac{1}{2} C_{\alpha\beta}\left(e_\alpha - e_\alpha^0\right)\left(e_\beta - e_\beta^0\right), \qquad e_\alpha^0 = Q_{\alpha ij} P_i P_j. \tag{2}$$

Engineering-shear conventions and the complete elastic elimination are specified in Sections S1–S4. In a uniform film, the unconstrained strains can be eliminated algebraically. In a domain pattern, the displacement field must remain compatible across the entire film. Replacing this field by independently relaxed strains at each point would remove the long-range elastic interaction responsible for much of the ferroelastic pattern selection [4–5,7–16,58]. The substrate is treated as mechanically rigid. Finite substrate compliance is a possible extension, rather than an implicit adjustable parameter.

The positive and negative species compete for the same population of surface sites. If $\theta_+$ and $\theta_-$ are their occupation fractions, the empty fraction is $\theta_0 = 1 - \theta_+ - \theta_-$. The chemical grand-potential density per area is

$$f_{ads} = N_s\{\sum_{i=\pm} \theta_i \,\Delta g_i + k_B T(\theta_0 \ln\theta_0 + \sum_{i=\pm} \theta_i \ln\theta_i)\}, \tag{3a}$$

$$\Delta g_i = \Delta g_i^0 - \nu_i k_B T \ln\rho, \qquad \sigma = N_s \sum_{i=\pm} q_i \,\theta_i. \tag{3b}$$

Here $N_s$ is the common site density, $q_i$ the signed ionic charge, $\rho$ a dimensionless reservoir activity, and $\nu_i$ the specified reaction stoichiometry. We use $\nu_+ = -1/2$ and $\nu_- = +1/2$. Activity is taken to be the primary control variable; assigning it to a particular gas pressure requires the corresponding surface reaction. A site is either empty or occupied by one species. The entropy therefore has three mutually exclusive states, rather than a product of two binary entropies. This is the common-site law used consistently in the scalar limit, multiaxial films, nonlinear domains, and surface relaxation problem.

Variation of Eq. (1) gives the displacement jump and competitive adsorption law,

$$\begin{gathered} \epsilon_f \,\partial_z \phi_f - \epsilon_e \,\partial_z \phi_e = P_z + \sigma, \\ w_i = \exp[-(\Delta g_i + q_i \psi)/(k_B T)], \quad \theta_i = \frac{w_i}{1 + w_+ + w_-}. \end{gathered} \tag{4}$$

The occupation constraint is automatic. The two species exchange with a reservoir, so neither the surface charge nor either occupation is constrained to have zero spatial mean. The bottom electrode supplies the compensating electrical charge. Equation (3) is convex in occupations in the interior of the site simplex; chemical screening alone cannot destabilize a fully convex polar and elastic energy. Its role is to change the cost of charged fluctuations and to select a self-consistent polar state. Additional lateral ionic attractions, site cooperativity, and composition-dependent bonding are not considered to be the part of this baseline model.

The baseline film calculations set the additional polarization surface energy $f_s$ to zero and use natural polarization boundary conditions. The gradient energy is positive, with an isotropic baseline coefficient. Consequently, the model does not insert a preferred lateral wavelength through a negative gradient term. A finite-wavevector instability, when present, results from competition among the local tangent, compatible strain relaxation, the electrostatic kernel, and the positive gradient penalty. Interface-induced modifications and flexoelectricity can change this competition, but require additional constitutive information [23,59–63].

The principal film comparison uses $h = 20$ nm, $\epsilon_b = 7.35$, $\epsilon_{e,r} = 1$, and $G = 10^{-10}$ J m$^3$ C$^{-2}$. The Landau, elastic, and electrostrictive coefficients are taken from the material parameterizations specified below and in Section S10 [64–68]. The isotropic gradient coefficient is a representative phase-field scale [15,56]; its use for every class is a controlled modeling choice, and its anisotropy is examined separately. The background permittivity is held at 7.35 for consistency with the companion uniaxial model; it excludes the soft polarization response already described by the Landau free energy. The exterior value represents an air-like dielectric. A 20 nm thickness provides a common nanoscale film benchmark, while the maps explicitly cover 1–300 nm rather than attributing universal significance to that single thickness.

The common site density is $N_s = 10^{18}$ m$^{-2}$, ionic charges are $q_\pm = \pm 2e$, and standard formation energies are $\Delta g_i^0 = 0.200$ eV. These values continue the uniaxial comparison and lie in the modeling range explored in ferroionic and multiaxial surface-screening studies [47,57]. They are not measured adsorption constants for every material or termination. The chosen density corresponds to one available site per square nanometer and a maximum single-sign ionic charge of approximately 0.320 C $m^{-2}$, making charge-capacity limitations explicit. Keeping this reservoir model common isolates changes due to the polar and elastic landscape. No thermal expansion correction to misfit is inserted during temperature sweeps, and every electrical kernel retains the exposed upper surface.

The tetragonal, orthorhombic, and rhombohedral parent landscapes, denoted T, O, and R, use the eighth-order Li–Cross–Chen $BaTiO_3$ potential at 300, 230, and 150 K [64–65]. These labels identify the stress-free parent landscape. The film orientation is recalculated under its actual mechanical and electrochemical constraints. A tetragonal-parent film can therefore select an in-plane diagonal state, and a rhombohedral-parent film need not retain equal Cartesian polarization components. The PZT comparison uses the 50/50 composition [66,68]. The $BiFeO_3$ calculation uses a polarization-only potential and is retained principally to examine ionic-capacity limitations [67,69]. Octahedral rotations and additional electronic compensation are not represented by that potential.

The weak-anisotropy film, W, is defined at 300 K by

$$f_W/f_* = -|\mathbf{p}|^2/2 + |\mathbf{p}|^4/4 + 0.002\sum_i p_i^4, \qquad \mathbf{p} = \mathbf{P}/P_*. \tag{5}$$

This model is chosen to represent the behavior of ferroelectric relaxors. We use $P_* = 0.300$ C m$^{-2}$, $a_* = 10^8$ J m C$^{-2}$, and $f_* = a_* P_*^2 = 9.00$ MJ m$^{-3}$. The elastic tensor is that of $BaTiO_3$. The baseline electrostriction is $0.1Q_{BTO}$, with a separate comparison at $Q_{BTO}$. A small angular anisotropy does not imply weak electrostriction or weak piezoelectric response: a soft rotational susceptibility can enhance the response even when the spontaneous strain is modest. The thermal extension of Eq. (5) is explicitly specified in Section S10. This clean model isolates rotational softness; it does not include quenched disorder or establish the response of a particular relaxor composition [70–75].

**3 Homogeneous phase selection under chemical and mechanical control**

The homogeneous problem must be solved before the spatial stability problem. We first eliminate the unconstrained strain components and obtain the coherent-film energy $f_{cl}(\mathbf{P}; u_m)$. Its quadratic and quartic corrections include the shear contribution associated with simultaneous in-plane components. We then solve the scalar electrochemical equation for each prescribed $P_z$,

$$C_f\psi = P_z + \sigma(\psi), \qquad C_f = \epsilon_f/h. \tag{6}$$

The equilibrium ionic response obeys $-d\sigma/d\psi = C_{chem} \geq 0$, so the left-minus-right side of Eq. (6) is strictly increasing. The potential is therefore unique for a specified normal polarization, even when the subsequent polarization minimization has several competing minima. This separates the numerical problem into a monotone electrochemical reduction and a nonconvex polarization minimization.

Elimination of the ions and potential defines a surface contribution $W(P_z)$ and the homogeneous grand-potential density,

$$\Phi_{hom} = f_{cl} + W/h, \qquad W' = \psi, \qquad W'' = \frac{1}{C_f + C_{chem}}, \tag{7a}$$

$$C_{chem} = \frac{N_s}{k_B T}\left[\sum_{i=\pm} q_i^2\,\theta_i - \left(\sum_{i=\pm} q_i\,\theta_i\right)^2\right]. \tag{7b}$$

This is the variance of the single-site charge, including the empty state. The cross term expresses competition for sites and is required in a linear response about a charged reference. At balanced activity and zero potential, $\theta_+ = \theta_- = w/(1+2w)$, where $w = \exp[-\Delta g^0/(k_B T)]$. Hence $C_{chem} = 2N_s(2e)^2 w/[k_B T(1+2w)]$. The homogeneous electrochemical contribution can be evaluated without integrating the response: $W = P_z\psi - C_f\psi^2/2 - N_s k_B T\ln(1 + w_+ + w_-)$. Differentiation gives Eq. (7a), with the potential at its stationary value. This expression is also the scalar reduction used to verify the vector implementation.

Equation (7) distinguishes the amount of compensating charge from its differential response. An almost saturated ionic population can supply substantial charge while providing little incremental screening. The largest capacitance occurs at intermediate occupation. Temperature and chemical activity therefore affect polarization through both the equilibrium charge and the local curvature of the adsorption free energy. The temperature-dependent surface-potential and retention measurements provide experimental precedents for treating these quantities separately [34–35,76–79].

A reduced two-amplitude energy makes the basic phase diagram topology transparent. Let $s$ and $z$ be the squared in-plane and normal amplitudes after rescaling the positive self-quartic coefficients. At zero bias,

$$F_4 = \frac{1}{2}As + \frac{1}{2}Cz + \frac{1}{4}(s^2 + z^2) + \frac{g}{2}sz, \qquad s \geq 0, \quad z \geq 0. \tag{8}$$

For $|g| < 1$, the mixed solution has $s = (gC - A)/(1 - g^2)$ and $z = (gA - C)/(1 - g^2)$ when both are positive. For the illustrated competitive case $g > 1$, a stable mixed interior minimum is absent and the pure branches compete across an energy-crossing line. This construction explains why a mixed-polarization region can either separate the in-plane and normal phases or be replaced by a first-order boundary. Note that this model explains the topology of the phase diagram, rather then defines a complete material polynomial.

Figure 2 combines this classification with vector minimization for the four angular landscapes. The vector calculation retains the in-plane azimuth, which distinguishes axis-like and diagonal states. The orthorhombic example includes a higher-order invariant: a purely fourth-order cubic potential cannot provide an isolated stable orthorhombic minimum with both rotational curvatures positive. The narrow mixed region in the weak-anisotropy example illustrates the sensitivity of orientation selection when radial ordering is strong but angular selection is weak. The coefficient maps were generated directly from the analytical candidate energies and vector minimization.

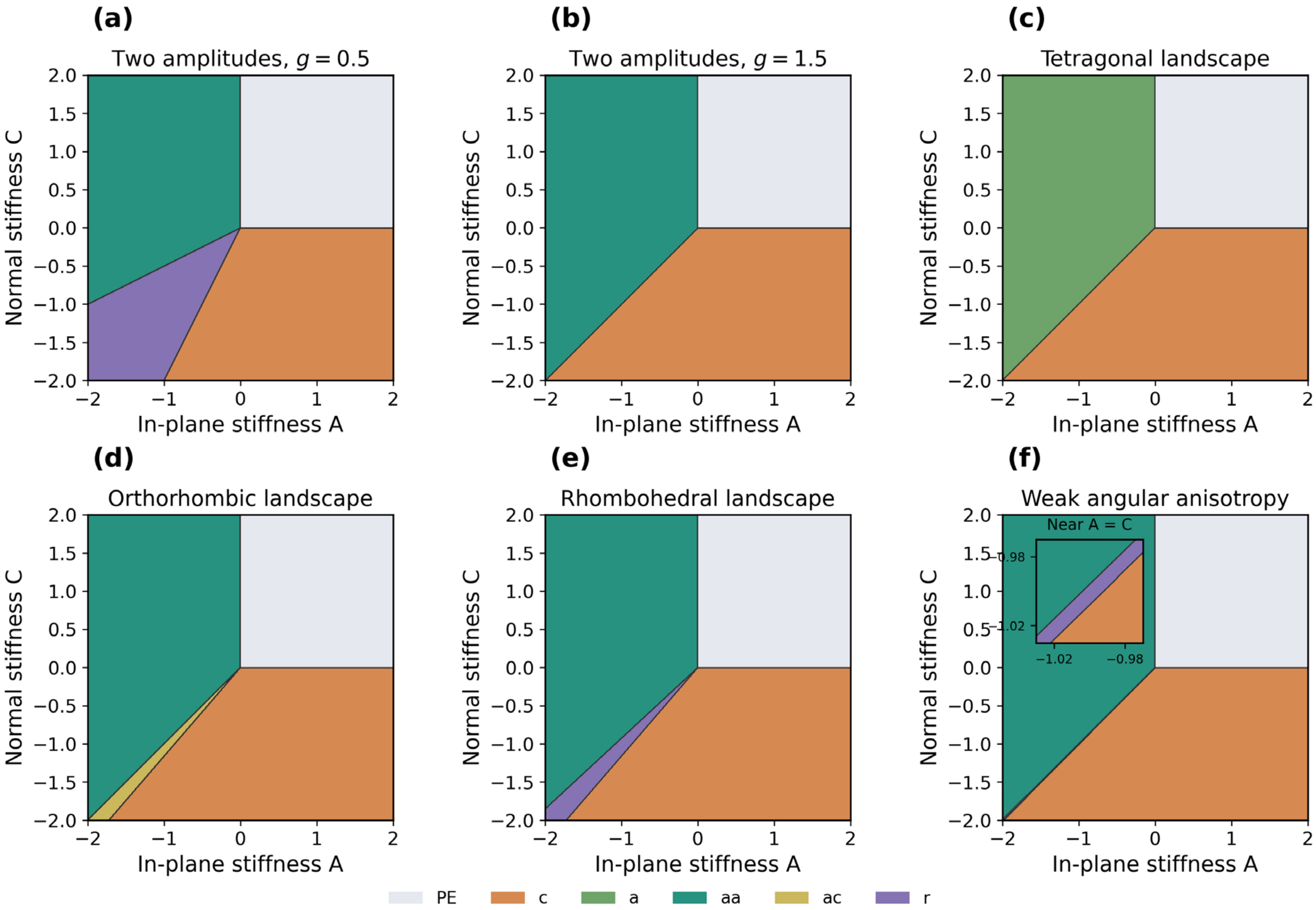

**Figure 2.** Homogeneous topology on a coefficient plane. (a,b) The two-amplitude quartic model at $g = 0.5$ and $g = 1.5$. (c–f) Vector landscapes with tetragonal, orthorhombic, rhombohedral, and weak angular preference, retaining the in-plane azimuth. The inset resolves the narrow mixed region near equal normal and in-plane stiffness in (f). Colors identify component patterns of uniform minima. These diagrams do not test spatial stability.

For the material maps, the in-plane minimization is performed algebraically using the invariants $s = P_x^2 + P_y^2$ and $v = P_x^2 P_y^2$, with axis, diagonal, and admissible interior azimuths retained. A resolved search over $P_z$ then enumerates and refines the competing minima. The reference is chosen by its grand potential, rather than by continuation of one favored orientation. This procedure is particularly important near first-order boundaries, where a stationary branch can persist after another branch has acquired lower energy. The algorithm, coefficient conventions, and independent minimization checks are given in Sections S10 and S11.

Figure 3 presents the complete $BaTiO_3$ temperature interval, 100–500 K, at the single fixed misfit $u_m = -0.100\%$. Each column corresponds to one activity; the top and bottom panels show the selected component pattern and its tilt from the normal. We define the tilt as $\vartheta = \mathrm{atan2}\left(\sqrt{P_x^2 + P_y^2}, |P_z|\right)$, so that $\vartheta = 0$ denotes normal polarization and $\vartheta = 90°$ denotes in-plane polarization. The angle is undefined when the polarization vanishes. The three parent temperature regimes are therefore contained in one continuous calculation. They are not assembled from separate calculations at different strains. The component labels $c$, $a$, $aa$, $ac$, and $r$ describe which film-axis components are nonzero. In particular, $r$ labels a tilted diagonal vector; it does not establish rhombohedral crystallographic symmetry throughout the region bearing that label.

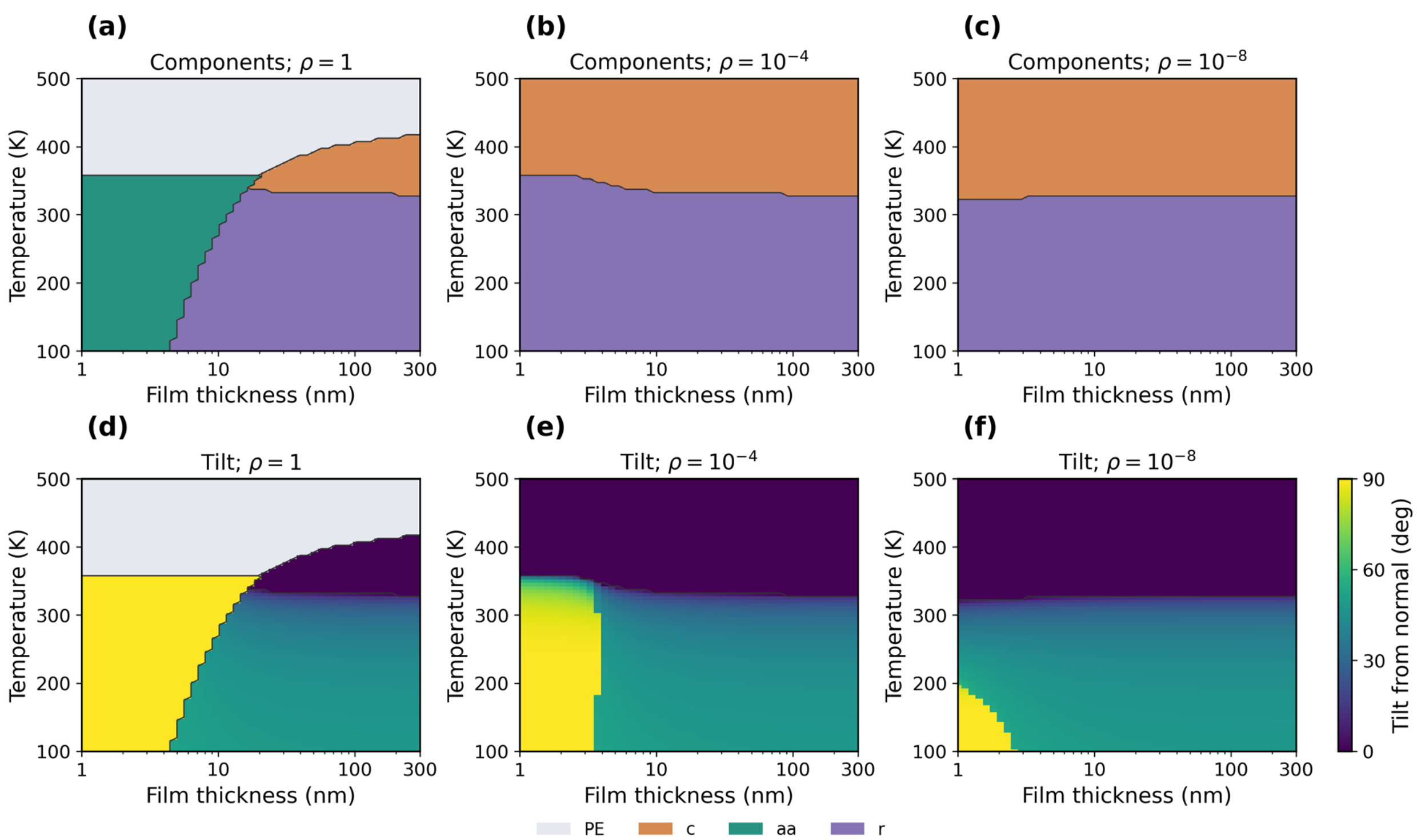

**Figure 3.** Unified $BaTiO_3$ temperature–thickness maps at $u_m = -0.100\%$. Columns correspond to activities $1$, $10^{-4}$, and $10^{-8}$. (a–c) Homogeneous component pattern. (d–f) Polarization tilt from the surface normal, defined in the text. Every panel spans 100–500 K and uses the same common-site chemistry and open exterior. Gray denotes the unpolarized candidate in the component map and its undefined tilt in the lower row. A biased normal component is not, by itself, evidence of spontaneous ferroelectric order.

The thickness dependence follows from the interfacial contribution per film volume, the changing electrostatic capacitance, and the competition with the clamped bulk energy. Chemical bias lifts the equivalence of opposite normal polarizations and changes the locations of orientation transitions. It need not produce a monotone increase of the normal component. Adsorption saturation and temperature-dependent Landau curvature can instead create narrow tilted regions or reentrant sequences. Such features should be traced to the equilibrium branch and its occupations before they are interpreted as new domain phases. Detailed component trajectories and the additional compressive-strain O/R window are retained in the supplement. Supplementary Movies 1–4 extend the homogeneous component and tilt maps through $10^{-12} \leq \rho \leq 10^{12}$ in half-decade steps for $BaTiO_3$, $BiFeO_3$, and the two weak-anisotropy electrostriction choices. These are equilibrium activity sweeps, not temporal simulations. Here $\rho$ is an effective reservoir activity; its conversion to pressure depends on the reaction and standard state. A different oxidizing agent can also change formation energies and reaction stoichiometry, so large $\rho$ is a chemical-potential sensitivity limit rather than a quantitative model of every oxidant.

Figure 4 tests transfer to PZT 50/50 and the polarization-only $BiFeO_3$ reference under balanced chemistry. The comparison emphasizes the role of the constitutive landscape and charge capacity. PZT near a compositionally soft angular landscape can support orientation competition that differs from $BaTiO_3$, while the larger polarization scale of $BiFeO_3$ can exceed the compensating capacity chosen for the common surface model. Figure 5 independently changes electrostriction in the weak-anisotropy model while keeping the same angular potential and reservoir conditions.

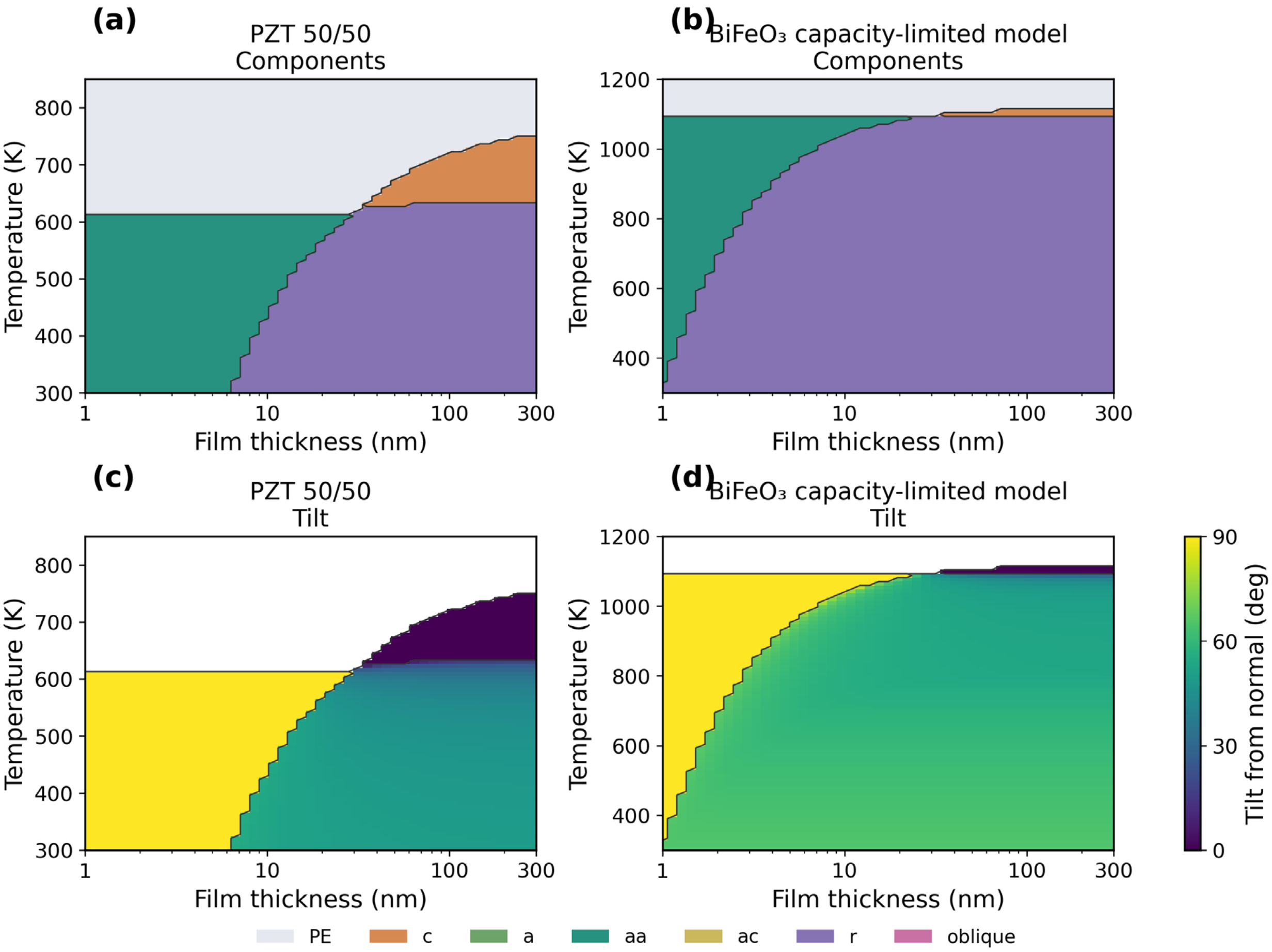

**Figure 4.** Material-transfer comparison at balanced activity and $u_m = -0.100\%$. (a,b) Uniform component maps for PZT 50/50 and the polarization-only $BiFeO_3$ model. (c,d) Corresponding tilts. $BiFeO_3$ includes capacity-limited conditions where additional electronic or structural physics would be needed for an experimental prediction. Its complete activity slices and the site-capacity audit appear in Figures S6 and S21.

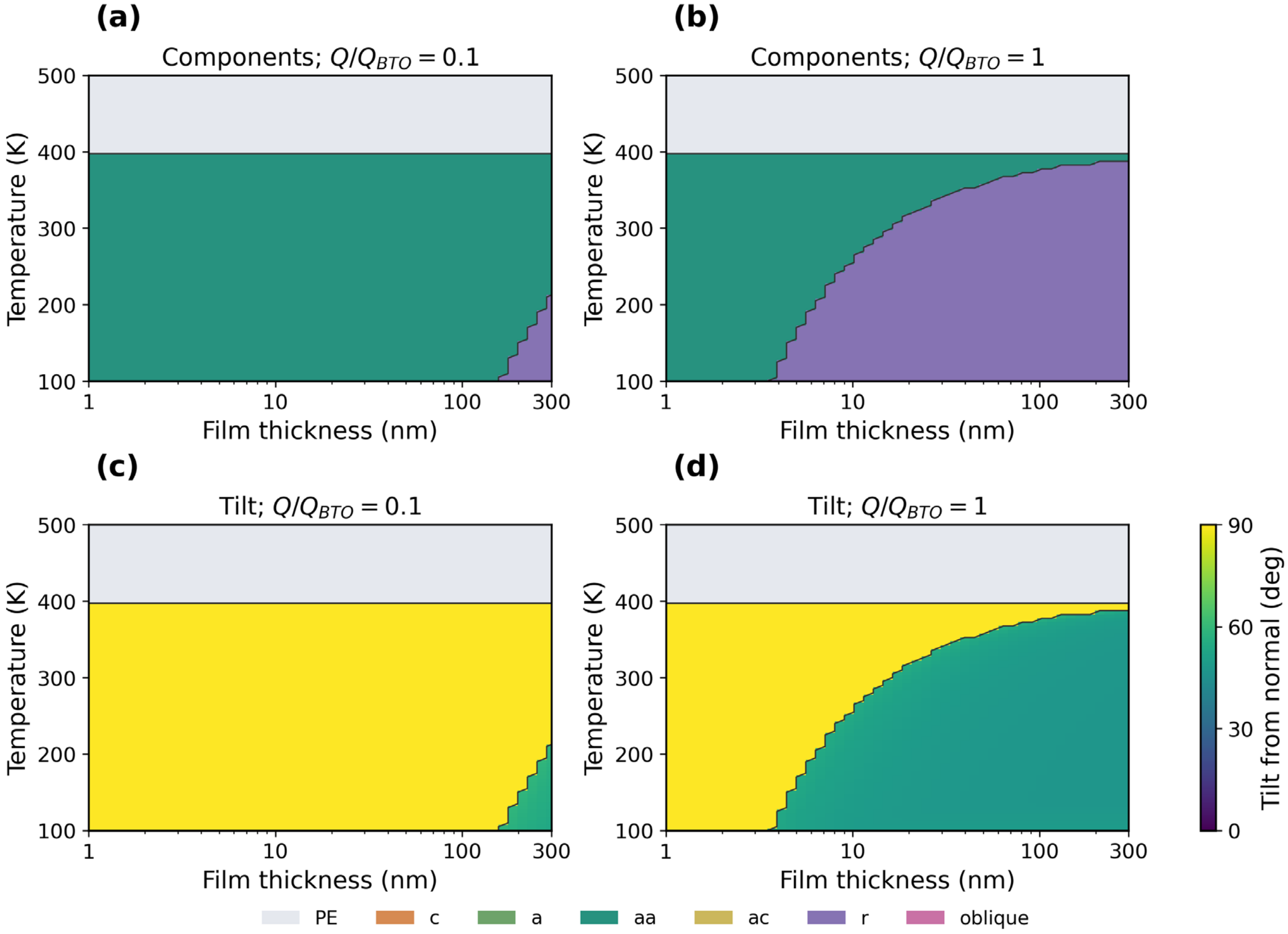

**Figure 5.** Electrostriction as an independent control in the weak-anisotropy film at balanced activity and zero misfit. (a,c) Component pattern and tilt for $Q = 0.1Q_{BTO}$. (b,d) The corresponding results for $Q = Q_{BTO}$. The polar anisotropy, chemical parameters, exterior, and thermal law are common to both columns.

The 20 nm reference states illustrate why parent symmetry and actual film orientation must be distinguished. At balanced activity, T at zero misfit selects approximately $(0.156{,}0.156{,}0)$ C m$^{-2}$. O and R at $u_m = -0.300\%$ select approximately $(0.0893{,}0.0893{,}0.268)$ and $(0.163{,}0.163{,}0.276)$ C m$^{-2}$. W at $u_m = -0.150\%$ selects approximately $(0.207{,}0.207{,}0)$ C m$^{-2}$. The corresponding chemical capacitances are 0.0216, 4.47, 5.94, and 0.0216 F m$^{-2}$. Identical adsorption parameters therefore do not imply identical differential screening. The zero-misfit W maps provide a separate comparison; they are not used as the reference for its nonlinear stripe. Frozen-ion and equilibrated-ion spectra below isolate incremental screening at fixed polarization and stress.

## 4 Linear instability and the initial polarization pattern

We now perturb a stationary homogeneous reference by $\delta\mathbf{P} = \mathbf{v}(z)\exp(i\mathbf{q}\cdot\mathbf{r}_{\parallel}) + c.c.$ and allow displacement, electrostatic potential, and surface occupations to respond. Both the magnitude and azimuth of the lateral wavevector $\mathbf{q}$ are varied. The depth dependence of $\mathbf{v}$ is retained in the numerical solution. This is a stability calculation about the actual film state found in Section 3, rather than about a prescribed bulk variant. A negative eigenvalue identifies an admissible fluctuation that lowers the grand potential to second order.

The local polarization tangent includes the prestress contribution,

$$A_{ij}^{0} = \frac{\partial^2 f_L}{\partial P_i \partial P_j} - \sigma_\alpha^0 \frac{\partial^2 e_\alpha^0}{\partial P_i \partial P_j}. \tag{9}$$

The remaining elastic term is obtained by minimizing the compatible displacement response. In operator form this is a Schur complement of the joint polarization–displacement Hessian. It is nonlocal in depth and depends on the lateral wavevector. The reference stress defined in Eq. (9) and the relaxed elastic operator are both required. Keeping only an electrostrictively renormalized homogeneous Landau polynomial would miss the difference between uniform strain relaxation and a spatially compatible deformation. Conversely, adding the full elastic operator to an already relaxed local polynomial without accounting for the eliminated terms would double-count part of the mechanical response.

The reduced electrical boundary problem is particularly useful because it exposes the role of surface chemistry without a special polarization ansatz. With $\varphi$ denoting the potential perturbation, elimination of the exterior and equilibrated ions gives

$$\epsilon_f(\partial_z^2 - q^2)\varphi = i\mathbf{q}\cdot\mathbf{v}_\parallel + \partial_z v_z, \qquad \varphi(0) = 0, \tag{10a}$$

$$\epsilon_f\varphi'(h) + [\epsilon_e q + C_{chem}]\varphi(h) = v_z(h). \tag{10b}$$

The exterior contribution tends to zero as $q \to 0$, while the grounded film retains its uniform capacitance $C_f$. Equation (10), derived in Section S4.3, retains both volume polarization charge and the normal surface component. The ionic perturbation is $\delta\sigma = -C_{chem}\varphi(h)$. A frozen-ion comparison sets this perturbation to zero while retaining the same equilibrium polarization, stress, and occupation. Re-equilibrating the reference with a different chemical condition would answer a different question.

We define the positive dielectric operator $\mathcal{A}_\phi$ by the quadratic form containing the film field energy and the boundary capacitance in Eq. (10). If $\mathcal{L}$ maps a polarization perturbation to its electrostatic source, elimination of the potential contributes $\mathcal{L}^\dagger\mathcal{A}_\phi^{-1}\mathcal{L}$. The complete relaxed polarization operator is therefore

$$\mathcal{H}(\mathbf{q}) = \mathbf{A}^0 + \mathcal{K}_G(\mathbf{q}) + \mathcal{K}_{el}(\mathbf{q}) + \mathcal{L}^\dagger\mathcal{A}_\phi^{-1}\mathcal{L}. \tag{11}$$

All terms are evaluated using one normalization of the polarization inner product. The operator is Hermitian for the reciprocal equilibrium model. Importantly, the reduced electrostatic contribution remains positive. Surface chemistry softens an electrical penalty; it does not create a negative electrostatic energy. An instability reflects the complete balance in Eq. (11), including any locally unfavorable polarization direction and the extent to which spatially varying strain can relax.

When adsorption affects electrostatics through the local net surface charge, allowing that charge to equilibrate changes the frozen-ion polarization stiffness by

$$\mathcal{H}_{rel} = \mathcal{H}_{fr} - \frac{C_{chem}}{1+C_{chem}R_q}\mathbf{w}_q\mathbf{w}_q^\dagger, \qquad R_q > 0. \tag{12}$$

Here $R_q$ is the bare surface-potential response to charge, and $\mathbf{w}_q$ measures the overlap of the polarization perturbation with the surface potential. Section S4.4 derives Eq. (12) by a rank-one inverse update. At a fixed reference, increasing a positive chemical capacitance cannot increase any ordered stiffness eigenvalue. For a normalized nondegenerate mode, its derivative is proportional to $-\left|\mathbf{w}_q^{\dagger}\mathbf{v}\right|^2/\left(1 + C_{chem}R_q\right)^2$. A mode with negligible surface-potential overlap is correspondingly insensitive to ionic relaxation even when the equilibrium surface charge is large.

This result does not establish a universal direction of chemical-activity-induced motion of a phase boundary. Along a self-consistent branch, activity changes $\mathbf{P}_0$, reference stress, occupation, capacitance, and the eigenvector. The total change of stiffness includes all of these effects. Chemical conditions can therefore select a more stable homogeneous orientation while simultaneously reducing the fluctuation penalty at that orientation. Distinguishing fixed-reference softening from branch selection resolves the apparent tension between chemical stabilization of polar order and chemical promotion of spatial modulation.

The uniform-depth Galerkin projection provides a closed analytical limit. Resolving polarization into longitudinal in-plane, transverse in-plane, and normal components gives

$$D_{es} = \begin{pmatrix} (1 - 2t/Q)/\epsilon_f + t^2 d_n & 0 & itd_n \\ 0 & 0 & 0 \\ -itd_n & 0 & d_n \end{pmatrix}, \qquad t = \tanh(Q/2), \quad Q = qh, \tag{13a}$$

$$d_n = \{h[\epsilon_f q\coth(qh) + \epsilon_e q + C_{chem}]\}^{-1}. \tag{13b}$$

The imaginary off-diagonal entries encode a spatial phase shift between normal and longitudinal polarization. Omitting them changes both the eigenvector and the eigenvalue. The transverse in-plane component is electrically neutral in this subspace, although elasticity can still couple it to other components. The full depth-dependent eigenproblem permits additional charge and strain relaxation. A stable uniform-depth projection consequently does not prove stability in the larger function space; an unstable projected mode already provides an admissible instability of that space.

The material classes differ through their local and compatible rotational responses. A high-symmetry tetragonal reference has an amplitude direction and two equivalent local rotations before the film boundaries split them. A rhombohedral reference similarly has two degenerate local rotational curvatures in the ideal parent setting. An orthorhombic reference has two inequivalent rotational channels, so their stiffnesses and gradient penalties must be retained separately. Once the actual film state is tilted, the Hessian is diagonalized directly rather than assigning it an idealized parent-mode basis. In the W limit, small angular stiffness makes elastic and electrical boundary terms especially effective selectors. Sections S5–S7 give the projections and the full boundary-value construction for each class.

Figure 6 shows the finite-wavevector periods of negative least-stiff modes on the broader temperature–thickness slices. At each temperature and thickness, we first determine the lowest-energy homogeneous polarization and its surface occupations. We then minimize the fluctuation stiffness over the sampled lateral wavevectors and directions, allowing a depth-dependent eigenfunction. Where that minimum is negative and occurs at $q > 0$, color gives $2\pi/q$ for this most weakly restored static mode. Moving across a colored region therefore follows changes in the reference state and its initial modulation; it does not track the period of an already equilibrated domain. The logarithmic color scale makes short and long modulation periods visible in the same panel. The calculation uses 49 × 49 control points for each of six cases and samples 31 nonzero wave-number magnitudes at eight azimuths, together with the zero-wavevector sector. Polarization has nine depth nodes; the electrical and displacement fields use an enriched 17-node grid. The denser control grid resolves the plotted boundaries more clearly but does not remove the finite wave-number or depth discretization. Their settings are exposed in the notebook, and separate refinement checks are reported in the supplement. Gray regions have no resolved negative least mode at nonzero wavevector under the sampling and tolerance used. An unhatched gray region is locally stable within the tested fluctuation space; it can still have a lower-energy finite-amplitude competitor. Blue crosshatched regions indicate that the most unstable sampled mode is spatially uniform in the film plane, $q = 0$, so no finite lateral period can be assigned to that selected mode. Other nonzero-wavevector modes may also be unstable there. Gray and crosshatched areas are therefore not missing calculations and are not interchangeable labels for a thermodynamic uniform phase.

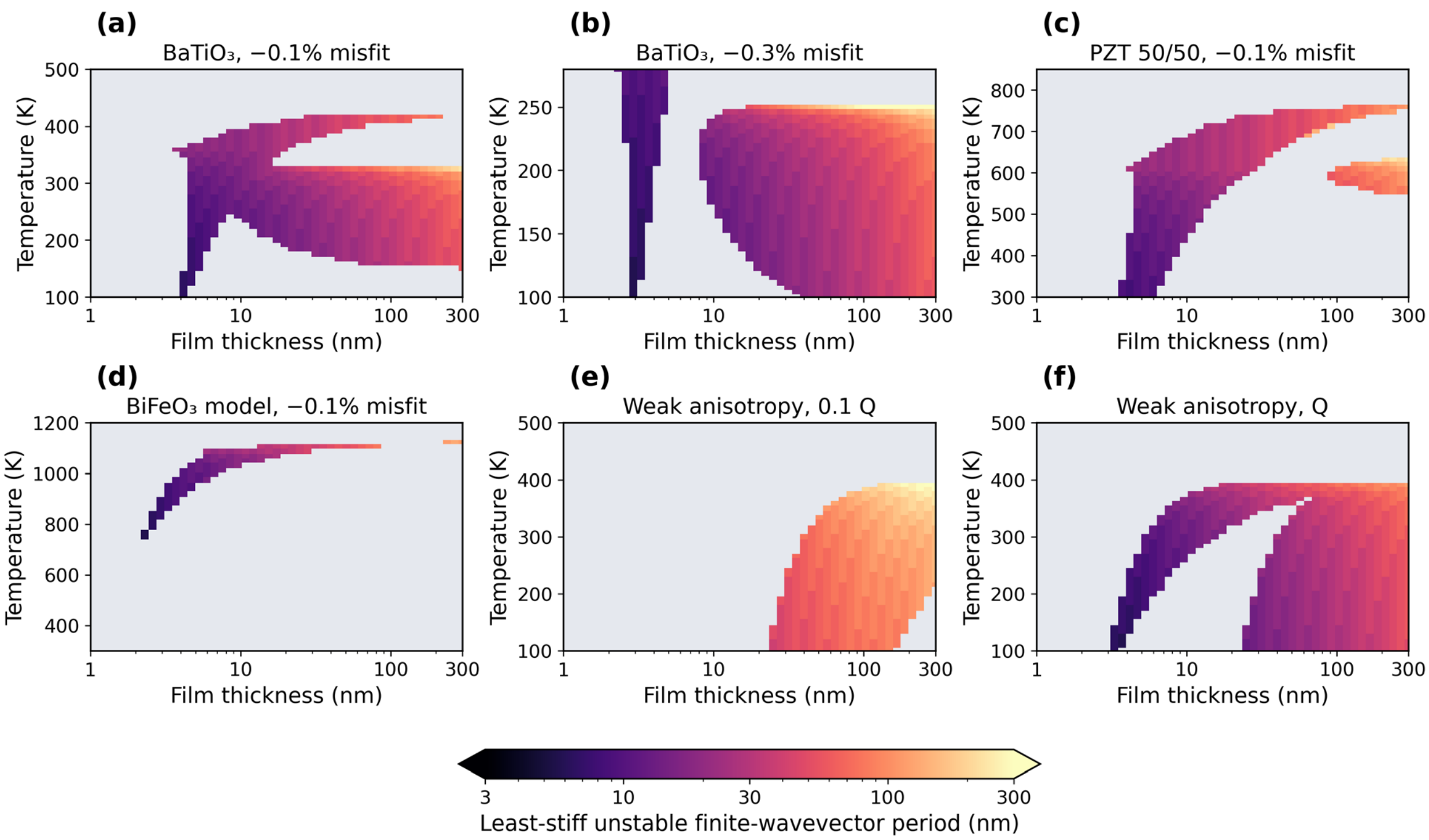

**Figure 6.** Dense maps of least-stiff unstable periods at balanced activity. (a,b) $BaTiO_3$ at misfits $-0.100\%$ and $-0.300\%$, the latter covering the O/R comparison window. (c,d) PZT 50/50 and $BiFeO_3$. (e,f) Weak anisotropy at reduced and ordinary electrostriction. Each panel has 49-by-49 control points, 31 nonzero wavevector magnitudes, eight azimuths, and a separate zero-wavevector test. Color is shown only for a negative least mode at nonzero wavevector. Gray indicates no negative sampled minimum at nonzero wavevector; it does not exclude a finite-amplitude domain state. Blue crosshatching identifies a negative minimum at $q = 0$, for which the selected lateral period is undefined. The tolerance is $10^{-5}a_*$; color clipping at the scale limits is indicated by the colorbar extensions.

Figure 7 places the baseline strain–activity component maps above their corresponding least-stiffness maps. Hatching marks a negative sampled spatial mode of the homogeneous reference represented by the underlying color. It is therefore an overlay of local instability, not a label for a particular equilibrated domain morphology. The lower row makes the sign and magnitude of this instability explicit. The mirrored activity axis follows the charge-sign symmetry of the charge-symmetric common-site model. It should not be interpreted as a general symmetry of an arbitrary experimental adsorption reaction.

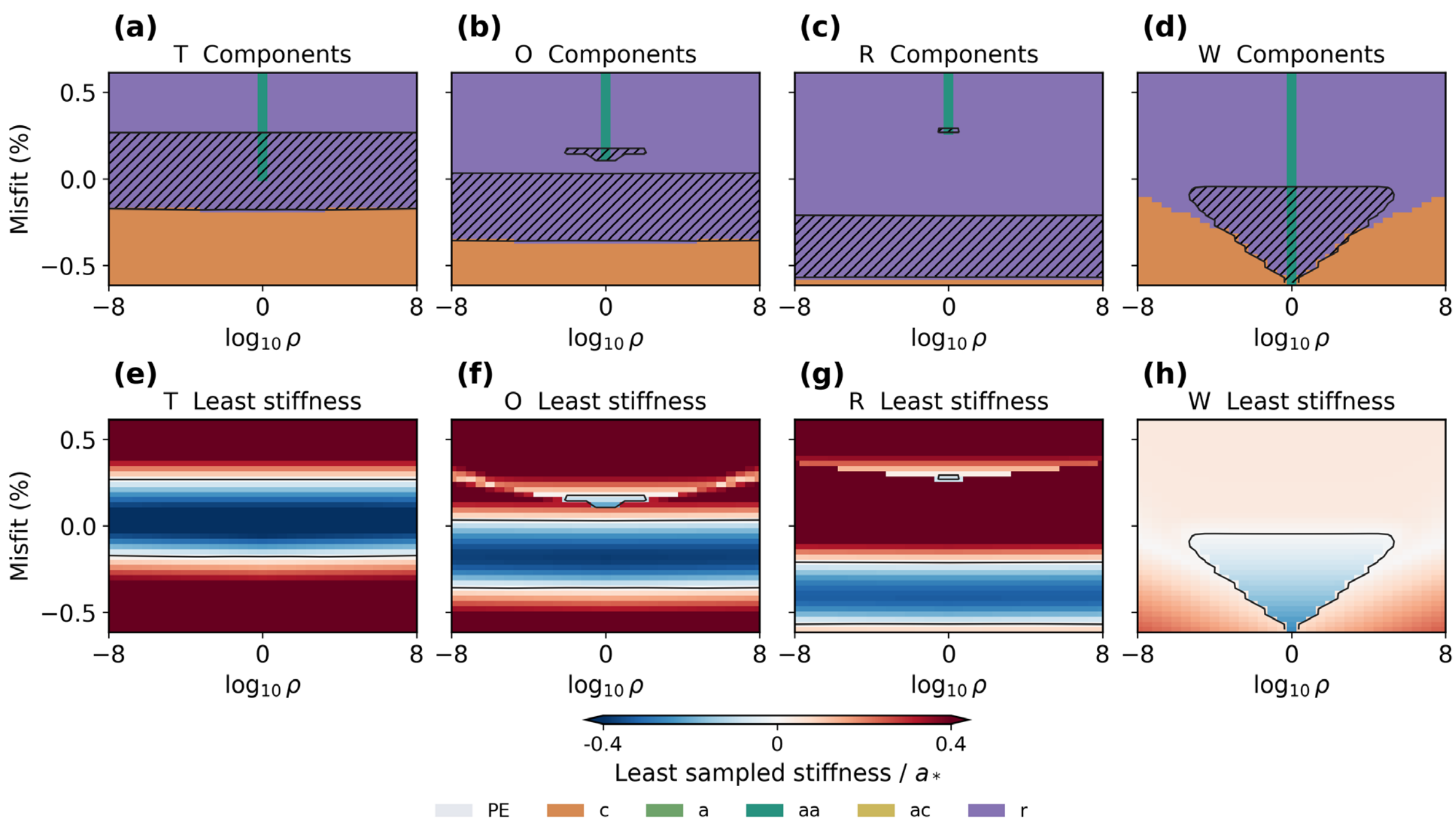

**Figure 7.** Matched uniform-state and spatial-stability maps for 20 nm films. (a–d) T, O, R, and W component maps, hatched where the sampled homogeneous spatial Hessian is negative. (e–h) Their corresponding least eigenvalues on the same coordinates. Black curves mark zero stiffness. The activity reflection follows the exact charge-inversion symmetry of the specified common-site reaction model. Columns pair each component map with its own stiffness map at identical controls. Hatching in the upper row means that this uniform reference has a negative sampled spatial mode; it does not identify the morphology reached after relaxation. Unhatched regions pass this local test but may still have finite-amplitude competitors. Negative values in the lower row identify onset, positive values identify local stability, and black curves mark zero stiffness. The underlying strain–activity sampling has not been increased in this revision.

At the stated reference conditions, the refined T, O, R, and W static spectra have negative minima near 15.7, 24.4, 16.0, and 41.7 nm. These are the least-stiff periods of already unstable homogeneous states. They are not the marginal periods found by tuning strain to the spinodal. The W reference here uses $u_m = -0.150\%$; its zero-misfit comparison belongs to the broader maps. The O and R modes primarily involve rotation and can have little electrical overlap. Nearly coincident frozen-ion and relaxed-ion spectra in such cases indicate weak incremental screening of that mode, while chemistry still helps select the reference polarization on which the mode exists. Figures S12–S15 retain the spectra, depth eigenfunctions, and corresponding strain–activity periods.

Finally, a static minimum is not generally the fastest growing mode after a finite quench. For passive reciprocal dynamics, the linear generator is $-\mathsf{M}\mathcal{H}$, with positive mobility operator $\mathsf{M}$ on the admissible variables. It is similar to a Hermitian operator, so the minimal equilibrium model has real growth rates and no Hopf bifurcation. The marginal wavevector is set by the zero of the static Hessian. Away from marginality, the fastest wavevector also depends on polarization, reaction, and diffusion mobilities. The static film plots consequently predict neither an absolute formation time nor a unique kinetic wavelength without those inputs [50,54,80–83].

**5 Nonlinear continuation from the spatial spinodal**

A linear eigenvector specifies the initial direction of change, but does not identify the amplitude or persistence of a developed domain. We therefore tune strain to a true spatial spinodal for each film class and expand the same reduced grand potential in the critical mode. The dimensionless eigenvector is normalized by $\int_0^1 |\mathbf{v}|^2 \, d(z/h) = 1$. Writing $\delta\mathbf{p} = A\mathbf{v}\exp(iq_c x) + c.c.$, we retain the mean and second-harmonic corrections generated at quadratic order. This is the first nonlinear Galerkin reduction; direct field minimization subsequently releases the truncation.

Using functional derivatives defined with the standard Taylor factors, the amplitude energy is

$$\Delta f/f_* = \lambda|A|^2 + g|A|^4 + O(|A|^6), \tag{14a}$$

$$g = \frac{1}{4}T_4(v, v, v^*, v^*) - \frac{1}{2}t_0^\dagger H_0^{-1} t_0 - \frac{1}{4}t_2^\dagger H_{2q_c}^{-1} t_2. \tag{14b}$$

The vectors $t_0 = T_3(v, v^*)$ and $t_2 = T_3(v, v)$ belong to their respective Fourier sectors. The inverses are taken only in stable admissible subspaces. The two negative corrections express relaxation of the slaved mean and second harmonic. They explain why the quartic coefficient of a domain-forming mode cannot generally be read from a local Landau coefficient. Compatible elasticity, the changing ionic occupations, and the electrical elimination all contribute to the nonlinear derivatives. The normalization and numerical extraction are documented in Section S9.

The selected T, O, R, and W spinodals have positive coefficients, approximately 3.01, 2.42, 3.04, and 0.0561 in this convention. Their marginal periods are approximately 13.8, 14.5, 15.1, and 49.4 nm. A positive coefficient gives a local branch with $|A|^2 = -\lambda/(2g)$ on the unstable side. It does not prove that the thermodynamic transition occurs at the spinodal: a separate finite-amplitude minimum can preempt the local bifurcation. Nor does it classify every path through the multidimensional phase diagrams. The result concerns the explicitly located branches and is tested by nonlinear continuation rather than extrapolated across the entire atlas.

Figure 8 compares the fundamental Fourier amplitude of the relaxed branch with Eq. (14). Agreement close to onset provides a check that the eigenvector, quartic coefficient, and nonlinear minimization use compatible conventions. Farther from onset, the branch develops a shifted mean, higher harmonics, and a changing preferred period. The departure from the quartic curve is therefore expected and carries physical information. A single sinusoidal approximation should not be used to infer sharp-wall profiles or final domain widths in that regime.

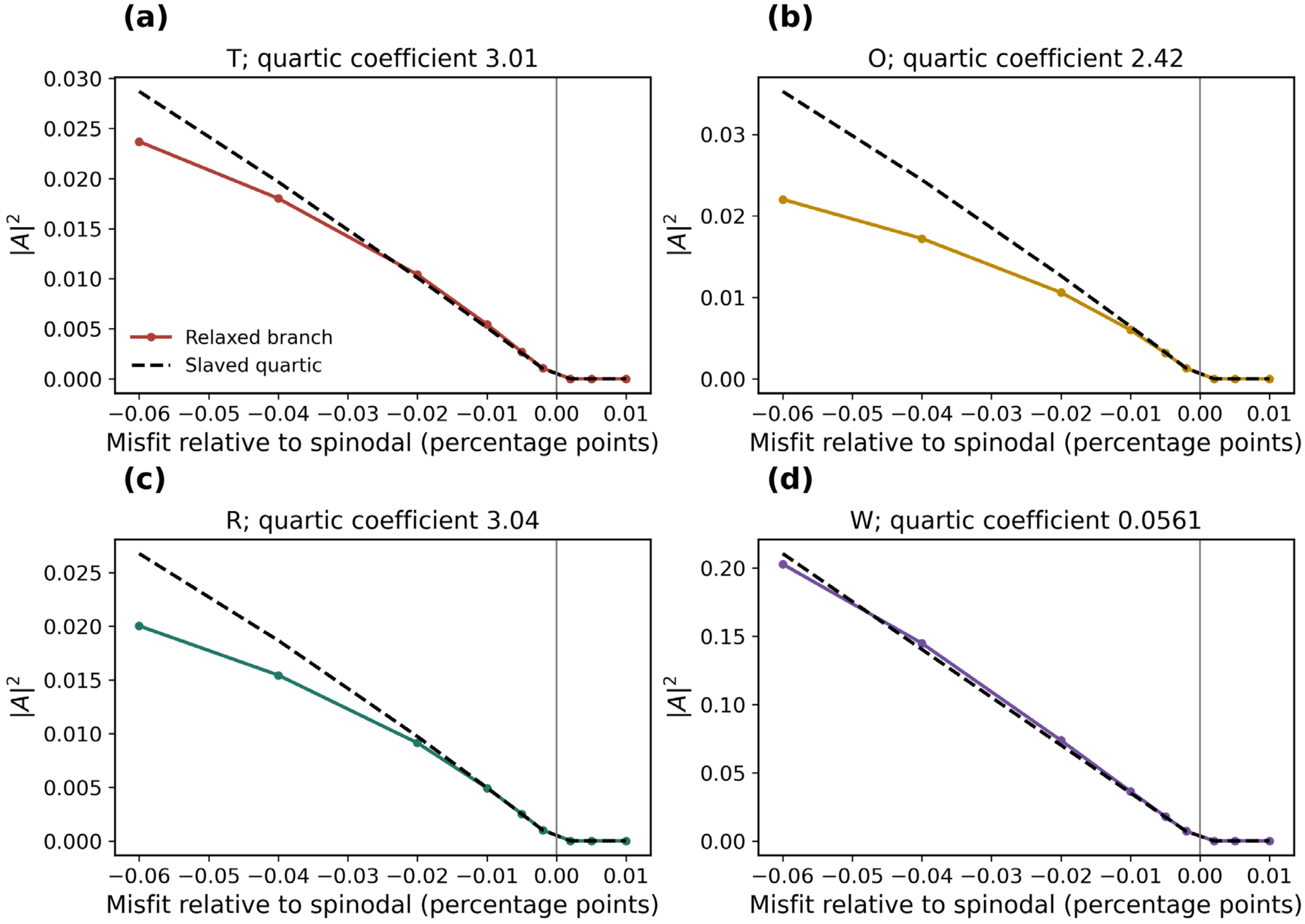

**Figure 8.** Nonlinear continuation from the selected film spinodals. (a–d) T, O, R, and W fundamental Fourier intensities compared with the slaved-quartic prediction $|A|^2 = -\lambda/(2g)$. The horizontal coordinate is misfit relative to the corresponding spinodal. The positive coefficients describe these local branches; they do not exclude a separate finite-amplitude competitor.

A second wavevector of the same magnitude need not be equally soft. At the selected spinodals, the orthogonal mode remains positively stiff. A stripe–checkerboard selection rule derived for two degenerate critical amplitudes is therefore not the controlling bifurcation here. The general two-amplitude coefficient is retained in the supplement for cases where the degeneracy conditions hold. This distinction prevents the use of crystal symmetry alone to infer a multidirectional onset pattern after the substrate, reference polarization, and chemical boundary have already selected an orientation.

Direct minimization permits all three polarization components and full depth profiles. The resulting candidates are shown in Figure 9. T, O, and R use their original reference conditions. For W, the imposed strain is changed to $-0.150\%$, where a negative spatial mode exists; the homogeneous reference and spectrum are recalculated at that same strain. Its selected wavevector lies along the other diagonal compared with the illustrated T/O/R branches. Comparing the W nonlinear candidate with the stable zero-misfit W spectrum would therefore be inconsistent.

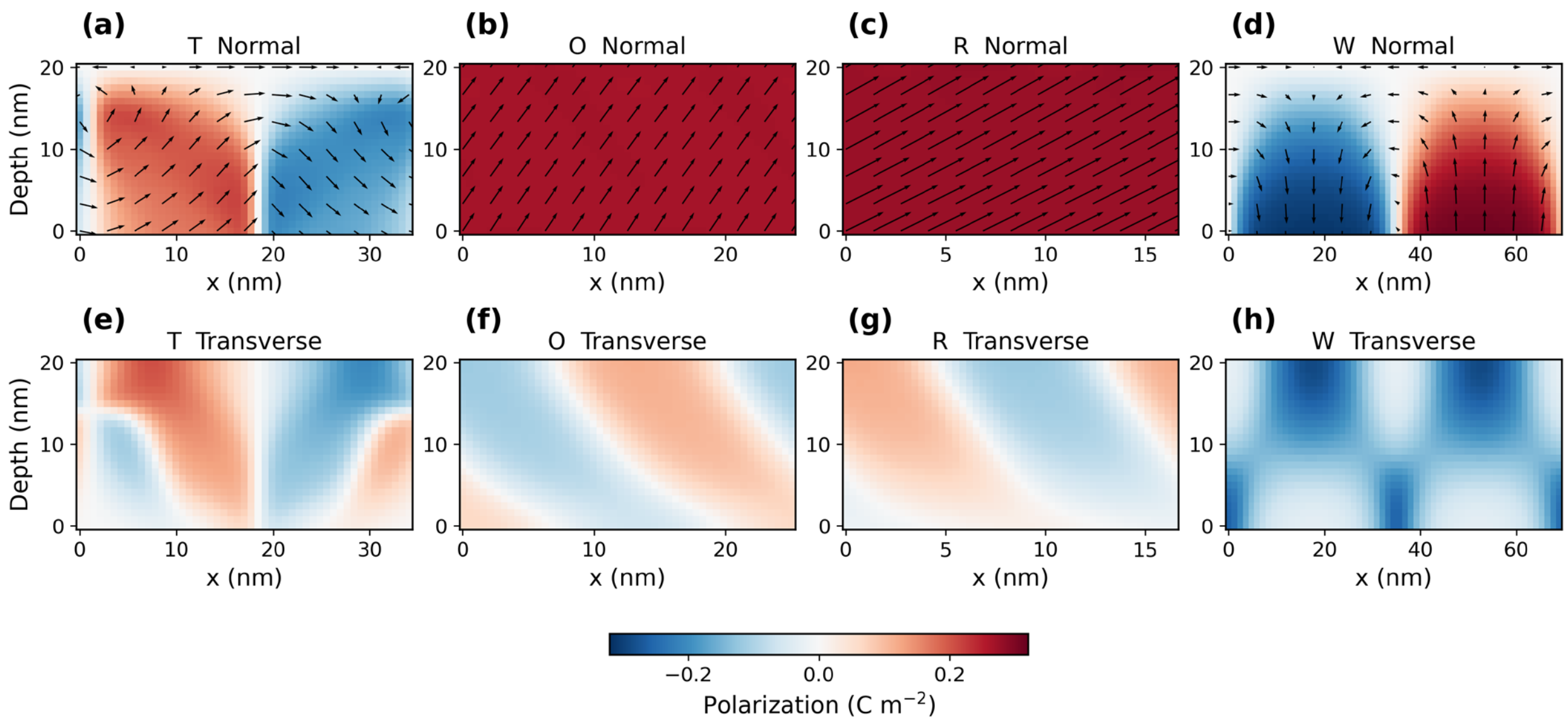

**Figure 9.** Recalculated developed periodic candidates. (a–d) Normal polarization of the T, O, R, and W stripes, with arrows showing the longitudinal–normal projection. (e–h) Their transverse in-plane components. The T reference has zero misfit, O and R have $-0.300\%$, and W has $-0.150\%$. Coordinates give the actual film thickness and cell repeat. The fields are energy-minimized candidates; their secondary stability is evaluated separately.

The T texture combines substantial normal reversal with in-plane rotation. Its lowest sampled primitive repeat is 34.7 nm. The O and R candidates primarily alternate an in-plane component while retaining a preferred sign of normal polarization; their lowest sampled repeats are 25.6 and 16.8 nm. The W repeat is 70.1 nm. Each value is an interior minimum of a refined period scan, rather than an optimum placed at an imposed box limit. Larger boxes can contain several copies of one texture. The primitive repeat is therefore checked using the full vector field and the Fourier harmonic number; neither the box length nor the strongest normal-polarization Fourier component is automatically the domain repeat. Tables in the supplement retain the unrounded values and the search intervals.

These results also distinguish a domain wall from a small-amplitude wave. Near onset, the polarization can remain within one orientation sector while its amplitude or direction varies smoothly. At larger amplitude, the trajectory may pass near several symmetry-related minima and acquire wall-like intervals. Whether that happens is decided by the nonlinear field, not by assigning a wall angle to the linear eigenvector. The field trajectory and its harmonic content are therefore retained alongside the domain images. The supplementary numerical tables give the reference states, marginal quantities, and nonlinear repeat scans separately.

## 6 Domain energetics and stability after formation

The comparison between onset and final states begins with two independent quantities: the least spatial stiffness of the homogeneous reference, $\lambda_{min}$, and the grand-potential difference between a computed pattern and that reference, $\Delta\Omega_{pat} = \Omega_{pat} - \Omega_{hom}$. Both are evaluated at the same temperature, strain, electrical boundary condition, and chemical potential. A negative $\lambda_{min}$ means that an infinitesimal admissible perturbation lowers the energy. A negative $\Delta\Omega_{pat}$ establishes that the particular computed pattern is energetically preferable to the reference. Neither statement alone proves that the pattern is stable against every perturbation or is the global minimum.

Figure 10 asks whether a modulation favored by the linear test also has a lower-energy finite-amplitude stripe descendant. At each white-dot control point, the temperature and thickness are fixed for that material class, while misfit and reservoir activity vary. The lowest-energy homogeneous state supplies the reference energy and the separate linear stability test. We then minimize the nonlinear energy from two initial polarization fields at each of three trial cell lengths. The color represents the largest energy reduction found in this search, $\Omega_{hom} - \Omega_{stripe}$, divided by the film volume. Thus, a brighter color indicates a larger energetic advantage of the computed stripe, rather than a larger polarization or a longer domain period.

The black line in each panel marks the homogeneous spatial spinodal, where the least linear stiffness changes sign. A colored region on the linearly stable side would identify a finite-amplitude domain candidate that cannot develop from infinitesimal noise; a nucleation event or another finite perturbation would be required. A colored region on the unstable side means that the uniform state can begin to modulate and that the nonlinear search has found an energetically favorable stripe. The latter still requires the developed-pattern stability test of Figure 11. Gray means only that the tested stripe family did not resolve an energy lowering. The display uses discrete calculations at the white dots; the colored cells and interpolated zero contour do not establish more precise boundaries than that sampling.

The four windows contain 180 control points and 540 cell-length conditions, with two seeds per condition. Within these windows and tolerances, the lower-energy stripe candidates occur on the spatially unstable side. We do not resolve a linearly stable interval containing a lower-energy stripe. This agreement is consistent with the positive quartic coefficients at the selected spinodals, but is established here only for the sampled nonlinear candidates. It is neither a classification of every competing texture nor a proof that a stripe reached dynamically will have the period selected by energy minimization.

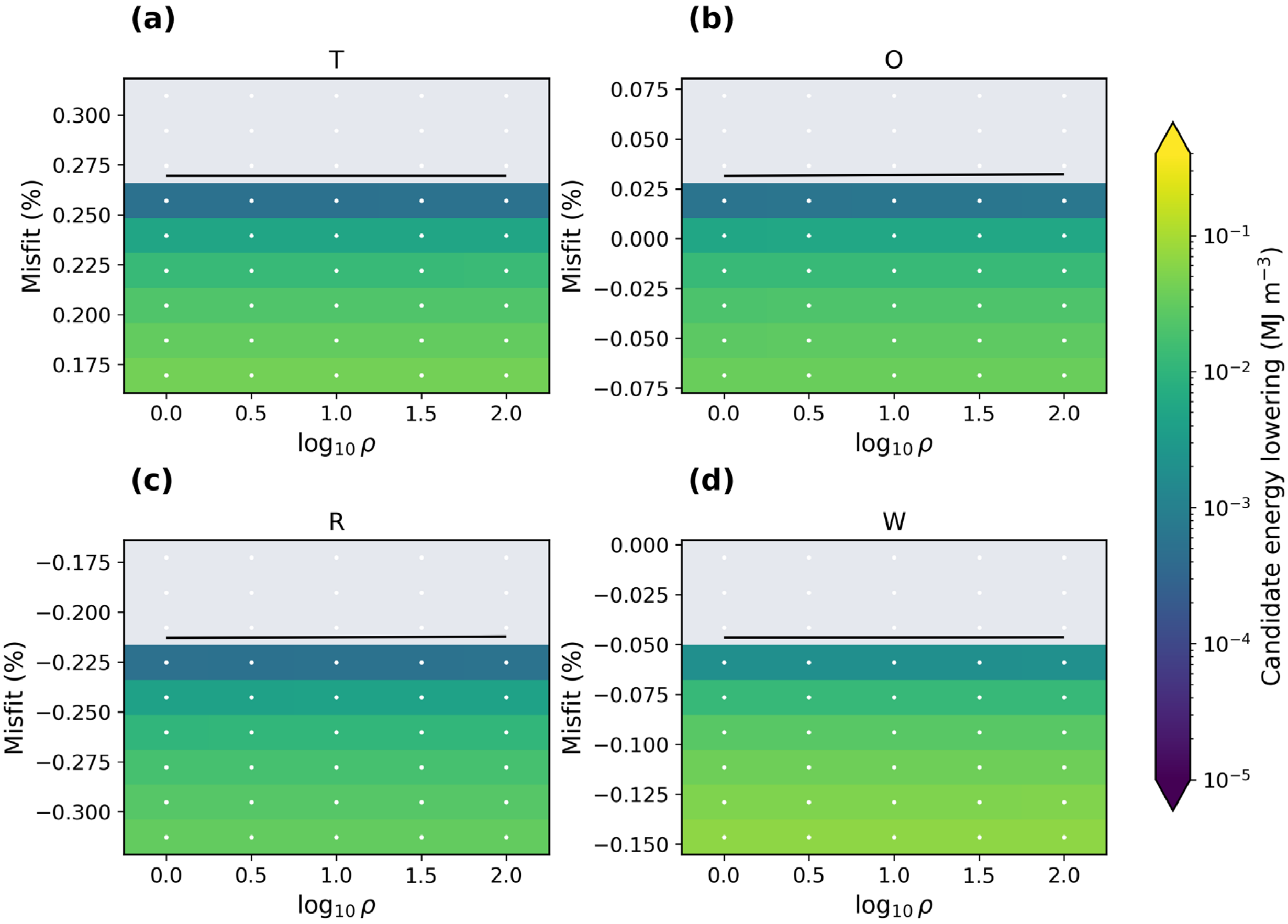

**Figure 10.** Candidate stripe energetics compared with the initial instability. (a–d) T, O, R, and W strain–activity windows. Color gives the greatest resolved energy lowering relative to the best uniform state from three tested cell lengths and two seeds per length. White dots mark calculated control points; black curves mark zero homogeneous spatial stiffness. Gray means no lower-energy stripe was found in the tested family, rather than a proof excluding all nonuniform states.

The possible mismatches have different interpretations. A lower-energy pattern with $\lambda_{min} > 0$ implies metastability of the homogeneous reference and a finite-amplitude route, such as nucleation. A negative $\lambda_{min}$ without a lower-energy result from the nonlinear search indicates that the search has not found the descending direction or its destination; it cannot demonstrate homogeneous stability. A lower-energy periodic candidate with a negative patterned Hessian is a saddle in the enlarged spatial state space. Finally, several locally stable periodic states may coexist with different energies and periods. This last situation permits memory of the preparation route even in a passive model.

Period selection introduces another distinction. Let $L_c = 2\pi/q_c$ denote the marginal onset period, $L_{stat}$ the period minimizing the static stiffness at a specified quenched reference, $L_{fast}$ the period maximizing the kinetic growth rate, and $L_{eq}$ the minimum-energy period within a computed nonlinear family. There is no general identity among these lengths. At a simple supercritical onset, the small-amplitude branch approaches $L_c$. Away from onset, nonlinear changes in the mean polarization, wall structure, elastic relaxation, and ionic occupations can move the minimum. The wavelength of the dominant Fourier peak can differ again from the primitive vector repeat when higher harmonics dominate.

The upper row of Figure 11 shows direct primitive-period scans of the developed stripe families. The energy is measured relative to the lowest sampled point of each scan, avoiding a comparison between unrelated absolute energy scales. A minimum in this curve establishes optimality with respect to the tested repeat changes within that family. It does not test a second lateral direction, changes of topology, or a competing homogeneous orientation. The original reference-spectrum minima and the separately determined marginal periods are therefore reported in the text and Tables S2–S4, rather than combined under one ambiguous label of domain size.

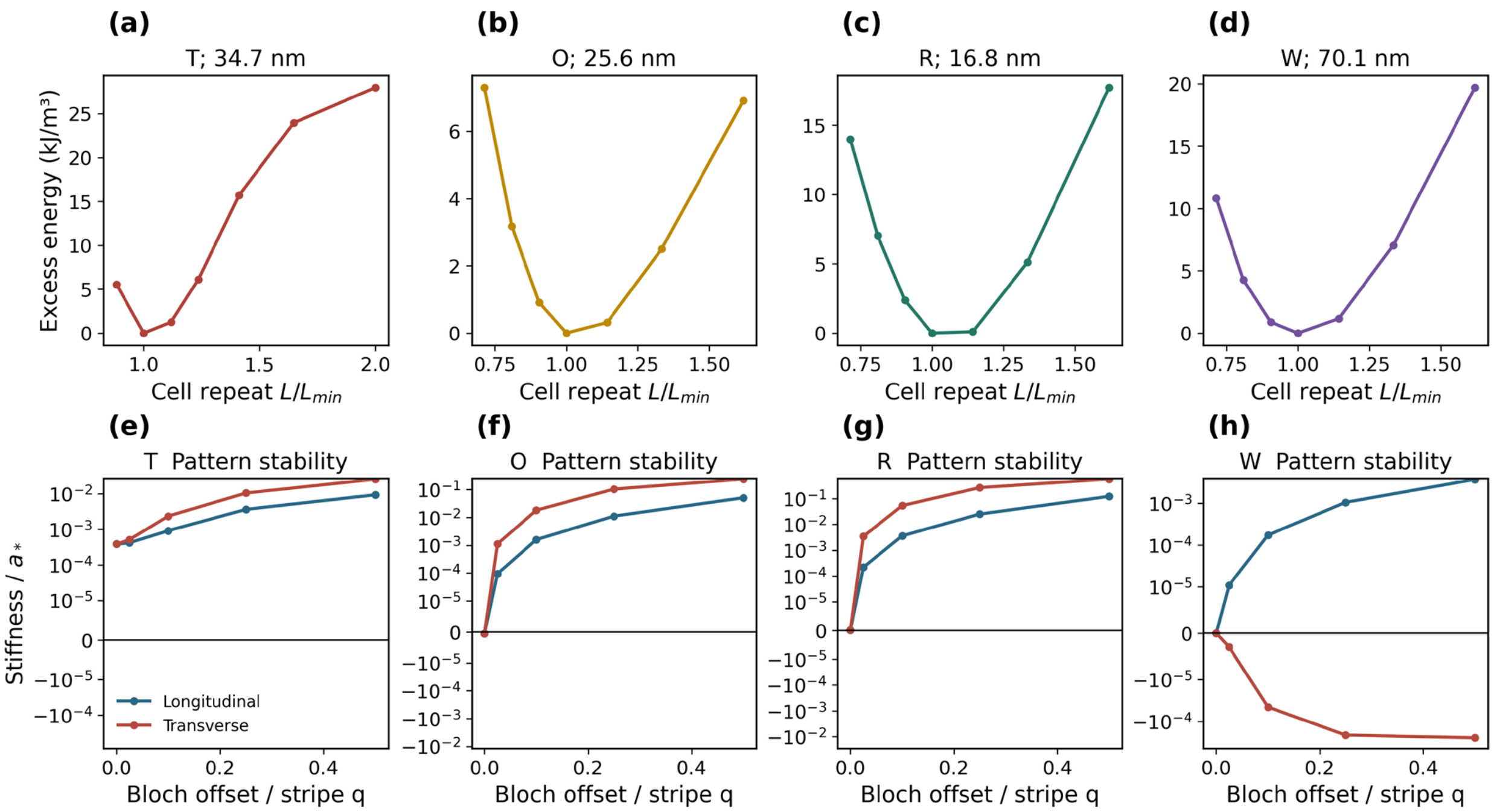

**Figure 11.** Developed-domain periods and sampled Bloch stability. (a–d) Refined primitive-repeat energy scans for T, O, R, and W, normalized by their lowest-energy sampled repeats. (e–h) Lowest Bloch eigenvalues about those developed patterns for longitudinal and transverse offsets on the refined 49-by-25 grid. Translation is neutral in the continuum; its finite-grid shift is discussed in Section S13. Nonnegative sampled points do not prove stability over the whole Bloch plane. Detuned and initial-period tests appear in Figure S27.

To test a periodic state after it forms, we evaluate Bloch perturbations,

$$\delta \mathbf{P} = e^{i(k_x x + k_y y)} \sum_m \mathbf{u}_m(z) e^{imqx}, \qquad H_{pat}(k_x, k_y)\mathbf{u} = \Lambda \mathbf{u}. \tag{15}$$

The periodic coefficients of $H_{pat}$ are evaluated on the nonlinear stripe. In particular, the chemical capacitance is now position dependent because the occupations vary across the surface. It cannot be replaced by the capacitance of the homogeneous reference. Longitudinal Bloch offsets test modulation of the spacing and phase along the stripe wavevector. Transverse offsets allow bending and redistribution into a second lateral direction. Translation supplies a neutral mode at zero offset, so small numerical eigenvalues must be assessed together with discretization error and eigensolver residuals.

The T, O, and R stripes have no resolved negative eigenvalue in the sampled nonzero Bloch sectors. Their zero-offset modes require a separate discretization audit: the O and R translation values are close to zero, while the sharper T stripe has appreciable numerical pinning that decreases on refinement. These are sampled stability results, not a proof covering the continuous Bloch plane. The W stripe has a decisive transverse instability: at one quarter of its stripe wavevector the refined stiffness is approximately $-2.14 \times 10^{-4} a_*$. Thus a true homogeneous instability can produce a lower-energy stripe that is itself unstable when a second lateral direction is admitted.

Releasing the second lateral direction permits a change in morphology as well as a change in repeat. Figure 12(a–c) shows a new minimization on a $1000 \times 1000$ nm square at the same W reference conditions. The initial straight reference contains fourteen primitive vector repeats of 71.4 nm; we continue its transverse unstable mode to a nonlinear periodic seed and then minimize on the full square with a weak broadband perturbation. The computed descendant is lower in energy by 1.4 kJ $\mathrm{m}^{-3}$ than the straight reference evaluated on the same 462-by-132 lateral grid and 9 polarization depth nodes. The surface-normal component is much smaller than the in-plane components; the separate scale in panel (c) resolves its spatial modulation. This does not imply that normal polarization is small throughout the film thickness. The energy decrease demonstrates an accessible descending pathway in the enlarged field of view, without identifying the finite-cell texture as the global minimum. Section S17 specifies the solver, residuals, and resolution check. The surface-ordering example in Figure 12(d–f), introduced in Section 7, provides a second route from noise to a multidirectional modulation.

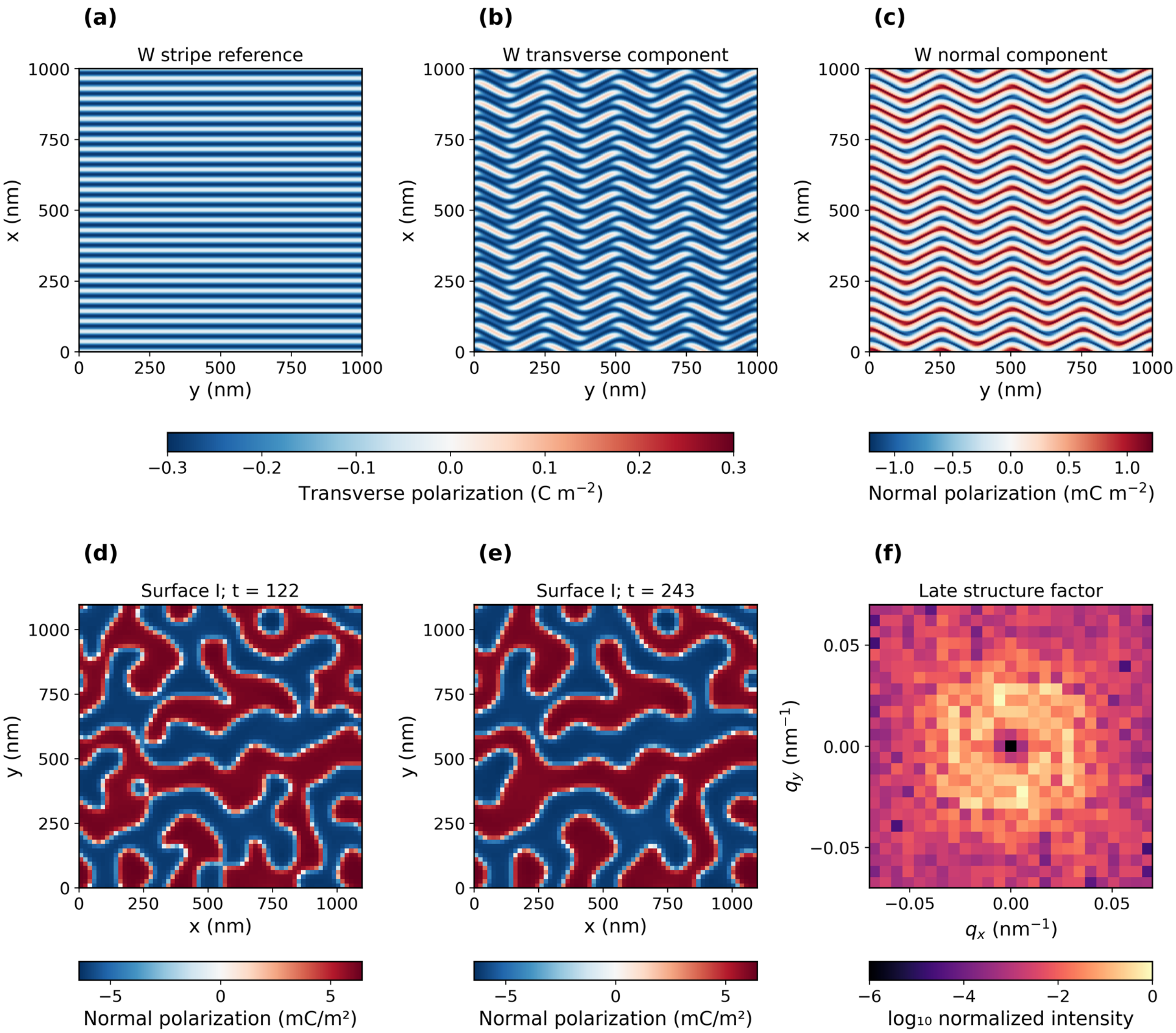

**Figure 12.** Two-dimensional polarization evolution in the two constitutive limits. (a–c) Straight W stripe reference and the transverse and normal polarization of its computed descendant in the newly calculated 1000-by-1000 nm square, with equal spatial aspect ratio. All three upper panels show the film surface. Panels (a,b) share a transverse-polarization scale; (c) uses a separate scale in millicoulombs per square meter to resolve the much smaller normal surface component. The upper row uses 462-by-132 lateral points and 9 polarization depth nodes in a physically square domain; every lateral Fourier component is free during minimization. (d,e) Normal surface polarization during noise-driven ordering of surface model I at 400 K and $\Delta a_s = -5\ \mathrm{m}^2\ \mathrm{F}^{-1}$ in the 300 nm Gaussian-bulk film. Its retained square has side 1097 nm. (f) The corresponding late structure factor. Both models use an exposed upper surface and grounded bottom electrode. The polar constitutive laws and colorbar units differ between the rows.

The initial-period comparison in Figure S27 adds a further distinction. The developed T stripe at the least-stiff reference period, about 15.7 nm, has a negative longitudinal mode, whereas its longer energy-preferred candidate passes the sampled nonzero-offset tests. The O and R initial-period candidates lie closer to their nonlinear minima and pass the corresponding sampled tests. For W, secondary stability changes with period: the tested initial-period candidate can remain locally stable although a different stripe repeat has lower energy and a transverse instability. A lower energy therefore does not imply an uninterrupted deterministic route from the initial pattern to that stripe.

Persistence consequently requires less than exact equality of $L_{fast}$ and $L_{eq}$, but more than negative homogeneous stiffness. An initial period lying inside a stable band can remain locally stable even if another period has slightly lower energy. It can then survive until defects, noise, boundaries, or a sufficiently large perturbation permit a change in repeat number. An initial period outside the stable band should instead undergo a secondary adjustment. In a multiaxial material this adjustment can involve rotation or transverse modulation, not only the creation or annihilation of reversal walls. The appropriate comparison is therefore between the initial mode, the nonlinear energy landscape, and the stability band of its descendants.

## 7 Surface ordering with a weakly anisotropic bulk

Rotational softness does not, by itself, favor a modulation normal to a surface. A nearly isotropic polar material can reduce its electrical energy by turning into the plane. The possibility of a surface pattern on a 100–300 nm scale therefore requires a separate examination of the normal ordering tendency, the electrical boundary condition, and the nonlinear competitors. We consider a stable Gaussian vector bulk together with an explicit normal surface preference. The upper surface is exposed to an unbounded dielectric, and a grounded electrode remains at the bottom of a 300 nm film. The adsorption free energy, site density, formation energies, charge numbers, and reservoir stoichiometry are exactly those of Sections 2–6. Net surface charge is allowed to change by exchange with the reservoir. In particular, a uniformly charged state is included in the thermodynamic comparison.

The positive bulk free energy is $[a|\mathbf{P}|^2 + G|\nabla\mathbf{P}|^2]/2$, with $a = 10^7$ m $\mathrm{F}^{-1}$, $G = 10^{-8}$ J $\mathrm{m}^3$ $\mathrm{C}^{-2}$, and $\epsilon_b = 7.35$. Its polarization correlation length is $\xi = \sqrt{G/a} = 31.6$ nm. The surface polarization scale is $P_s^* = 0.0500$ C $\mathrm{m}^{-2}$ and the surface energy scale is $E_s^* = 0.00250$ J $\mathrm{m}^{-2}$, giving $K_s^* = E_s^*/(P_s^*)^2 = 1$ $\mathrm{m}^2$ $\mathrm{F}^{-1}$. These scales differ from the ferroelectric-film normalization. The Gaussian bulk is a controlled description of a stable, soft susceptibility; it is not a microscopic model of polar nanoregions, random fields, or frequency dispersion in a named relaxor [70–75]. The surface ordering term is the source of nonconvexity in this limit.

We write the normal surface coefficient as

$$a_{s,n}(T) = [-3.18 + 0.0200(T - 300\,\mathrm{K})/\mathrm{K}]\,\mathrm{m}^2\mathrm{F}^{-1} + \Delta a_s. \tag{16a}$$

The independently varied parameter $\Delta a_s$ measures additional normal preference. The tangential quadratic surface coefficient is zero. At $\Delta a_s = 0$, the common chemical model does not give a linear normal-ordering instability over the tested 200–500 K interval. This negative result matters: retaining a previously obtained wavelength while changing the chemical free energy would not constitute a prediction of the revised model. We therefore calculate the required normal coefficient directly, rather than adjusting an ionic capacitance to reproduce a desired period.

Three nonlinear completions share the same quadratic surface energy. With $\mathbf{p}_s = \mathbf{P}_s/P_s^*$, their dimensionless surface energies are

$$\frac{f_s}{E_s^*} = \frac{a_{s,n}}{2K_s^*} p_{s,z}^2 + \frac{B}{4}|\mathbf{p}_s|^4 + \frac{V}{6}|\mathbf{p}_s|^6 + \frac{B_t}{4}\left|\mathbf{p}_{s,\parallel}\right|^4, \tag{16b}$$

where $(B, V, B_t)$ is $(100,1,0)$ for I, $(1,1,0)$ for II, and $(100,1,-102)$ for III. Completion I limits the amplitude strongly; II permits larger amplitudes; III supplies a finite-amplitude in-plane competitor. The latter has tangential quartic coefficient $-1/2$ in the energy, with the positive sixth-order term ensuring boundedness. These are explicitly different nonlinear surface constitutive laws. The chemical law and electrical geometry are common to all three. Consequently, their identical linear spectrum about the unpolarized reference cannot determine which nonlinear state has the lowest energy.

The Gaussian bulk and exterior fields can be eliminated exactly for prescribed surface polarization and ionic charge. Let $\mathbf{u}_s = (P_{s,x}, P_{s,y}, P_{s,z}, \sigma)$. Their contribution is $\frac{1}{2}\sum_{\mathbf{q}} \mathbf{u}_s^\dagger K(\mathbf{q}; h)\mathbf{u}_s$. The matrix is Hermitian and retains the imaginary coupling between longitudinal polarization and the other charged variables. For nonzero wavevector the depth solution contains electrical, transverse-polarization, and longitudinal-polarization inverse lengths $q$, $\sqrt{q^2 + a/G}$, and $\sqrt{q^2 + (a + 1/\epsilon_f)/G}$. Both signs of each exponent are required in a finite film; bottom and top boundary conditions fix their coefficients. Section S14 gives the corresponding six-amplitude boundary problem and its uniform limit.

The zero-wavevector sector is retained explicitly. It describes uniform charge exchange balanced by the grounded electrode, and cannot be removed by a projection on zero mean charge. Chemical equilibration at a specified local charge minimizes the common-site entropy over the two occupations subject to that charge. Subsequent minimization over charge gives precisely the competitive Langmuir law used in the film calculation. This two-stage reduction is useful numerically, but introduces neither an additional ionic species nor a different chemical ensemble. It also allows the nonlinear local capacitance to change across a domain pattern.

At balanced activity, the unpolarized state has $\theta_+ = \theta_-$ and zero net charge. Eliminating its stable tangential and charge channels defines a normal stiffness $\Lambda_n(q) = a_{s,n} + \Gamma(q, T)$. The critical coefficient and marginal period follow from

$$a_{s,n}^c(T) = -\min_{q\geq 0}\Gamma(q, T), \qquad L_c(T) = 2\pi/q_c, \qquad q_c \in \arg\min\Gamma. \tag{17}$$

For the stated parameters, the required normal coefficients are approximately $-9.73$, $-5.52$, and $-3.28$ m$^2$ F$^{-1}$ at 300, 400, and 500 K. Their marginal periods are approximately 64, 272, and 257 nm, respectively. Thus, a wavelength in the proposed experimental interval is accessible, but only together with a sufficiently strong normal preference. The 31.6 nm bulk correlation length alone neither fixes that wavelength nor establishes the instability. Temperature changes the ionic susceptibility substantially because the formation energy is held at 0.200 eV throughout.

Figure 13 presents the nonlinear candidate maps on the temperature–normal-preference plane. The calculation minimizes a vector first-harmonic surface field, its uniform component, and the period, and compares its energy with separately minimized uniform normal and in-plane states. All ionic occupations remain physically bounded. These competitors represent the main ways to avoid or accommodate normal surface charge. The unpolarized state avoids the cost of polar order, uniform normal polarization benefits from the negative normal surface coefficient while requiring net compensation, and a uniform in-plane state avoids normal bound charge. A periodic vector state can combine the normal ordering preference with spatial charge and polarization relaxation. Comparing all four is essential because a normal instability alone does not establish that the final state remains periodic. The retained family does not exhaust possible two-dimensional textures. The upper row identifies the lowest energy among these candidates. The lower row relates that choice to the unpolarized normal-mode spinodal. Yellow marks periodic candidates below every tested uniform competitor while the unpolarized state remains linearly stable. Purple marks normal onset with a lower-energy in-plane competitor. The dashed boundary is common because Eq. (16b) changes no quadratic coefficient. The solid boundary is the resolved crossing of candidate energies, rather than a separately proven global coexistence curve.

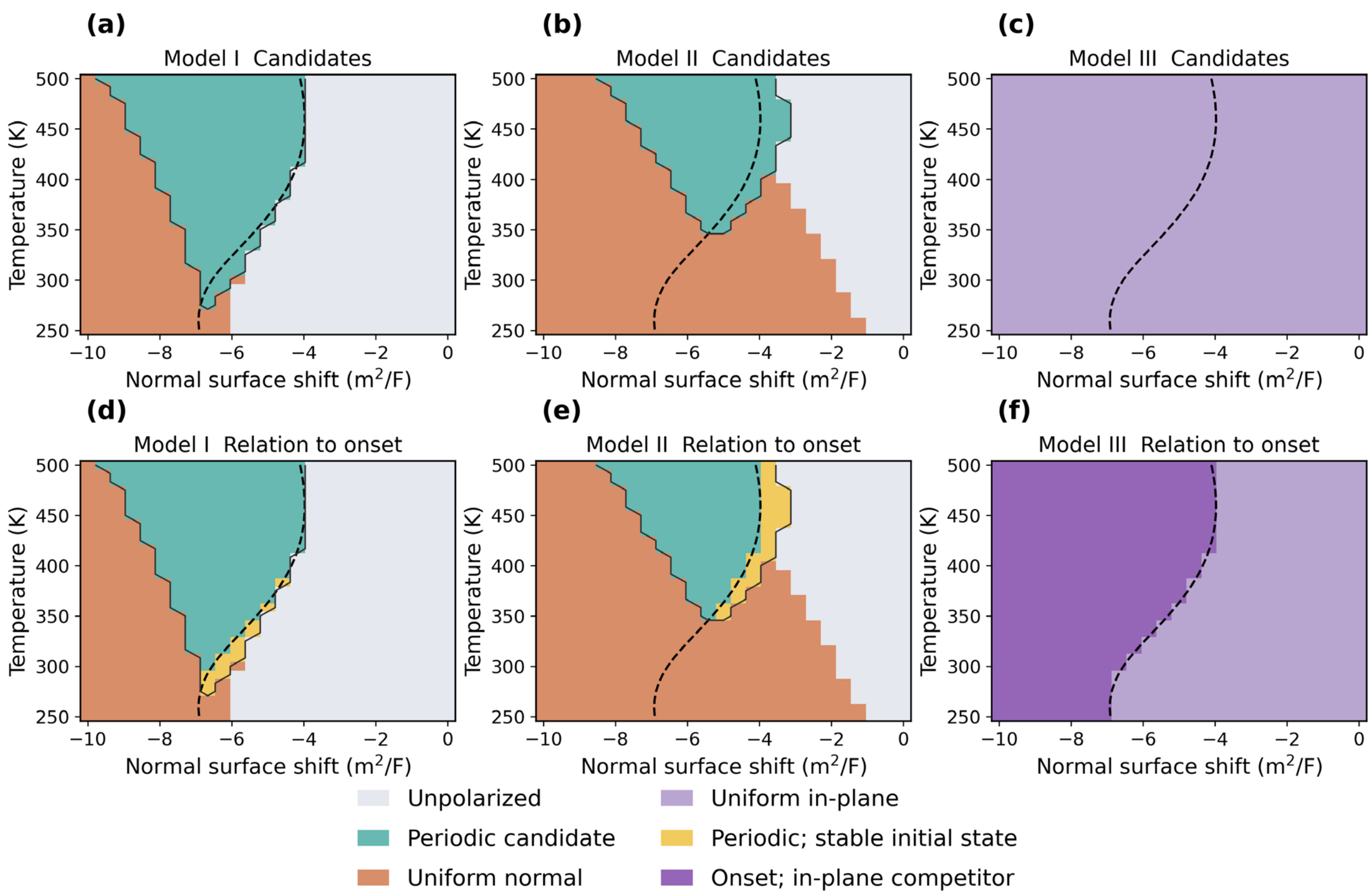

**Figure 13.** Surface-ordering candidate maps with the common chemical model. (a–c) Lowest energy among uniform competitors and the optimized vector first-harmonic family for surface models I, II, and III. (d–f) Their relation to the common unpolarized normal-mode spinodal. Yellow identifies a lower-energy periodic candidate while the initial reference remains linearly stable. Purple identifies normal onset with a lower-energy in-plane competitor. Dashed curves are the common spinodal; solid curves are resolved candidate-energy boundaries. The mean polarization and ionic charge are free. Higher-harmonic refinements and Bloch tests in Section S14 limit the interpretation of these candidate boundaries.

The full-field calculations in Section S14 release the first-harmonic restriction. They show why the truncation must be stated. At the illustrative shift $\Delta a_s = -5\ \mathrm{m^2\ F^{-1}}$ and 400 K, surface model I has the lowest sampled periodic repeat near 312 nm, compared with a linear static scale near 272 nm. Completion II prefers a sampled repeat near 369 nm, while a finite-amplitude periodic state is also lower than its uniform competitors at 350 K, where infinitesimal disturbances of the unpolarized state decay. Completion III relaxes to its uniform in-plane competitor for the representative initializations. The first-harmonic map can miss a narrow finite-amplitude branch: surface model I at 350 K supplies such an example after higher harmonics are released. Accordingly, the atlas is labeled as a candidate map, and its limitations are directly tested rather than hidden by assigning a single equilibrium label to every point.

These examples separate three questions that a measured spacing cannot answer simultaneously. The initial susceptibility selects a marginal or growing mode. Nonlinear polarization and adsorption determine whether a lower-energy patterned state exists and which repeat minimizes its energy within a chosen family. Bloch perturbations determine whether a particular developed repeat survives disturbances of longer wavelength or another lateral orientation. Full-period energy and Bloch calculations are provided in Figures S24 and S25. Several energy-preferred surface stripes retain negative transverse Bloch modes on the refined grid. Their optimized stripe periods therefore do not establish the spacing of a fully stable two-dimensional equilibrium. The selected repeats are interior points of the sampled energy scans for nonuniform candidates; the apparent period of a solution that has become uniform has no physical meaning.

The origin of the normal preference remains a constitutive question. Broken surface symmetry allows different normal and tangential quadratic coefficients. Orientation-dependent bonding, reconstruction, or a modified near-surface structural order parameter can generate this difference [37–43,59]. A localized stress changes the polar quadratic energy through $-\sigma_{ij}Q_{ijkl}P_kP_l$; integrating a modified layer through its depth gives an effective surface coefficient. Strain gradients and flexoelectricity provide further contributions whose reduction depends on the mechanical and polarization boundary conditions [61–63,84]. Adsorption can change bonding and therefore the surface anisotropy, but that interaction is additional to the electrostatic coupling retained here. An isotropic competitive Langmuir entropy does not itself supply a freely adjustable negative normal coefficient.

Equation (17) converts this uncertainty into a measurable requirement. A calculation or measurement of the surface anisotropy can be compared with the threshold, while the same adsorption parameters predict the associated charge response. The shift used for the representative patterns is a transparent sensitivity parameter, not an inferred property of a particular relaxor. The resulting 100–300 nm scale is therefore a conditional ferroionic mechanism. It is not evidence that every surface pattern of that size has the same origin. Local relaxation measurements have resolved mesoscopic dynamic heterogeneities on related length scales [85], but spatial heterogeneity alone does not demonstrate a periodic spinodal.

The relaxation calculation uses the same empty, positive, and negative site states. Forward and backward surface reaction rates satisfy the equilibrium Langmuir ratio, and polarization follows dissipative relaxation of the same grand potential. There is no artificial subtraction of the mean ionic charge. Figure 12(d–f) shows growth from noise in a two-dimensional exposed surface, and Figure S26 compares noise, finite-amplitude, and in-plane initializations. Time is reduced by an unspecified polarization mobility; the calculation predicts neither seconds nor an experimentally calibrated relaxation spectrum. A finite-wavevector peak can resemble spinodal decomposition, but polarization is nonconserved and the surface exchanges charge with a reservoir. The conservation law is consequently different from that of a closed binary mixture [80–83].

The distinction between thermodynamic and kinetic wavelengths persists in this surface limit. Finite reaction rates change the fastest growing wavevector without changing the equilibrium instability threshold. A slow chemical channel can delay charge compensation and redistribute the growth among polar modes. Under reciprocal passive dynamics all linear rates remain real, so a lag between ionic and polarization responses does not by itself imply an oscillatory instability. The full chemical entropy and reversible reaction law are retained in the kinetic Hessian; replacing them by a fitted capacitance would lose the occupation constraints and would prevent a consistent comparison with the nonlinear equilibria.

**8 Applicability and experimental comparison**

The calculations connect several levels of description, but each has a defined scope. Homogeneous maps compare fully relaxed uniform candidates. Linear maps test infinitesimal perturbations of those candidates over a sampled function space. Nonlinear stripe maps compare a finite set of relaxed periodic solutions. Bloch calculations test selected perturbations of the developed stripes. Two-dimensional relaxations establish additional accessible states in finite cells. Calling all of these results phase diagrams without specifying their state space would obscure the central distinction of the work. Their agreement is meaningful only when the ensemble, reference state, and constitutive parameters are held consistent.

The ionic-capacity audit is particularly important for the $BiFeO_3$ example. At 300 K, 20 nm, and $u_m = -0.100\%$, the model selects $P_z \approx 0.330$ C m$^{-2}$ while the available single-sign surface charge is approximately 0.320 C m$^{-2}$. The calculated surface potential is about 3.06 V, corresponding to an average film field of about 153 MV m$^{-1}$. This is the exposed-surface film field; there is no gap field in the revised geometry. Ionic saturation gives negligible differential screening. Electronic charge transfer, band bending, and additional structural response must be assessed before this constitutive reference can be treated as a physical prediction. The common-site-density sweep in Figure S21 quantifies the sensitivity to available capacity. Carrier screening cannot generally be absorbed into a constant background permittivity [51–52,86–88].

Gradient anisotropy affects direction selection, wall width, and absolute wavelength. The isotropic baseline is useful for separating angular and mechanical effects, but is not a universal material property. A positive depth-to-lateral gradient ratio is varied in the supplement, without introducing a negative gradient stiffness. Flexoelectricity requires similar care: in a freely relaxing scalar example, eliminating strain from $C\epsilon^2/2 + fP\epsilon' + G(P')^2/2$ gives an effective coefficient $G - f^2/C$. In a clamped vector film, the correction is nonlocal and tensorial, and surface terms must be handled consistently [61,84]. An inferred small gradient coefficient can therefore reflect a particular mechanical reduction; it should not be transferred unchanged between geometries.

The absence of explicit octahedral rotations is another limitation of the $BiFeO_3$ transfer. Coupled structural order can modify polarization anisotropy and wall properties, while strain and electrostatic fields can affect local conduction [69,89–92]. Likewise, the W model omits the random fields, spatially distributed local preferences, and broad relaxation spectrum that characterize real relaxors [70–75]. It tests one consequence of a small angular stiffness. Polarization rotation is well established as an important response mechanism in soft ferroelectric landscapes [22,93], but rotational softness alone does not identify a material as a relaxor or determine the sign of its surface anisotropy.

Experimental tests should distinguish the initial mode from the developed pattern. Time-resolved vector-sensitive measurements can determine whether the first modulation is primarily normal reversal, in-plane rotation, or a mixed motion. Surface-potential imaging can test the predicted overlap with the electrical channel, including modes with weak potential contrast.

Piezoresponse imaging and local switching methods provide relevant observables, but their electromechanical and electrostatic contributions must be interpreted consistently [50,94–98]. A domain image, a surface-potential image, and a topographic pattern need not represent the same field. Measurements should therefore report how the contrast is connected to polarization and ionic redistribution.

Controlled activity changes and strain variations provide complementary tests. Activity affects both branch selection and differential screening; strain changes the compatible rotational pathways. Their combined variation can distinguish an electrically active normal mode from an almost charge-neutral ferroelastic mode. Initial wavelength, primitive repeat, dominant Fourier wavelength, and their evolution should be reported separately. A pattern that retains its initial spacing may lie within a stable band without being the minimum-energy member of that band. A later change of spacing or morphology can indicate secondary instability, while persistence under a linearly stable initial condition suggests a finite-amplitude preparation route.

The calculations preserve these distinctions in both the manuscript and the accompanying code. The numerical maps, spectra, branches, and images are generated from constitutive inputs. Section S16 compares the exposed surface with a dielectric layer terminated by a top electrode and explains why electrically neutral rotational modes and normal surface ordering have different sensitivities. Section S17 documents the square-domain calculation and the activity-sweep movies. The notebook contains readable source cells followed by separate main-paper and supplementary calculation sections. It has no saved outputs, encoded internal package, supplied numerical arrays, or dependency on an external source archive. Running the numerical stages creates intermediate result files within the session; figure cells read those freshly computed results. Grid sizes, wavevector sampling, period searches, and solver tolerances remain explicit. An independent verification stage checks site balance, electrostatic charge balance, the matched scalar limit, analytical kernels, energy gradients, Hessian actions, and nonlinear-to-linear consistency before the extensive calculations are started.

## 9 Conclusions

Surface electrochemistry controls multiaxial domain formation through selection of the stationary polarization–strain state and through relaxation of fluctuation charge about that state. A common variational formulation separates these effects and retains compatible elasticity, vector rotation, and depth-dependent electrical fields. The homogeneous phase atlases, including a unified $BaTiO_3$ temperature interval, identify the reference branches. Their spatial Hessians then determine the admissible initial instabilities. At a fixed reference, positive chemical capacitance gives a negative semidefinite correction to the frozen-ion stiffness; its effect along an equilibrium branch also includes changes of polarization and stress.

Nonlinear continuation from four selected film spinodals gives positive slaved quartic coefficients and locally continuous branches. The developed repeats need not equal the marginal wavelength or the fixed-reference spectral minimum. Primitive-period energy scans and stability tests about the developed patterns are therefore separate parts of the analysis. For a stable, weakly anisotropic bulk, the common adsorption law does not by itself produce the previously presumed surface-ordering instability: a sufficiently strong normal surface preference is required. Its calculated threshold admits conditional wavelengths of order 100–300 nm, while different nonlinear surface constitutive laws with the same onset spectrum give different candidate thermodynamics. An initial instability specifies a route away from a homogeneous state; persistence depends on the energy and stability of the structures reached along that route.

### Data and calculation availability

The accompanying self-contained Colab notebook contains all model and solver code and separates main-paper and supplementary calculations. It is delivered without execution outputs and calculates its numerical arrays from the equations. The source archive additionally preserves the parameter audit, freshly calculated results, figure scripts, and verification records used for this manuscript. No experimental images are synthesized, no old phase labels are reused, and no numerical phase boundary is reconstructed from a raster figure.

### Author and AI contributions

Sergei V. Kalinin conceived the scientific problem, specified the material classes and physical boundary conditions, directed the adoption of a common surface-chemical model, and formulated the comparison between initial polarization instabilities and the thermodynamics and stability of developed domain structures. He introduced the extension to weakly anisotropic materials and relaxor-like surface ordering, guided the physical interpretation, and critically reviewed and revised the analysis and manuscript. OpenAI ChatGPT/Codex assisted with literature organization, developed analytical derivations under the author's direction, implemented and executed the numerical calculations, generated phase diagrams and figures, performed numerical consistency and resolution checks, and prepared and revised the manuscript, supplementary material, and reproducible notebooks. Fable 5.1 was used as an independent checker of selected analytical derivations and manuscript consistency. The author retains responsibility for the physical assumptions, interpretation, scientific conclusions, and final manuscript.

### Acknowledgments

This work (S.V.K.) was supported by the DOE BES project DE-SC0026253, "Deciphering Electrochemical Transformation Pathways on the Nanometer Scale: Advancing Fundamental Discovery for Material Innovation."

## Supplementary material

### Scope and calculation sequence

This supplement specifies the functional, ensemble, analytical reductions, material coefficients, numerical discretization, and recalculated results used in the main paper. The sequence is: choose the constitutive energy and reservoir; minimize homogeneous candidates; calculate their full spatial Hessian; continue a critical mode to finite amplitude; optimize a periodic family; test that family against Bloch perturbations; and, where indicated, release a second lateral direction. Every comparison retains the same mechanical and electrical boundaries. The figures distinguish a homogeneous component map, a linear-instability map, a nonlinear candidate-energy map, and a stability test about a developed pattern. These objects have different meanings even when their boundaries nearly coincide.

The revised electrical problem has a grounded bottom electrode and an exposed upper surface adjoining an unbounded dielectric. The ionic model has one common population of sites, with empty, positive, and negative states. The same model is retained in the surface-ordering and reaction-kinetic calculations. All numerical figures in this document were regenerated from these assumptions. A standalone notebook contains visible solver code and computes all intermediate arrays. The accompanying archive preserves the results for numerical inspection, but the notebook does not require them as input.

### S1 Geometry, variables, and thermodynamic ensemble

The ferroelectric occupies $0 < z < h$ and the exterior dielectric occupies $z > h$. The bottom electrode sets $\phi(0) = 0$. We write $\mathbf{r} = (\mathbf{r}_\parallel, z)$, $\mathbf{E} = -\nabla\phi$, and $\mathbf{D}_f = \epsilon_f \mathbf{E} + \mathbf{P}$, with $\epsilon_f = \epsilon_0 \epsilon_b$. In the exterior $\mathbf{D}_e = \epsilon_e \mathbf{E}$. The background permittivity excludes the soft polarization represented explicitly by $\mathbf{P}$. Adding a measured low-frequency ferroelectric permittivity to that background would double count the soft response. No mobile volume charge is included in the baseline film. Surface adsorbates are distinct from the bulk-defect fields considered in a different model.

For a lateral mode $\exp(i\mathbf{q} \cdot \mathbf{r}_\parallel)$ with $q > 0$, the exterior solution is $\phi_e(z) = \psi \exp[-q(z - h)]$. At $q = 0$, its field vanishes and its potential is the constant $\psi$. The latter is not set independently to zero. Imposing both $\phi(0) = 0$ and a zero exterior potential on this uniform sector would describe a different electrical constraint. The grounded electrode supplies the countercharge for a nonzero mean surface charge. An open upper surface is consequently compatible with reservoir exchange of net ionic charge.

The imposed coherent strains are $e_{xx} = e_{yy} = u_m$ and $2e_{xy} = 0$ for a uniform state. In a nonuniform state the perturbation displacement is zero at the substrate and free at the top, with zero surface traction. Natural polarization conditions $G\, \partial_z P_i = 0$ apply on both faces in the baseline isotropic-gradient film. A specified polar surface energy modifies the corresponding condition by its derivative. The mechanical model treats the substrate as rigid; it does not silently integrate out a finite substrate whose elastic constants are unspecified.

Temperature, reservoir activity, site density, standard formation energies, and bottom electrode potential are fixed in each energy comparison. Adsorption can change both ionic occupations and their spatial averages. Polarization is not conserved. The equilibrium ensemble is therefore grand canonical with respect to surface reactions. Introducing a zero-average surface-charge constraint would add a Lagrange multiplier and change both the uniform competitors and the nonlinear energies. No such constraint is used here.

The labels T, O, and R identify the stress-free $BaTiO_3$ parent landscapes at 300, 230, and 150 K. They do not impose the corresponding bulk direction on the film. The labels $c$, $a$, $aa$, $ac$, and $r$ on a map identify zero and nonzero Cartesian components of a calculated homogeneous vector. In particular, $r$ denotes a tilted diagonal component pattern, not proof of the crystallographic space group of an unconstrained crystal. W denotes a small angular anisotropy in an explicitly stated polynomial. Reduced electrostriction is a second, separately controlled assumption.

## S2 Common-site chemical free energy and variational equations

### S2.1 Cubic polar energy and strain convention

The cubic Landau polynomial is written without hidden factorials:

$$f_L = a_1 \sum_i P_i^2 + a_{11} \sum_i P_i^4 + a_{12} \sum_{i<j} P_i^2 P_j^2 + a_{111} \sum_i P_i^6 + a_{112} \sum_{i\neq j} P_i^4 P_j^2 + a_{123} P_x^2 P_y^2 P_z^2 + f_8, \quad \text{(S1)}$$

$$f_8 = a_{1111} \sum_i P_i^8 + a_{1112} \sum_{i\neq j} P_i^6 P_j^2 + a_{1122} \sum_{i<j} P_i^4 P_j^4 + a_{1123} P_x^2 P_y^2 P_z^2 \sum_i P_i^2. \quad \text{(S2)}$$

The sums over unequal indices are ordered; the sums over $i<j$ are unordered. These conventions matter for transferring published coefficients. The baseline gradient term is $G \sum_{ij} \left(\partial_j P_i\right)^2 /2$. The elastic energy is $f_{el} = (\mathbf{e} - \mathbf{e}^0)^T C (\mathbf{e} - \mathbf{e}^0)/2$, where the engineering-strain vector is $\mathbf{e} = \left(e_{xx}, e_{yy}, e_{zz}, 2e_{yz}, 2e_{xz}, 2e_{xy}\right)$. Normal eigenstrains are $e_{ii}^0 = Q_{11} P_i^2 + Q_{12} \sum_{j\neq i} P_j^2$, and engineering shears are $e_4^0 = Q_{44} P_y P_z$, $e_5^0 = Q_{44} P_x P_z$, and $e_6^0 = Q_{44} P_x P_y$. Using tensor shear in one term and engineering shear in another changes the compatibility energy.

### S2.2 Competitive adsorption and its domain

One site has charge 0, $q_+$, or $q_-$. Its fractions satisfy $\theta_a \geq 0$ and $\sum_{a=0,+,-} \theta_a = 1$. The chemical grand-potential density is

$$g = N_s\left[\theta_+ \Delta g_+ + \theta_- \Delta g_- + k_B T \sum_{a=0,+,-} \theta_a \ln\theta_a\right], \qquad \Delta g_i = \Delta g_i^0 - \nu_i k_B T \ln\rho. \quad \text{(S3)}$$

The convention $0\ln 0 = 0$ gives a finite entropy at the boundary; its derivative prevents a finite equilibrium potential from producing an occupation exactly outside the simplex. The surface charge is $\sigma = N_s(q_+\theta_+ + q_-\theta_-)$. Variation of $g + \sigma\psi$ with respect to the two independent occupations gives

$$\Delta g_i + q_i \psi + k_B T \ln(\theta_i/\theta_0) = 0. \quad \text{(S4)}$$

Thus $\theta_i = w_i/Z$, $\theta_0 = 1/Z$, $w_i = \exp[-(\Delta g_i + q_i\psi)/(k_B T)]$, and $Z = 1 + w_+ + w_-$. The common-site denominator is the mathematical expression of exclusion. It must be the same in equilibrium, response, and kinetics. A product of two denominators would describe two independent site families. Such a model can be physically appropriate for distinct sublattices, but it is not the common-site assumption studied here.

The occupation Hessian before electrostatic elimination is

$$H_{ij}^{chem} = N_s k_B T \left(\frac{\delta_{ij}}{\theta_i} + \frac{1}{\theta_0}\right), \qquad i, j \in \{+, -\}. \tag{S5}$$

For an interior occupation this matrix is positive definite. Its inverse is the covariance matrix $[\mathrm{diag}(\boldsymbol{\theta}) - \boldsymbol{\theta}\boldsymbol{\theta}^T]/(N_s k_B T)$. Therefore

$$\delta\boldsymbol{\theta} = -\frac{\mathrm{diag}(\boldsymbol{\theta}) - \boldsymbol{\theta}\boldsymbol{\theta}^T}{k_B T}\mathbf{q}_s\,\delta\psi, \qquad C_{chem} = \frac{N_s}{k_B T}\mathbf{q}_s^T[\mathrm{diag}(\boldsymbol{\theta}) - \boldsymbol{\theta}\boldsymbol{\theta}^T]\mathbf{q}_s, \tag{S6}$$

where $\mathbf{q}_s = (q_+, q_-)^T$ is a charge vector, distinct from the lateral wavevector. The empty-site fluctuations are already included through the covariance. In a highly occupied, nearly single-species state the available charge variation can be small, although the mean charge is large. This distinction is why a constant fitted chemical capacitance cannot replace the nonlinear isotherm in a domain calculation.

### S2.3 Electrical saddle and boundary conditions

At fixed bottom potential the functional is

$$\mathcal{G} = \int_f \left[f_L + f_G + f_{el} + \mathbf{P}\cdot\nabla\phi - \frac{\epsilon_f}{2}|\nabla\phi|^2\right]dV - \int_e \frac{\epsilon_e}{2}|\nabla\phi|^2 dV + \int_s (g + \sigma\psi + f_s)\,dA. \tag{S7}$$

The potential is stationary and concave in this representation; material variables are minimized. Eliminating the potential gives a positive electrostatic quadratic form in the polarization and charge sources. One must not minimize the negative field-energy term as though it were a material energy. Conversely, adding a second positive field energy after eliminating this saddle would count the electrical interaction twice.

Variation gives $\epsilon_f\nabla^2\phi = \nabla\cdot\mathbf{P}$ in the film, $\nabla^2\phi = 0$ outside, continuity of potential, and $\epsilon_f{\phi'}_f - \epsilon_e{\phi'}_e = P_z + \sigma$ at the surface. Mechanical variation gives $\partial_j\sigma_{ij} = 0$ and the stated displacement/traction boundaries. Polarization variation gives $\partial f_L/\partial P_i - \sigma_\alpha\,\partial e_\alpha^0/\partial P_i - G\nabla^2 P_i = -\partial_i\phi$. Together with the isotherm, these equations are the nonlinear stationary conditions. Eliminating their auxiliary fields before or after differentiation gives the same reduced derivatives when the fields are solved consistently.

## S3 Homogeneous reduction and uniform phase classification

### S3.1 Eliminating the free strains

Partition strain indices into clamped $a = (1,2,6)$ and free $b = (3,4,5)$. Let $\mathbf{m}_a = \mathbf{e}_a^{misfit} - \mathbf{e}_a^0$. Zero homogeneous stress on the free components gives $\mathbf{m}_b = -C_{bb}^{-1}C_{ba}\mathbf{m}_a$. Hence

$$f_{cl}(\mathbf{P}; u_m) = f_L(\mathbf{P}) + \frac{1}{2}\mathbf{m}_a^T C^{cl}\mathbf{m}_a, \qquad C^{cl} = C_{aa} - C_{ab}C_{bb}^{-1}C_{ba}. \tag{S8}$$

The shear entry associated with $P_x P_y$ remains in this expression. Dropping it makes an in-plane diagonal state artificially inexpensive and changes the competition among $a$, $aa$, and tilted branches. The derivative of this expression agrees with differentiation of the original elastic energy at its stationary free strains, which provides an independent implementation check.

**S3.2 Monotone electrical reduction**

The uniform potential is $\phi(z) = \psi z/h$. Charge balance gives $C_f\psi = P_z + \sigma(\psi)$, with $C_f = \epsilon_f/h$. Its derivative with respect to $\psi$ is $C_f + C_{chem} > 0$, so there is a unique root at fixed $P_z$. Bisection is robust even close to ionic saturation. The reduced electrical and chemical energy per area is

$$W(P_z) = P_z\psi - \frac{1}{2}C_f\psi^2 - N_s k_B T \ln Z(\psi), \qquad W' = \psi, \qquad W'' = \frac{1}{C_f + C_{chem}}. \tag{S9}$$

Only an additive constant independent of polarization can be dropped in energy comparisons at the same reservoir. The uniform grand-potential density is $\Phi = f_{cl} + W/h$. Its stationary equation is $\nabla_P f_{cl} + \psi\hat{z}/h = 0$. Its Hessian includes the exact chemical curvature above. The derivative formula is useful for resolving minima near phase boundaries and for checking a numerical interpolation of the electrochemical energy.

At balanced activity, the free energy is invariant under simultaneous reversal of polarization and charge. For the symmetric standard formation energies and opposite stoichiometries used here, inversion of activity is accompanied by this reversal. Maps of component magnitudes can therefore be reflected about $\log_{10}\rho = 0$. Signed polarization and surface potential must be reflected with the correct sign. This symmetry is a property of the specified reaction model and does not apply to an arbitrary experimental pair of surface species.

**S3.3 Exhaustive in-plane minimization and angular topology**

At fixed $P_z$, introduce $X = P_x^2$, $Y = P_y^2$, $S = X + Y$, and $V = XY$. Cubic invariants and equal biaxial clamping give a polynomial in $S, V, P_z^2$. The allowed interval is $0 \le V \le S^2/4$. Its endpoints describe an in-plane axis and diagonal. Interior stationary values describe a general azimuth. The solver tests admissible endpoint and interior roots, minimizes the radial in-plane variable, and then searches all resolved normal-polarization intervals. Normal minima are polished with the exact potential root and analytic polar derivatives. This construction is more reliable than launching a single Cartesian minimization at an assumed bulk variant.

The squared-component reduction does not predetermine the final phase. At a nonzero chemical bias, a component that was symmetry-forced to vanish can become small and finite. Component labels use stated numerical thresholds and are accompanied by continuous tilt maps. A change in label near a threshold can therefore be less significant than a discontinuity of the minimizing vector or an energy crossing. Residuals and full vectors are preserved with each map.

For the analytical two-amplitude map, $F = As/2 + Cz/2 + (s^2 + z^2)/4 + gsz/2$ with $s, z \ge 0$. Candidates are the origin, $(s, z) = (-A, 0)$, $(0, -C)$, and $[(gC - A)/(1 - g^2), (gA - C)/(1 - g^2)]$. Each candidate must satisfy positivity and local stability in the allowed cone, and its energy is compared with all others. For $g > 1$, the interior quartic stationary point is a saddle, and the two pure minima cross at $A = C < 0$. For $|g| < 1$, an admissible mixed minimum can separate them. The vector maps in Figure 2 retain the in-plane azimuth and higher-order invariant required to support an isolated orthorhombic minimum. They are topology illustrations rather than a replacement for the material-specific polynomial.

For the vector coefficient-plane examples, the dimensionless topology polynomial is $F = A(p_x^2 + p_y^2)/2 + Cp_z^2/2 + |\mathbf{p}|^4/4 + \delta \sum_i p_i^4 + \kappa p_x^2 p_y^2 p_z^2$. The pairs $(\delta, \kappa)$ are $(-0.04,0)$, $(0.04,1)$, $(0.04,0)$, and $(0.002,0)$ for T, O, R, and W. The positive sixth-order invariant penalizes simultaneous occupation of three components in the orthorhombic illustration. Candidate endpoint and interior radial roots are compared explicitly. Independent Cartesian differential-evolution minimizations at 32 test points verify the selected energies to within $7 \times 10^{-12}$ in the stated dimensionless normalization. The displayed coefficient grid has 601 points on each axis.

## S4 Full linearization and the effect of chemical screening

### S4.1 Reference stress and compatible displacement

Let a homogeneous stationary state be $(\mathbf{P}_0, \boldsymbol{\theta}^0, \psi_0, \boldsymbol{\sigma}^0)$. Perturb polarization by $\mathbf{v}(z)e^{i\mathbf{q}\cdot\mathbf{r}_\parallel}$, displacement by $\mathbf{u}(z)e^{i\mathbf{q}\cdot\mathbf{r}_\parallel}$, and potential by $\varphi(z)e^{i\mathbf{q}\cdot\mathbf{r}_\parallel}$. The eigenstrain derivative is $D_{\alpha i} = \partial e_\alpha^0 / \partial P_i$, and its second derivative is $D_{\alpha ij}^{(2)}$. The local tangent before compatible relaxation is $A_{ij}^0 = f_{L,ij} - \sigma_\alpha^0 D_{\alpha ij}^{(2)}$. The second term is the prestress correction. It remains even when the first-order displacement is eliminated.

With engineering strain, the Fourier displacement operator is

$$B_q \mathbf{u} = \left(iq_x u_x, iq_y u_y, u'_z, u'_y + iq_y u_z, u'_x + iq_x u_z, iq_y u_x + iq_x u_y\right)^T. \qquad \text{(S10)}$$

The remaining quadratic elastic energy is $\frac{1}{2}\int \left(B_q\mathbf{u} - D\mathbf{v}\right)^\dagger C\left(B_q\mathbf{u} - D\mathbf{v}\right)dz$. Compatible relaxation solves the displacement equilibrium and produces

$$K_{el} = D^T C D - D^T C B_q \left(B_q^\dagger C B_q\right)^{-1} B_q^\dagger C D, \qquad \text{(S11)}$$

with the prescribed displacement and natural traction boundaries incorporated in the inverse. This is positive semidefinite as a squared-mismatch minimum. It is nonlocal in depth and depends on wavevector direction. Together with the possibly negative prestress term, it gives the complete mechanical tangent. The inverse denotes an operator solve, not a pointwise local compliance.

### S4.2 Polarization operator and inner product

The natural continuum normalization is $\langle \mathbf{v}, \mathbf{v} \rangle = h^{-1} \int_0^h |\mathbf{v}|^2 \, dz$. The polarization Hessian is the sum of the local tangent, positive gradient operator, compatible elastic contribution, and electrostatic contribution. Its coefficients can be represented as complex matrices because a Fourier mode retains spatial phase. The physical field is real after adding its complex conjugate. Hermiticity, rather than reality of every entry, is the relevant symmetry. Omitting imaginary couplings would impose an extra phase constraint and alter the permitted fluctuations.

### S4.3 Open-surface electrostatic and ionic Schur complement

The linearized chemical charge is $\delta\sigma = -C_{chem}\varphi(h)$. Eliminating the exterior gives

$$\epsilon_f(\partial_z^2 - q^2)\varphi = iq_x v_x + iq_y v_y + v'_z, \quad \varphi(0) = 0, \quad \epsilon_f\varphi'(h) + (\epsilon_e q + C_{chem})\varphi(h) = v_z(h). \text{(S12)}$$

The exterior contribution is $\epsilon_e q$, including its zero limit at $q = 0$. It is not $\epsilon_e q\coth(qd)$. The latter would require a second electrode at distance $d$ and would retain a finite uniform exterior capacitance. This difference changes the spectrum at every wavelength, most strongly in the long-wave sector.

Define the positive potential quadratic form

$$\mathcal{A}_\phi[\varphi] = \int_0^h \epsilon_f\,(|\varphi'|^2 + q^2|\varphi|^2)dz + (\epsilon_e q + C_{chem})|\varphi(h)|^2. \tag{S13}$$

The source operator $\mathcal{L}$ is defined by the corresponding weak polarization–potential coupling. Electrical elimination gives $K_{es} = \mathcal{L}^\dagger\mathcal{A}_\phi^{-1}\mathcal{L}$. This form is valid for a depth-dependent vector polarization and includes both volume and surface bound charge. It is positive semidefinite. Surface chemistry changes the inverse operator through its boundary term. The nonlinear calculation differentiates this response at the actual local occupations, so the resulting domain Hessian uses a spatially varying chemical susceptibility.

### S4.4 Fixed-reference softening theorem

Let $A_0$ be the positive potential operator with frozen ionic fluctuations, and let $b$ evaluate the surface potential. Chemical relaxation changes it to $A_0 + C_{chem}bb^\dagger$. The rank-one inverse identity gives

$$H_{rel} = H_{fr} - \frac{C_{chem}}{1+C_{chem}R_q}\, w_q w_q^\dagger, \quad R_q = b^\dagger A_0^{-1} b > 0, \quad w_q = \mathcal{L}^\dagger A_0^{-1} b. \tag{S14}$$

At the fixed reference, $\partial H/\,\partial C_{chem} = -w_q w_q^\dagger/\left(1 + C_{chem}R_q\right)^2$ is negative semidefinite. Every ordered eigenvalue is nonincreasing with chemical capacitance, with the usual one-sided interpretation at degeneracy. For a normalized nondegenerate eigenvector, its derivative is $-\left|w_q^\dagger v\right|^2/\left(1 + C_{chem}R_q\right)^2$. A charge-neutral rotational mode can therefore respond weakly to ionic relaxation even when its homogeneous reference carries substantial screening charge.

The theorem compares fluctuations at the same polarization, stress, and equilibrium occupations. It does not identify the total derivative along a new equilibrium branch as activity or temperature changes. Such a derivative contains changes in the Landau tangent, prestress, eigenstrain derivative, and eigenvector as well as in the capacitance. Chemical control can select a more stable orientation while lowering the electrostatic cost of fluctuations about that orientation. The two effects are compatible and are evaluated separately in the calculations.

### S5 Closed-form columnar electrostatics and the matched scalar limit

Take a polarization perturbation independent of depth and resolve it into longitudinal, transverse, and normal components $(p_l, p_t, p_n)$ relative to the lateral wavevector. Set $Q = qh$ and $t = \tanh(Q/2)$. Solving the inhomogeneous Poisson equation and the open-surface Robin condition gives the electrostatic stiffness per film volume

$$D_{es} = \begin{pmatrix} (1-2t/Q)/\epsilon_f + t^2 d_n & 0 & itd_n \\ 0 & 0 & 0 \\ -itd_n & 0 & d_n \end{pmatrix}, \qquad d_n = \frac{1}{h[\epsilon_f q \coth(qh) + \epsilon_e q + C_{chem}]}. \quad \text{(S15)}$$

The sign of the imaginary pair follows the Fourier convention $e^{iqx}$ and the chosen longitudinal axis; changing that convention conjugates the matrix without changing its eigenvalues. The ratio between longitudinal and normal components can therefore be imaginary even for a real physical stripe. A sine-like longitudinal response screens the derivative of a cosine-like normal response. The transverse component is electrically neutral in this projection, but remains subject to local, gradient, and elastic stiffnesses.

For $q \to 0$, $t \sim Q/2$, $D_{nn} \to 1/\left(\epsilon_f + hC_{chem}\right)$, and the longitudinal and cross terms vanish. Setting both in-plane components to zero reproduces the scalar uniaxial film kernel exactly. The common-site chemical variance and homogeneous energy also reproduce the scalar formulation. These identities are checked before numerical phase maps are generated. At large $q$, the positive gradient contribution eventually dominates, so an instability driven by a negative local or elastic-renormalized polar curvature occupies a finite band rather than persisting to arbitrarily small scales.

The positive electrostatic matrix alone cannot create a negative mode. It changes how much of a negative polar tendency remains unscreened at each wavevector. A finite-wavevector minimum results when electrical and elastic costs decrease sufficiently with increasing wavevector before the positive gradient penalty becomes dominant. This mechanism differs from a chemical spinodal caused by a nonconvex adsorption entropy or an imposed negative gradient coefficient. Neither is used here.

Figure S1 compares this analytical projection with the variational field solution and illustrates capacitances obtained from the common-site law at different temperatures. The full-depth polarization calculation subsequently enlarges the admissible space. A negative eigenvalue in the restricted columnar space establishes an instability; a positive eigenvalue only establishes stability within that restriction.

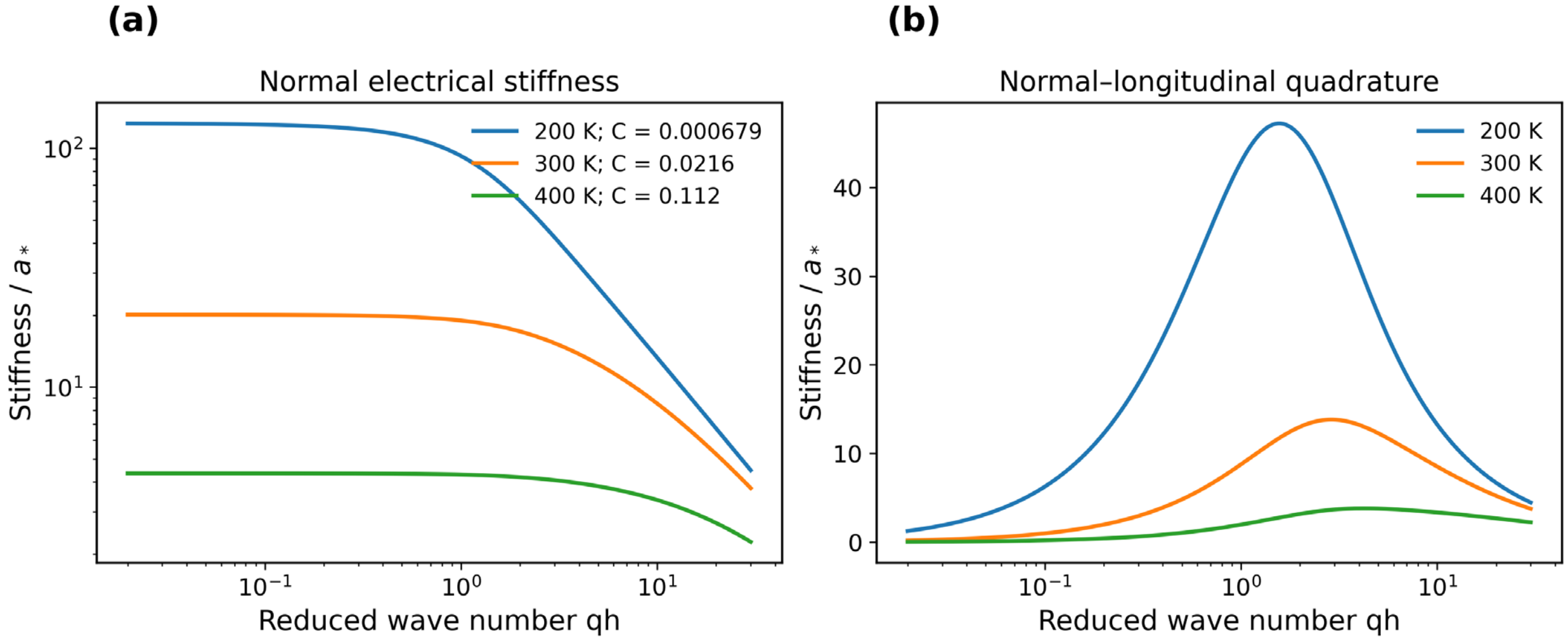

**Figure S1.** Analytical open-surface electrical kernel for a columnar perturbation in a 20 nm film. (a) Normal stiffness. (b) Imaginary normal–longitudinal coupling. Each chemical capacitance is evaluated from the common-site law at zero potential and the indicated temperature. Other material scales are held fixed to isolate the electrical contribution. Stiffnesses are divided by $a_*$.

## S6 Analytical linear modes for the separate crystal classes

### S6.1 General amplitude and rotational decomposition

At a polarized stationary point, let $\mathbf{e}_A = \mathbf{P}_0/|\mathbf{P}_0|$ and choose two unit vectors $e_1$ and $e_2$ perpendicular to $e_A$. A perturbation is

$$\mathbf{p} = p_A\mathbf{e}_A + p_1\mathbf{e}_1 + p_2\mathbf{e}_2. \tag{S16}$$

The angular perturbations are $\delta\theta_j = p_j/P_s$. The curvature with respect to angle is $P_s^2$ times the Cartesian rotational stiffness. Cartesian curvature and angular curvature therefore have different dimensions and should not be interchanged when reporting anisotropy.

The following explicit formulas refer first to stress-free, zero-field stationary points of the sixth-order cubic polynomial. Their role is to identify the symmetry channels. In a strained, chemically biased film, the actual stationary vector and the complete tangent must be recomputed; the numerical analysis does exactly this. Eighth-order coefficients are included by direct differentiation of $f_L + f_8$.

### S6.2 Tetragonal reference and the release of a variants

For the c reference, $\mathbf{P}_0 = (0,0,P_s)$, let $u = P_s^2$. Its stationary equation is

$$a_1 + 2a_{11}u + 3a_{111}u^2 = 0. \tag{S17}$$

The amplitude and two equal bulk rotational curvatures are

$$a_A^T = 8a_{11}u + 24a_{111}u^2, \tag{S18}$$

$$a_R^T = 2u[(a_{12} - 2a_{11}) + (a_{112} - 3a_{111})u]. \tag{S19}$$

For $q$ along a mirror direction, one in-plane transverse polarization mode decouples. In the (l,n) sector, write the projected matrix as

$$\mathbf{K}_{ln} = \begin{pmatrix} A & B + iC \\ B - iC & D \end{pmatrix}. \tag{S20}$$

Here $A$, $B$, $C$, and $D$ contain the Landau, gradient, compatible elastic, and electrical terms. In the simple strain-free c reference, $B = 0$, $A = a_R^T + Gq^2 + d_l$, $D = a_A^T + Gq^2 + d_n$, and $C = d_c$. A film elastic kernel can supply additional coupling and must be retained when present.

The two exact projected eigenvalues are

$$\lambda_\pm = \frac{A+D}{2} \pm \frac{1}{2}\sqrt{(A-D)^2 + 4(B^2 + C^2)}. \tag{S21}$$

An eigenvector can be chosen with

$$\frac{p_l}{p_n} = -\frac{B+iC}{A-\lambda}. \tag{S22}$$

For $A > 0$ and $D > 0$, instability occurs when $AD < B^2 + C^2$. The independent transverse mode must also be tested. At $q = 0$, a negative in-plane rotational stiffness corresponds to a homogeneous reorientation route. A finite-$q$ instability can instead mix rotation, normal amplitude, and compatible strain. Neither eigenvector by itself demonstrates a fully developed 90-degree twin.

An a reference is obtained by orienting $\mathbf{P}_0$ along an in-plane tetragonal axis and recomputing the clamped equilibrium. Waves parallel and perpendicular to this axis exchange amplitude and rotational roles in their in-plane channel. The normal rotational component remains sensitive to electrostatic compensation. Both a1/a2 and a/c nonlinear competitors must consequently be admitted.

**S6.3 Rhombohedral reference**

For $\mathbf{P}_0 = \sqrt{u}(1,1,1)$, choose

$$\mathbf{e}_A = \frac{(1,1,1)}{\sqrt{3}}, \quad \mathbf{e}_1 = \frac{(1,-1,0)}{\sqrt{2}}, \quad \mathbf{e}_2 = \frac{(1,1,-2)}{\sqrt{6}}. \tag{S23}$$

Define $U_R = a_{11} + a_{12}$ and $V_R = 3a_{111} + 6a_{112} + a_{123}$. The stationary equation and local curvatures are

$$a_1 + 2U_R u + V_R u^2 = 0, \tag{S24}$$

$$a_A^R = 8u(U_R + V_R u), \tag{S25}$$

$$a_R^R = 4u[2a_{11} - a_{12} + (6a_{111} - a_{123})u]. \tag{S26}$$

Both transverse curvatures equal $a_R^R$ in this ideal bulk state. The local Hessian is $a_R^R\mathbf{I} + (a_A^R - a_R^R)\mathbf{e}_A\mathbf{e}_A^T$. In a (001) film and for $q$ along $[110]$, $e_1$ is perpendicular to the sagittal plane spanned by $q$ and $z$. In the unstrained reference, the local sagittal entries are

$$A_L = \frac{2a_A^R + a_R^R}{3}, \quad D_L = \frac{a_A^R + 2a_R^R}{3}, \quad B_L = \frac{\sqrt{2}}{3}(a_A^R - a_R^R). \tag{S27}$$

Adding the projected elastic, gradient, and electrostatic matrices gives the same two eigenvalues derived above. The transverse mode is independent on this symmetry line. The nonzero real coupling $B_L$ arises from observing an inclined polar axis in film coordinates; the imaginary coupling arises from the spatial electrostatic phase relation. These two mechanisms should not be conflated.

In the actual strained film, the stationary vector generally becomes $(p, p, w)$, with scalar components $p \neq w$. The x-y exchange symmetry still separates the $[1\bar{1}0]$ fluctuation for $q$ along $[110]$, but the two bulk rotation modes are no longer degenerate. At arbitrary azimuth the full three-component matrix, or the full depth-dependent operator, must be diagonalized.

**S6.4 Orthorhombic reference**

For $\mathbf{P}_0 = \sqrt{u}(1,1,0)$, choose $\mathbf{e}_A = (1,1,0)/\sqrt{2}$, $\mathbf{e}_I = (1,-1,0)/\sqrt{2}$, and $\mathbf{e}_Z = (0,0,1)$. Put $U_O = 2a_{11} + a_{12}$ and $V_O = a_{111} + a_{112}$. Then

$$a_1 + U_O u + 3V_O u^2 = 0, \tag{S28}$$

$$a_A^O = 4U_O u + 24V_O u^2, \tag{S29}$$

$$a_I^O = 4u[2a_{11} - a_{12} + (6a_{111} - 2a_{112})u], \tag{S30}$$

$$a_Z^O = 2u[a_{12} - 2a_{11} + (a_{123} - 3a_{111} - a_{112})u]. \tag{S31}$$

The I mode rotates within the plane of the two nonzero components; the Z mode introduces the previously zero component. For an in-plane $[110]$ reference, $q$ parallel to $\mathbf{P}_0$ couples the amplitude channel to $p_Z$ through electrostatics, whereas $q$ parallel to $[1\bar{1}0]$ couples the I rotational channel to $p_Z$. In the simple diagonal-gradient, zero-elastic-coupling projection, the two sagittal matrices are

$$\mathbf{K}_\parallel = \begin{pmatrix} a_A^O + Gq^2 + d_l & id_c \\ -id_c & a_Z^O + Gq^2 + d_n \end{pmatrix}, \tag{S32}$$

$$\mathbf{K}_\perp = \begin{pmatrix} a_I^O + Gq^2 + d_l & id_c \\ -id_c & a_Z^O + Gq^2 + d_n \end{pmatrix}. \tag{S33}$$

The unused local channel is independent in each case. These expressions explicitly show why the relative rotational stiffnesses affect orientation selection. They are not sufficient to choose the film's final domain-wall family because elasticity and finite-amplitude compatibility also enter.

A tilted $[101]$ orthorhombic reference is another distinct film variant. In its $x$–$z$ sagittal plane, the amplitude and in-plane rotational directions are $(1,0,1)/\sqrt{2}$ and $(1,0,-1)/\sqrt{2}$. Their local matrix has equal diagonal entries $(a_A^O + a_I^O)/2$ and real off-diagonal entry $(a_A^O - a_I^O)/2$. The missing-axis mode along $y$ has curvature $a_Z^O$. The same electrostatic kernel can therefore be used after the correct basis transformation.

At fourth order only, $a_I^O = 4u(2a_{11} - a_{12})$ and $a_Z^O = 2u(a_{12} - 2a_{11})$ have opposite signs unless both vanish. A generic isolated stable orthorhombic minimum is impossible in that cubic quartic potential. A radial quartic fit along [110] can conceal this failure. Sixth- or higher-order anisotropy, or additional structural fields, is needed.

### S6.5 Weak angular anisotropy and weak ferroelastic distortion

Write a generic nearly isotropic potential as

$$f_L = f_{rad}(|\mathbf{P}|) + K\, w(\widehat{\mathbf{P}}), \qquad |K| \ll f_{rad} \text{ barrier scale.} \tag{S34}$$

At a stationary direction, the Hessian has a radial curvature $a_A$ and two small angular curvatures $a_1$ and $a_2$. The symmetry-specific matrix is obtained from the same projection used above, with these small curvatures. As $K$ approaches zero, a homogeneous stress-free bulk state has two rotational zero modes. Positive gradient energy gives them a cost at nonzero wave vector.

The film does not become isotropic in this limit. The normal component still generates surface charge, a longitudinal in-plane component still generates volume bound charge, and electrostriction still couples orientation to the substrate. A uniform in-plane polarization can therefore be the selected equilibrium. Smooth closure or vortex-like textures are possible competitors, but are not a necessary consequence of small $K$.

The numerical weak-anisotropy polynomial is deliberately explicit:

$$\frac{f_L}{a_* P_*^2} = -\frac{1}{2}|\mathbf{p}|^2 + \frac{1}{4}|\mathbf{p}|^4 + 0.002\sum_i p_i^4, \qquad \mathbf{p} = \mathbf{P}/P_*. \tag{S35}$$

The angular term favors $\langle 111 \rangle$ weakly. It is a controlled model, not a parameterization of PMN, PZN, or another relaxor. Quenched random fields, random anisotropy, their spatial correlations, and relevant dynamics would be needed for a specific relaxor theory [70–71]. A possible extension is

$$f_{dis} = -\mathbf{h}(\mathbf{r}) \cdot \mathbf{P} + \frac{1}{2}\delta a_{ij}(\mathbf{r}) P_i P_j. \tag{S36}$$

Disorder changes the reference state into a spatially heterogeneous one. A single-wave-vector linearization around a homogeneous state is then a reference calculation, not the full disorder-averaged stability problem.

The supporting calculation is shown in Figure S2.

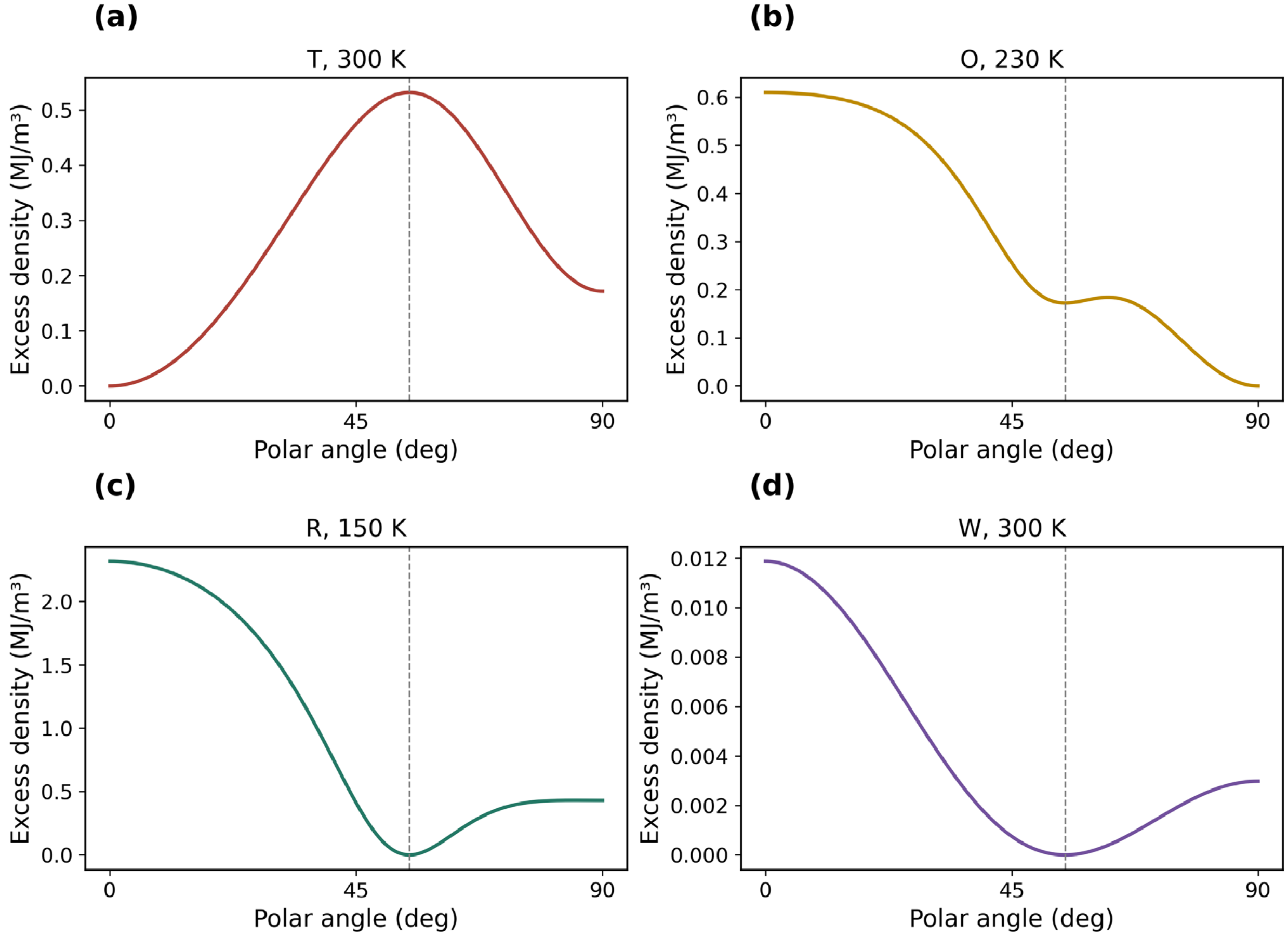

**Figure S2.** Stress-free angular energy after minimizing polarization magnitude. (a–c) The eighth-order $BaTiO_3$ polynomial at 300, 230, and 150 K. (d) The W polynomial at 300 K. The dashed angle marks the [111] direction. The rotation plane connects [001], [111], and [110]. Film strain and chemical energies are absent from this parent-landscape check.

## S7 Full depth-dependent analytical solution

### S7.1 Why the uniform-depth solution is not the final stability test

The projected solution in Section S5 is exact within its assumed polarization subspace. A film can nevertheless reduce its energy by bending polarization near a surface or relaxing strain over a distance comparable to the lateral wavelength. These variations require $\mathbf{p}(z)$, even when the homogeneous reference is constant in depth. The variational principle implies that the lowest eigenvalue of the full operator cannot exceed the lowest eigenvalue in a restricted polarization subspace. A positive projected eigenvalue therefore does not prove stability of the full film.

For constant coefficients, the depth dependence has an explicit exponential construction. Define

$$\mathbf{d}_\kappa = \left(iq_x, iq_y, \kappa\right)^T, \qquad \mathbf{B}_\kappa = \mathcal{B}_q|_{\partial_z=\kappa}, \qquad \mathbf{J}_\kappa = \mathbf{B}_\kappa^T. \tag{S37}$$

Here J is an algebraic divergence matrix. The transpose does not conjugate iq; this is a strong-form exponential substitution, whereas the weak-form adjoint in Section S4 includes integration by parts. Confusing these two operations reverses signs in the characteristic equation.

For the isotropic gradient energy $G$ times the squared Cartesian gradients, write the mode as $(\mathbf{p}, \mathbf{v}, \varphi) = (\mathbf{p}_k, \mathbf{v}_k, \phi_k)\exp(\kappa z)$. The static eigenproblem $H\mathbf{p} = \lambda\mathbf{p}$ becomes

$$\mathbf{M}(\kappa; q, \lambda)\begin{pmatrix}\mathbf{p}_k \\ \mathbf{v}_k \\ \phi_k\end{pmatrix} = 0, \tag{S38}$$

where the polarization block is

$$\mathbf{M}_{PP} = \mathbf{A}^0 + \mathbf{D}^T\mathbf{C}\mathbf{D} + [G(q^2 - \kappa^2) - \lambda]\mathbf{I}, \tag{S39}$$

and

$$\mathbf{M} = \begin{pmatrix}\mathbf{M}_{PP} & -\mathbf{D}^T\mathbf{C}\mathbf{B}_\kappa & \mathbf{d}_\kappa \\ -\mathbf{J}_\kappa\mathbf{C}\mathbf{D} & \mathbf{J}_\kappa\mathbf{C}\mathbf{B}_\kappa & 0 \\ -\mathbf{d}_\kappa^T & 0 & \epsilon_f(\kappa^2 - q^2)\end{pmatrix}. \tag{S40}$$

The first row is the polarization equation, the second is mechanical equilibrium, and the third is Gauss's law. For nondegenerate coefficients there are fourteen characteristic roots, counting multiplicity, because the seven fields obey a second-order system. Repeated roots are treated with generalized solutions, such as $z\exp(\kappa z)$, or a first-order matrix exponential. A direct unscaled product of large growing and decaying exponentials can be ill-conditioned; the variational discretization below is more robust for numerical evaluation.

### S7.2 Boundary determinant and eigenfunctions

Let $y_m$ be the null vector associated with $\kappa_m$. Then

$$\mathbf{y}(z) = \sum_{m=1}^{14} c_m\, \mathbf{y}_m e^{\kappa_m z}. \tag{S41}$$

The fourteen scalar boundary conditions are three zero gradient fluxes at each film face, three zero displacement perturbations at the bottom, three zero traction perturbations at the top, zero potential at the bottom, and the electrochemical Robin condition at the top. Applying them gives

$$\mathbf{B}_{bc}(q, \lambda)\mathbf{c} = 0, \qquad \det\mathbf{B}_{bc}(q, \lambda) = 0. \tag{S42}$$

This determinant is an analytical implicit dispersion relation for each of the tetragonal, rhombohedral, orthorhombic, and weak-anisotropy cases. Their different Landau tangents and eigenstrain derivatives are inserted into the same matrix. The eigenfunction follows from its null vector. The film spinodal is the first parameter value for which this determinant vanishes at $\lambda = 0$ for some $q$ and azimuth.

There is no general elementary formula for the selected $q$ after the finite-thickness electrical and elastic boundaries have been imposed. Hyperbolic functions of $qh$ appear even before mechanical relaxation is included. Reporting an implicit boundary determinant and solving its variational equivalent is therefore a complete linear solution, rather than a replacement of the film by a bulk dispersion formula.

### S7.3 Variational evaluation

The numerical solution uses continuous piecewise-linear polarization functions and enriched displacement and potential fields. If polarization has $n_z$ nodes, auxiliary fields use $2n_z - 1$ nodes. The exterior is represented exactly by its Fourier boundary term $\epsilon_e q$. Bottom displacement and potential variables are removed. Auxiliary elimination gives a Hermitian polarization matrix and a positive mass matrix, and the generalized problem is $H_q v = \lambda M v$. The local polynomial and elastic tangent use the same projected quadrature as the nonlinear cell. This equality of function spaces and quadrature is checked by a nonlinear sinusoidal second variation. The reported physical stiffness is the dimensionless eigenvalue times $a_*$.

The broad atlas uses nine polarization nodes; selected spectra and onset points are refined separately. Increasing the auxiliary resolution prevents an unresolved potential or displacement variation from making a retained polarization perturbation artificially soft. A dense grid of temperature and thickness controls does not replace a depth-convergence check. The boundary determinant above and its variational discretization solve the same linear problem, but the latter avoids exponentially ill-conditioned determinant evaluation at large depth rates.

## S8 Reciprocal dynamics and reduced control parameters

### S8.1 Static stiffness and growth rates are different spectra

The static problem minimizes the grand potential over displacement, potential, and equilibrium occupations. For slow polarization dynamics with rapidly equilibrated ions, its linearized equation is $\dot{\mathbf{v}} = -L_P H_{rel} \mathbf{v}$. If the ionic occupations have finite reaction times, they must instead remain explicit dynamical variables. Let $\mathbf{x} = (\delta \mathbf{P}, \delta \boldsymbol{\theta})$ and let $H_{full}$ be the joint Hessian after mechanical and electrostatic elimination. A reciprocal gradient dynamics has $\dot{\mathbf{x}} = -M H_{full} \mathbf{x}$, with $M$ positive on the admissible variables. Surface diffusion, if included, introduces a positive wavevector-dependent mobility; it does not change the equilibrium Hessian.

For positive $M$, the generator is similar to the Hermitian matrix $-M^{1/2} H_{full} M^{1/2}$. Its eigenvalues are real. The location of a static instability is governed by loss of positive definiteness of the Hessian, whereas the fastest growing mode after a finite parameter change depends on the mobility as well. A slow reaction does not generate a Hopf bifurcation in this passive reciprocal model. Driven reactions, nonreciprocal coupling, or a different nonequilibrium reference would require their own analysis. Conserved limits introduce zero modes and must be handled on the appropriate constrained subspace rather than by assigning negative dissipation.

The least-stiff film wavelength plotted in the static maps is therefore not labeled as a kinetic prediction. The marginal wavelength is found by tuning a control to the first zero eigenvalue and minimizing over wavevector. The fixed-reference least-stiff wavelength is evaluated after the homogeneous reference is already unstable. The fastest kinetic wavelength maximizes the growth rate of the full dynamical generator. The developed period is a fourth quantity obtained by minimizing the nonlinear energy of a candidate family. These four operations coincide only in special limiting situations.

### S8.2 Scales and dimensionless groups

For the film calculations, $P_* = 0.300$ C m$^{-2}$, $a_* = 10^8$ J m C$^{-2}$, and $f_* = a_* P_*^2 = 9.00$ MJ m$^{-3}$. We use $\mathbf{p} = \mathbf{P}/P_*$, $\zeta = z/h$, and $Q = qh$. The dimensionless background permittivity is $a_* \epsilon_f$, the gradient scale is $\gamma_G = G/(a_* h^2)$, the chemical capacitance is $\hat{C} = a_* h C_{chem}$, and the exterior coefficient is $a_* \epsilon_e Q$. The normal columnar electrical stiffness in these units is

$$\hat{d}_n = \left[a_* \epsilon_f Q \coth Q + a_* \epsilon_e Q + \hat{C}\right]^{-1}. \tag{S43}$$

The dimensionless chemical charge capacity is $N_s |q_i| / P_*$. The formation parameters are $\Delta g_i^0 / (k_B T)$ and $\nu_i \ln \rho$. The ratio of a typical electrostrictive strain $Q_{ij} P_*^2$ to imposed misfit controls mechanical branch selection, while combinations $C_{\alpha\beta} (Q P_*)^2 / a_*$ measure the tangent elastic scale. Radial and rotational eigenvalues of the local Landau Hessian divided by $a_*$ distinguish amplitude-dominated and rotation-dominated regimes. An angular stiffness should not be replaced by a radial curvature taken along a single crystal axis.

In a screened long-wave regime, $\hat{C}$ exceeds the dielectric term and the electrical penalty is weakly wavevector dependent at small $Q$. In a poorly screened regime, the finite-thickness field cost is substantial and finite-wavevector charge reduction is more competitive. A nearly neutral rotational mode can remain weakly dependent on either regime. Ionic saturation is governed by the complete occupation law, not simply by the value of $\hat{C}$: a large mean charge near capacity can coexist with a small incremental response. These reduced parameters organize the comparisons without assuming one universal wavelength scaling for all crystal classes.

## S9 Weakly nonlinear reduction and domain interpretation

### S9.1 Isolated-mode expansion

Let $v(z)$ be a critical eigenvector normalized by $\int_0^1 v^\dagger\, v\, d\zeta = 1$. Write the real polarization perturbation as $A v e^{i q_c x} + A^* v^* e^{-i q_c x}$ and retain its induced mean and second harmonic at order $|A|^2$. Translational symmetry makes the energy depend on $|A|$. With the standard Taylor convention for derivatives of the full reduced functional,

$$\Delta f / f_* = \lambda |A|^2 + g |A|^4 + O(|A|^6), \tag{S44}$$

$$g = 1/4\, T_4(v, v, v^*, v^*) - 1/2\, t_0^\dagger H_0^{-1} t_0 - 1/4\, t_2^\dagger H_{2q_c}^{-1} t_2, \tag{S45}$$

where $t_0 = T_3(v, v^*)$ and $t_2 = T_3(v, v)$. The inverse operators act on stable admissible sectors. If either sector has another zero or negative eigenvalue, a single complex amplitude is not an adequate reduction. The negative terms are energy lowering by slaved modes. They include compatible strain and the nonlinear change of ionic occupations because the derivatives are taken after their consistent elimination. A local bulk quartic coefficient alone cannot determine the onset type.

For $g > 0$, the small-amplitude minimum has $|A|^2 = -\lambda/(2g)$ on the unstable side. For $g < 0$, higher-order terms are required, and a subcritical branch can preempt the spinodal. Even $g > 0$ does not exclude a disconnected finite-amplitude minimum elsewhere in state space. If two lateral wavevectors are simultaneously critical, the expansion contains $g_1 |A_1|^4 + g_2 |A_2|^4 +$ $g_{12}|A_1|^2|A_2|^2$ and symmetry-allowed resonances. Such a two-mode selection rule requires an actual degeneracy of the strained film, not merely a symmetry of the stress-free parent polynomial.

### S9.2 Numerical extraction and branch continuation

For each selected film spinodal the minimum over wavevector magnitude and azimuth is continued in misfit and bracketed by opposite signs of the least eigenvalue. A scalar root solve locates the spinodal. The primary calculation uses 17 polarization nodes and is repeated with 25 nodes. At the critical wavevector, the mean and second-harmonic Hessians are checked for positive definiteness before slaving is attempted.

Directional differences of the exact nonlinear gradient determine the quadratic forcing using step sizes 0.0015, 0.003, and 0.006 in dimensionless polarization. The energy is then evaluated at amplitudes 0.003, 0.005, 0.007, 0.010, and 0.013 including the slaved correction. Even averaging over positive and negative amplitudes cancels odd truncation errors. Extrapolating the remaining quartic estimate in $|A|^2$ gives the reported coefficient. The calculation stores all step-dependent estimates, rather than only the extrapolated sign. The continuous branches in Figure 8 are separately obtained by full nonlinear relaxation with 25 lateral points and 17 polarization depth nodes at nine misfits around each spinodal.

A near-onset sinusoid can remain within one polarization-orientation sector. It should not automatically be called a reversed-domain state or assigned a crystallographic wall angle. At finite amplitude, the full vector trajectory can approach several minima, change the sign of the normal component, or rotate principally in the plane. The developed profiles and their occupations are retained in Figures 9 and S16 to identify which of these possibilities actually occurs.

### S9.3 Compatibility and admissible wall families

An ideal planar wall with normal $\mathbf{n}_w$ is charge neutral when $\mathbf{n}_w \cdot \left(\mathbf{P}^{(1)} - \mathbf{P}^{(2)}\right) = 0$. Mechanical compatibility of two uniform spontaneous strains requires their difference to have the appropriate symmetrized rank-one form. These are useful guides for sharp-wall limits, but the present diffuse-interface calculation enforces compatibility through a single displacement field and computes electrostatics without imposing wall neutrality. Charged or curved wall segments can therefore appear if their cost is offset by other energy reductions. The ability to construct a locally compatible wall is not a proof that a periodic array of those walls has the lowest film energy.

## S10 Material coefficients and parameter audit

The polynomial conventions are those of Section S2. Coefficients below are in SI units appropriate to the indicated power of polarization. The $BaTiO_3$ potential is the eighth-order Li–Cross–Chen parameterization [64–65]. The PZT reference is the audited 50/50 composition [66,68], rather than an extrapolation of a scalar tetragonal polynomial to arbitrary vector directions. $BiFeO_3$ is a polarization-only comparison with cubic quartic anisotropy [67,69]. It omits an independently relaxing antiferrodistortive order parameter. The common adsorption parameters are assumptions for controlled comparison across materials; they are not claimed to be measured for each surface termination.

The numerical coefficients are collected in Table S1.

**Table S1.** Landau coefficients in the convention of Section S2; temperature is in kelvin. Values identify the parameterization; full input precision is retained in the executable source.

| **Coefficient** | $BaTiO_3$ | **PZT 50/50** | $BiFeO_3$ |
|---|---|---|---|
| $a_1$ | $4.124 \times 10^5(T - 388)$ | $1.33028 \times 10^5(T - 665.515)$ | $4.9 \times 10^5(T - 1103)$ |
| $a_{11}$ | $-2.097 \times 10^8$ | $4.764 \times 10^7$ | $6.5 \times 10^8$ |
| $a_{12}$ | $7.974 \times 10^8$ | $1.735 \times 10^8$ | $1.0 \times 10^8$ |
| $a_{111}$ | $1.294 \times 10^9$ | $1.336 \times 10^8$ | 0 |
| $a_{112}$ | $-1.950 \times 10^9$ | $6.128 \times 10^8$ | 0 |
| $a_{123}$ | $-2.500 \times 10^9$ | $-2.894 \times 10^9$ | 0 |
| $a_{1111}$ | $3.863 \times 10^{10}$ | 0 | 0 |
| $a_{1112}$ | $2.529 \times 10^{10}$ | 0 | 0 |
| $a_{1122}$ | $1.637 \times 10^{10}$ | 0 | 0 |
| $a_{1123}$ | $1.367 \times 10^{10}$ | 0 | 0 |

For $BaTiO_3$, the elastic compliance triplet $(s_{11}, s_{12}, s_{44})$ is $(8.30, -2.70, 9.24) \times 10^{-12}$ $\mathrm{Pa}^{-1}$ and $(Q_{11}, Q_{12}, Q_{44}) = (0.110, -0.0430, 0.0590)$ $\mathrm{m^4\ C^{-2}}$. PZT uses $(10.5, -3.70, 28.7) \times 10^{-12}$ $\mathrm{Pa}^{-1}$ and $(0.0966, -0.0460, 0.0819)$ $\mathrm{m^4\ C^{-2}}$. The $BiFeO_3$ stiffnesses are $(C_{11}, C_{12}, C_{44}) = (302, 162, 68.0)$ GPa and its electrostriction triplet is $(0.0320, -0.0160, 0.0200)$ $\mathrm{m^4\ C^{-2}}$. These are engineering-shear conventions. The source code constructs the full tensors from these triplets and differentiates the polynomial directly.

For W, $f_W/f_* = \alpha(T)|\mathbf{p}|^2/2 + |\mathbf{p}|^4/4 + 0.002 \sum_i p_i^4$, with $\alpha(T) = (T - 400\,\mathrm{K})/(100\,\mathrm{K})$. At 300 K this reproduces the reference polynomial in the main paper. The thermal law is an explicit model choice. W uses the $BaTiO_3$ elastic tensor and either $0.1Q_{BTO}$ or $Q_{BTO}$, with the same angular coefficient. No material name is attached to this artificial thermal extension. Its quadratic zero at 400 K does not imply that all strained-film phase boundaries occur there.

Every baseline film uses $G = 10^{-10}$ $\mathrm{J\ m^3\ C^{-2}}$, $\epsilon_b = 7.35$, $\epsilon_{e,r} = 1$, $N_s = 10^{18}$ $\mathrm{m^{-2}}$, $q_\pm = \pm 2e$, $\Delta g_i^0 = 0.200$ eV, and $\nu_\pm = \mp 1/2$. Finite-strain and finite-temperature maps do not alter these chemical coefficients. The stable Gaussian bulk used in Section S14 changes the polar and gradient constitutive law, but retains this same chemical parameter set and the same exposed-surface/grounded-bottom geometry.

The main reference states and selected spinodals are listed separately below. Energies and periods in the nonlinear table correspond to the specified reference strains, not to a state infinitesimally beyond each spinodal. This distinction prevents a comparison of wavelengths taken at different physical controls from being presented as a single continuous evolution.

Table S2 gives the equilibrated reference states; Table S3 gives the separately tuned spinodals.

**Table S2.** Recalculated homogeneous reference states and their least-stiff periods. T, O, R, and W temperatures are 300, 230, 150, and 300 K.

| Case | $u_m$ (%) | $\mathbf{P}_0$ (C/m$^2$) | $\psi$ (V) | $C_{chem}$ (F/m$^2$) | $L_{min}$ (nm) |
|---|---|---|---|---|---|
| T | 0 | (0.156,0.156,0) | 5.37e-18 | 0.0216 | 15.7 |
| O | -0.3 | (0.0893,0.0893,0.268) | 0.116 | 4.47 | 24.4 |
| R | -0.3 | (0.163,0.163,0.276) | 0.112 | 5.94 | 16 |
| W | -0.15 | (0.207,0.207,0) | 5.37e-18 | 0.0216 | 41.7 |

**Table S3.** Selected isolated-mode spinodals. The final column is the stiffness at the orthogonal wavevector of the same magnitude, confirming that a second lateral amplitude is not simultaneously critical.

| Case | $u_m^c$ (%) | $L_c$ (nm) | $g$ | $L_c$, 25 nodes (nm) | $\lambda_\perp/a_*$ |
|---|---|---|---|---|---|
| T | 0.269 | 13.8 | 3.01 | 13.8 | 0.991 |
| O | 0.0315 | 14.5 | 2.42 | 14.5 | 1.94 |
| R | -0.213 | 15.1 | 3.04 | 15.1 | 2.74 |
| W | -0.0465 | 49.4 | 0.0561 | 49.4 | 0.0406 |

### S11 Complete homogeneous and instability atlas

#### S11.1 Numerical grids and what each map means

The homogeneous temperature–thickness atlas uses 81 temperatures and 49 logarithmically spaced thicknesses from 1 to 300 nm at each of three activities, 1, $10^{-4}$, and $10^{-8}$. $BaTiO_3$ is evaluated on one continuous 100–500 K interval at $u_m = -0.100\%$. The more compressive O/R comparison uses a common 100–280 K calculation at $-0.300\%$; Figures S3 and S4 are overlapping temperature windows of this calculation. PZT 50/50 spans 300–850 K, $BiFeO_3$ spans 300–1200 K, and W spans 100–500 K. These are constitutive-model sweeps, including regions where a real film could require additional physics.

At each point, the normal-polarization search starts from 1201 samples over the resolved interval and polishes candidate minima after exact in-plane reduction. The stored result includes the minimizing vector, component label, energy, surface potential, ionic charge, chemical capacitance, and residual. For the sign-biased activity slices, the code uses the exact charge-inversion symmetry while restoring the appropriate sign of normal polarization and potential. Magnitude maps alone do not display that sign reversal.

The spatial atlas samples 49 temperatures and all 49 thicknesses for each of the six cases and three activities. At each point it tests zero wavevector together with 31 nonzero magnitudes spanning $qh = 0.0800$ to 80.0 at eight azimuths in $[0, \pi)$. Polarization uses nine depth nodes and the auxiliary electrical and displacement fields use 17. Hatching on a component map denotes a negative sampled eigenvalue about the uniform state identified by its color. It does not select a nonlinear morphology. Color in a period map denotes the least-stiff sampled nonzero mode only where its eigenvalue is negative. Gray therefore means that no such finite-wavevector negative mode was selected; it does not establish global homogeneous equilibrium.

The strain–activity maps use 41 misfits from $-0.600\%$ to $+0.600\%$ and 17 values of $\log_{10}\rho$ from zero to eight. The opposite activity half is supplied by the exact model symmetry. T, O, R, W, PZT, and $BiFeO_3$ are each recalculated; these maps are not overlays imported from a different electrical geometry. Their finite-wavevector periods are retained separately from their homogeneous component labels in Figures S14 and S15.

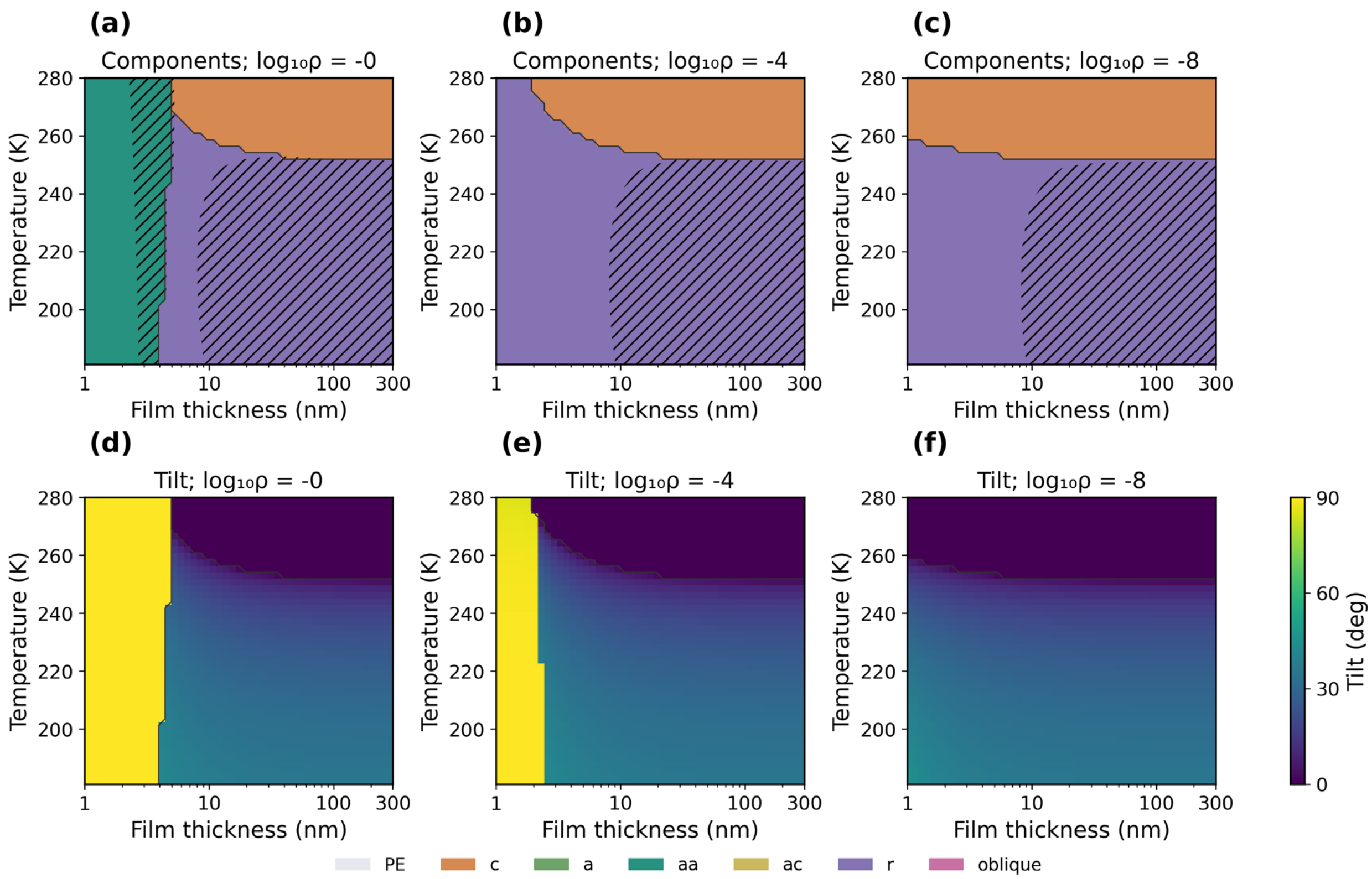

**Figure S3.** Orthorhombic-parent temperature window of the common $BaTiO_3$ calculation at −0.300% misfit. (a–c) Component patterns with hatching for negative sampled spatial modes at the three indicated activities. (d–f) Corresponding continuous tilts. This is a window of one unrestricted calculation, not a restriction to an orthorhombic polarization axis.

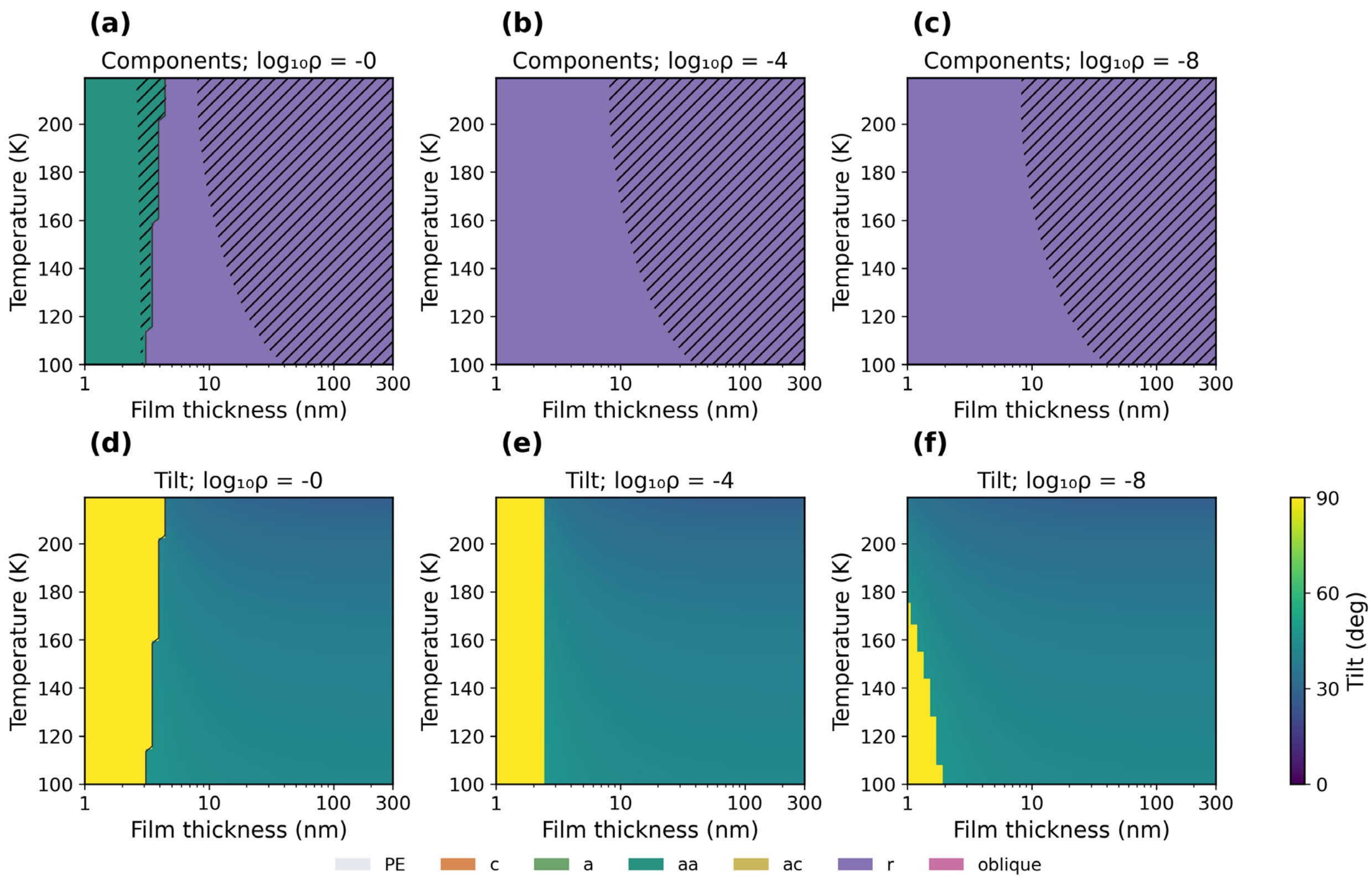

**Figure S4.** Rhombohedral-parent window of the same compressively strained $BaTiO_3$ calculation. Panel organization and hatching have the meaning of Figure S3. The two temperature windows overlap intentionally; neither imposes a bulk direction on the actual film minimum.

The complete PZT, $BiFeO_3$, and two weak-anisotropy atlases are shown in Figures S5–S8.

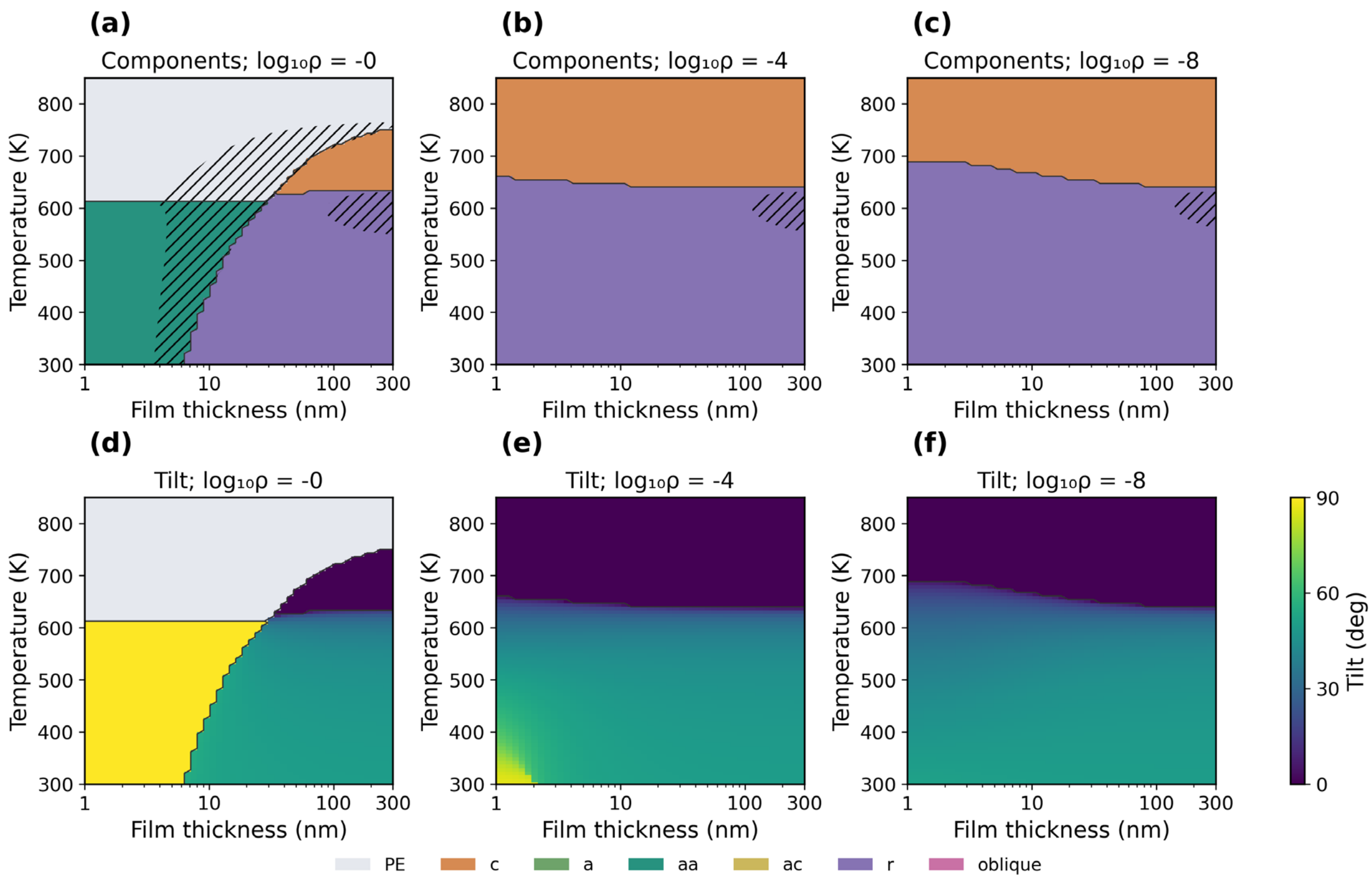

**Figure S5.** Full PZT 50/50 atlas at −0.100% misfit. (a–c) Component patterns and sampled-instability hatching. (d–f) Tilts, at activities 1, $10^{-4}$, and $10^{-8}$. The entire polynomial and common chemical law are reevaluated at each temperature and thickness.

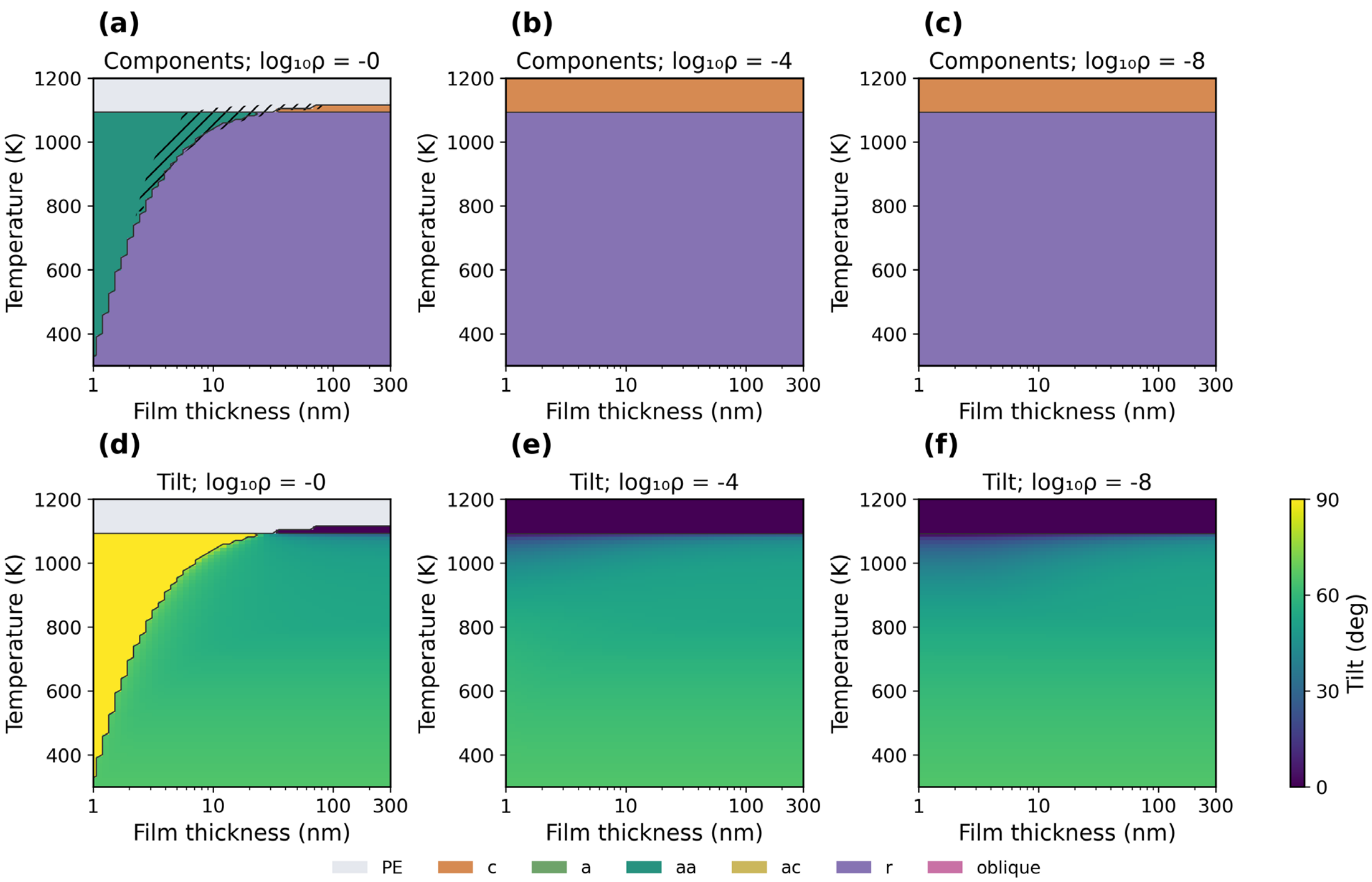

**Figure S6.** Full polarization-only $BiFeO_3$ atlas at $-0.100\%$ misfit. (a–c) Component patterns with instability hatching. (d–f) Tilts. These are constitutive-reference maps; ionic saturation and missing electronic or octahedral-rotation responses constrain their physical interpretation.

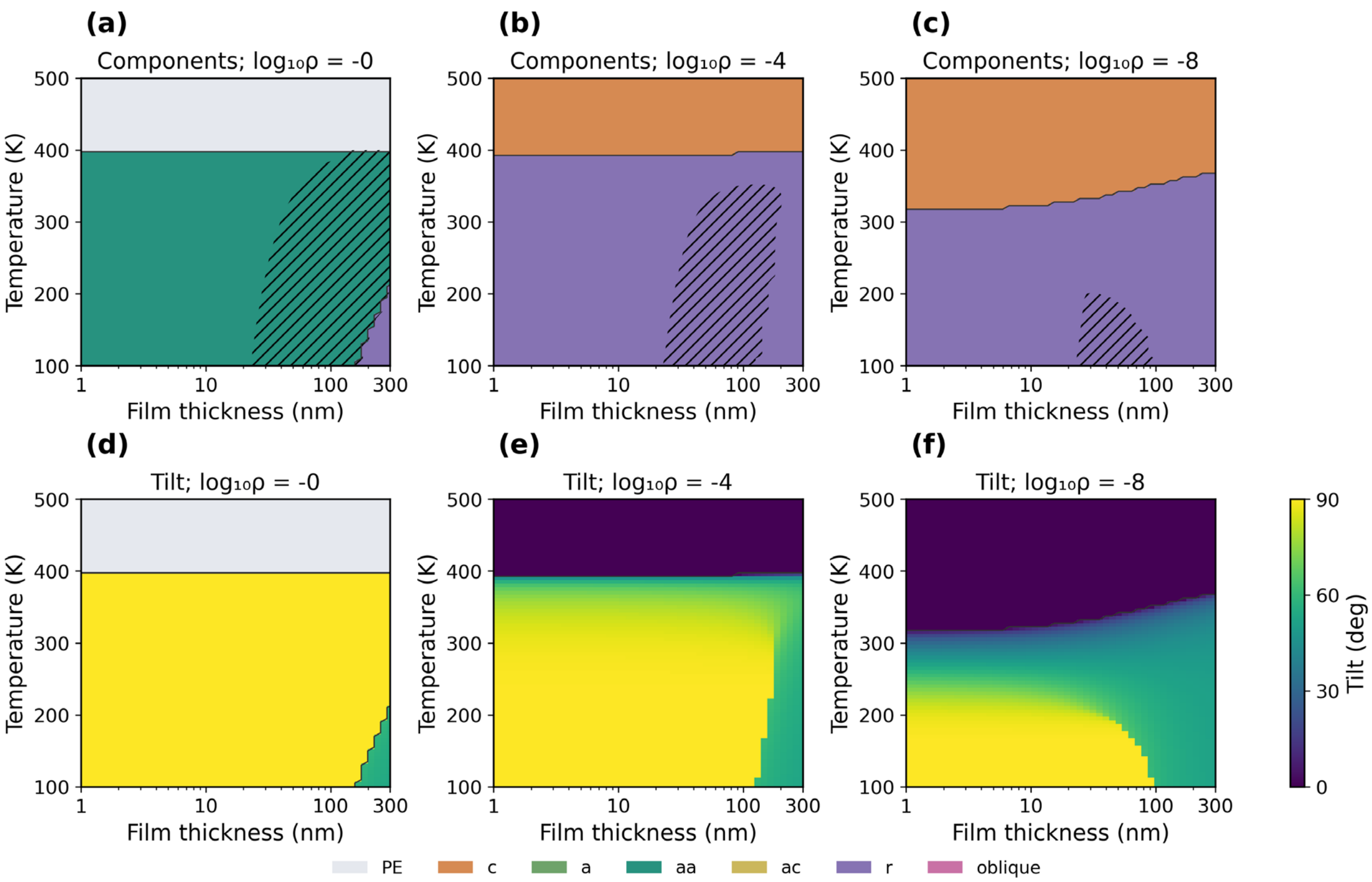

**Figure S7.** Weak-anisotropy atlas with $Q = 0.1Q_{BTO}$ and zero imposed misfit. (a–c) Component patterns with sampled-instability hatching. (d–f) Tilts. The specified thermal quadratic law is an idealized control, not a fit to a named relaxor.

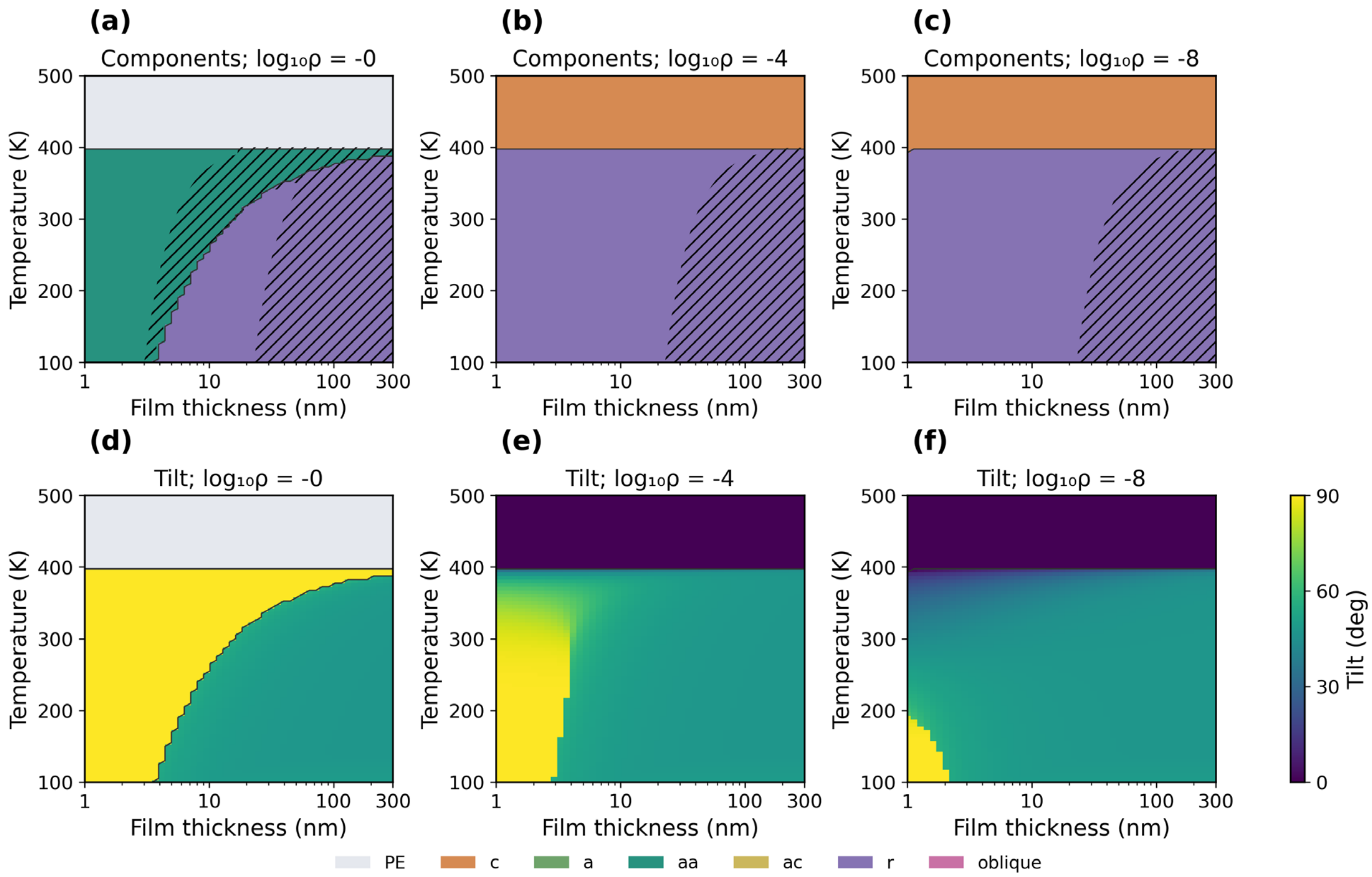

**Figure S8.** The same weak angular landscape with $Q = Q_{BTO}$. Panel organization matches Figure S7. Comparing the two figures isolates the change of electrostriction without changing the chemical model or polar anisotropy.

### S11.2 Boundary complexity and numerical resolution

Several physical effects can give a complicated component map: a first-order crossing of polar branches, continuous rotation of a tilted minimum, ionic saturation, or a change between axis-like and diagonal in-plane order. A finite-wavevector instability can also enter a region without changing its homogeneous component label. These are different boundaries. In addition, a component threshold can introduce a visually narrow region near a continuous zero. The full vectors and energies are needed to distinguish these cases. The plots use nearest-cell coloring; smoothing an integer phase label would create intermediate regions that were never calculated.

Figure S9 tests representative points close to a sampled boundary and at the most unstable point of each atlas case. It compares the baseline spatial grid with 17 polarization nodes, 33 auxiliary nodes, 61 nonzero magnitudes, and 16 azimuths. This is a local resolution audit, not a claim that every boundary cell has undergone that refinement. Figure S19 independently follows the four reference spectra through 9, 17, 25, 33, and 49 polarization nodes. Small shifts near zero stiffness should be interpreted at the resolution of these tests. The notebook exposes the map and wavevector grids so that a selected boundary can be refined without changing the model.

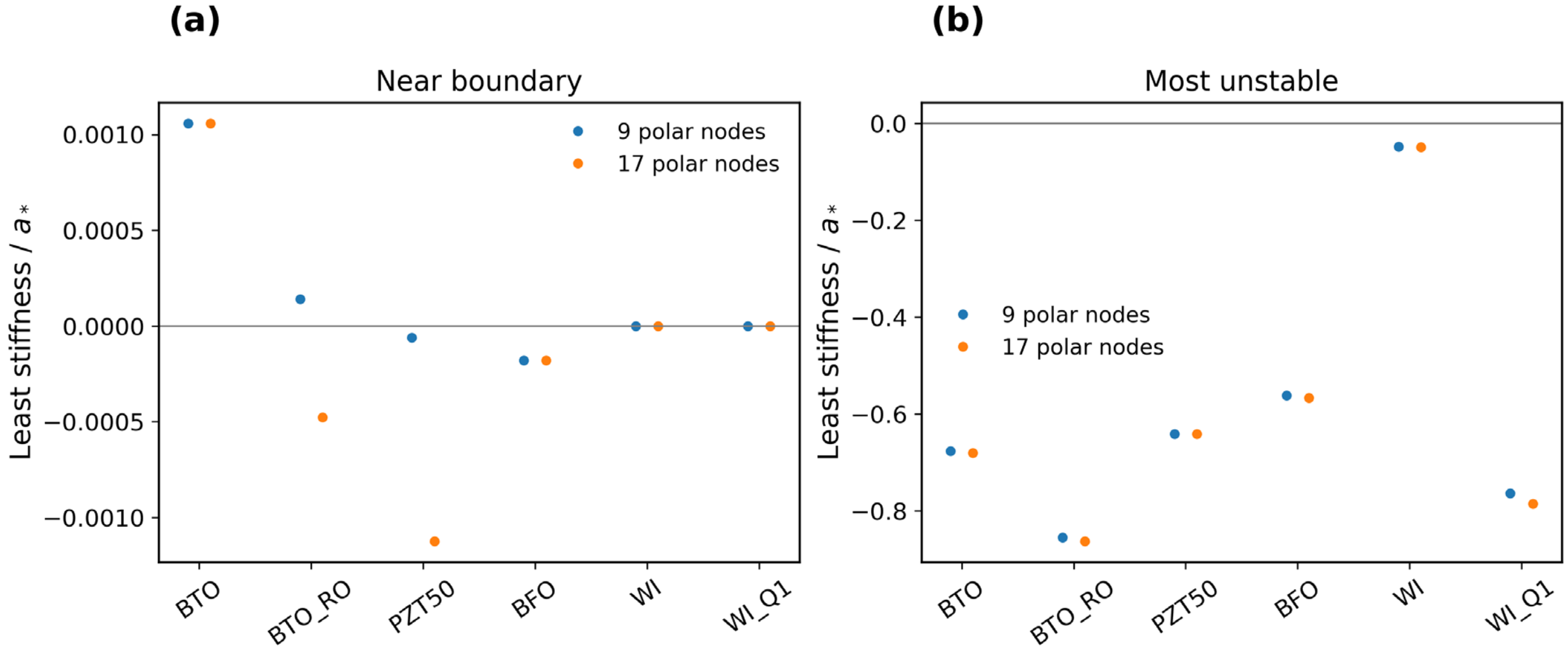

**Figure S9.** Representative refinement audit of the dense spatial atlas. (a) Points close to the sampled stability boundary. (b) The most unstable sampled points. The baseline has nine polarization nodes, 31 nonzero magnitudes, and eight azimuths; the refined calculation has 17 nodes, 61 magnitudes, and 16 azimuths. Auxiliary-field grids are enriched in both cases. This is a local audit rather than full refinement of every boundary cell.

The component-versus-temperature cuts in Figure S10 use the nearest thickness in the homogeneous atlas, which is explicitly stated in each panel. A separate 361-point temperature sweep at exactly 20 nm resolves the balanced $BaTiO_3$ branch and its chemical capacitance in Figure S11. It follows the self-consistent minimum; it is not a calculation at fixed polarization. Consequently, changes of its capacitance and spatial stability cannot be interpreted using the fixed-reference theorem alone. If a branch changes discontinuously, a numerical derivative across that crossing is not the susceptibility of either metastable branch.

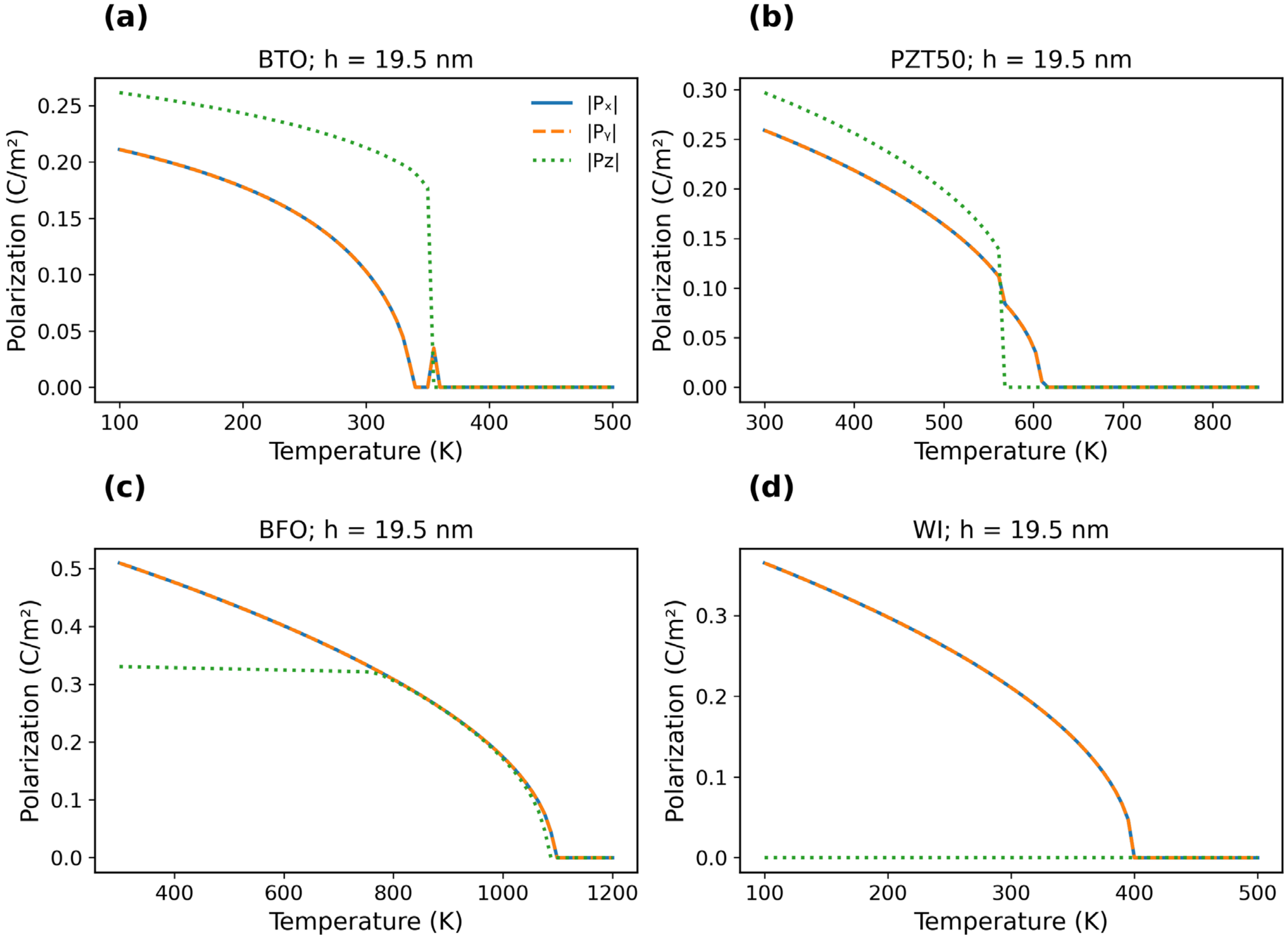

**Figure S10.** Continuous polarization components versus temperature on near-20 nm atlas slices. (a–d) $BaTiO_3$, PZT 50/50, $BiFeO_3$, and W at balanced activity and their map misfits. The exact sampled thickness is printed in each panel. Overlapping curves correspond to equal component magnitudes.

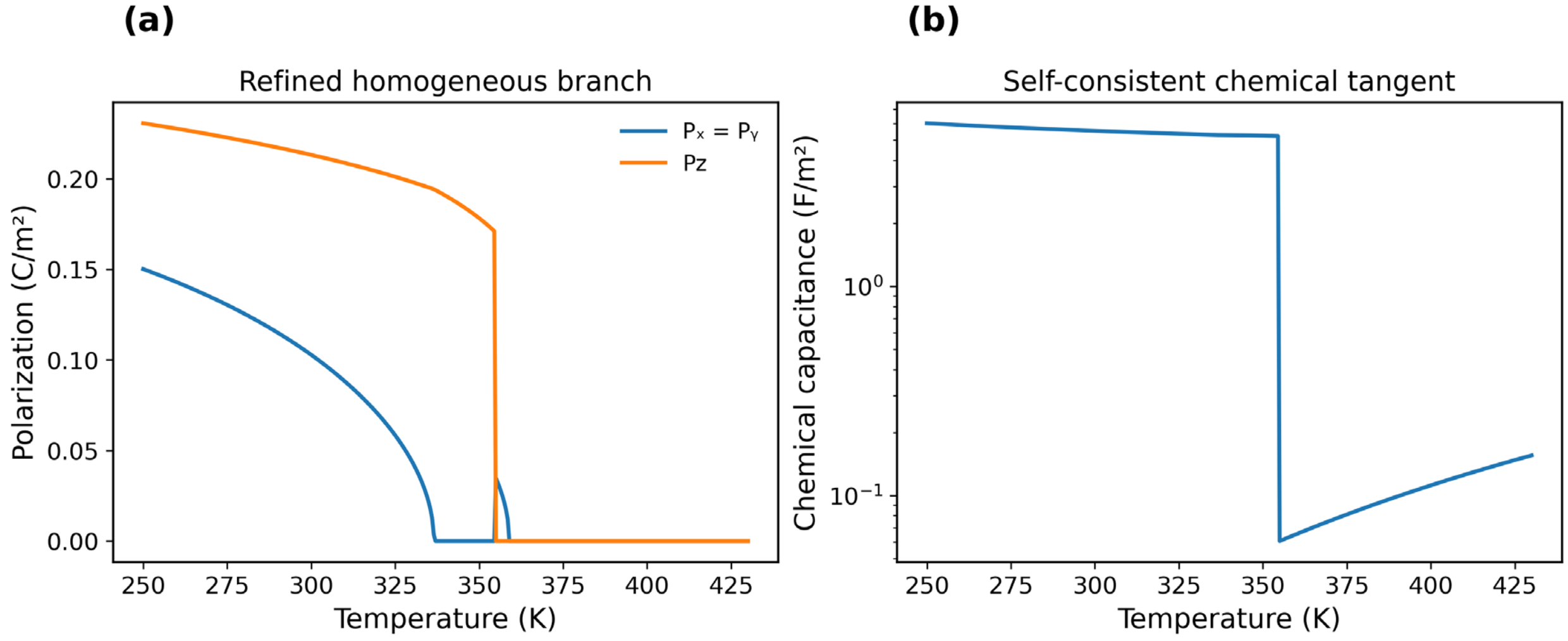

**Figure S11.** Refined balanced $BaTiO_3$ branch at exactly 20 nm and −0.100% misfit. (a) Polarization components. (b) Self-consistent chemical capacitance. The 361-point temperature sweep resolves the uniform branch and does not hold polarization fixed during the capacitance comparison.

### S12 Reference spectra, mode phases, and spatial orientation

The reference calculation first minimizes over wavevector magnitude and azimuth, then evaluates the three lowest eigenvalues along the selected azimuth on a 92-point magnitude grid including zero. Frozen-ion spectra retain the same equilibrium polarization, stress, and occupations while suppressing only the ionic fluctuation. Columnar spectra restrict polarization to be constant through the thickness while permitting the corresponding fields to relax. Comparing these calculations isolates chemical fluctuation screening and depth relaxation, respectively. Changing the reference chemistry between the curves would not isolate either effect.

The depth eigenvector is evaluated with 33 polarization nodes. A common complex phase is chosen so that its largest component is real, and both real and imaginary parts are plotted. This gauge choice changes no observable. A nearly imaginary component represents spatial quadrature relative to the dominant component rather than a time delay. The eigenvectors are normalized in the continuum mass inner product, not by the largest sampled value. Their relative sizes can therefore be compared within each mode, while converting a mode into a physical polarization requires specifying an amplitude.

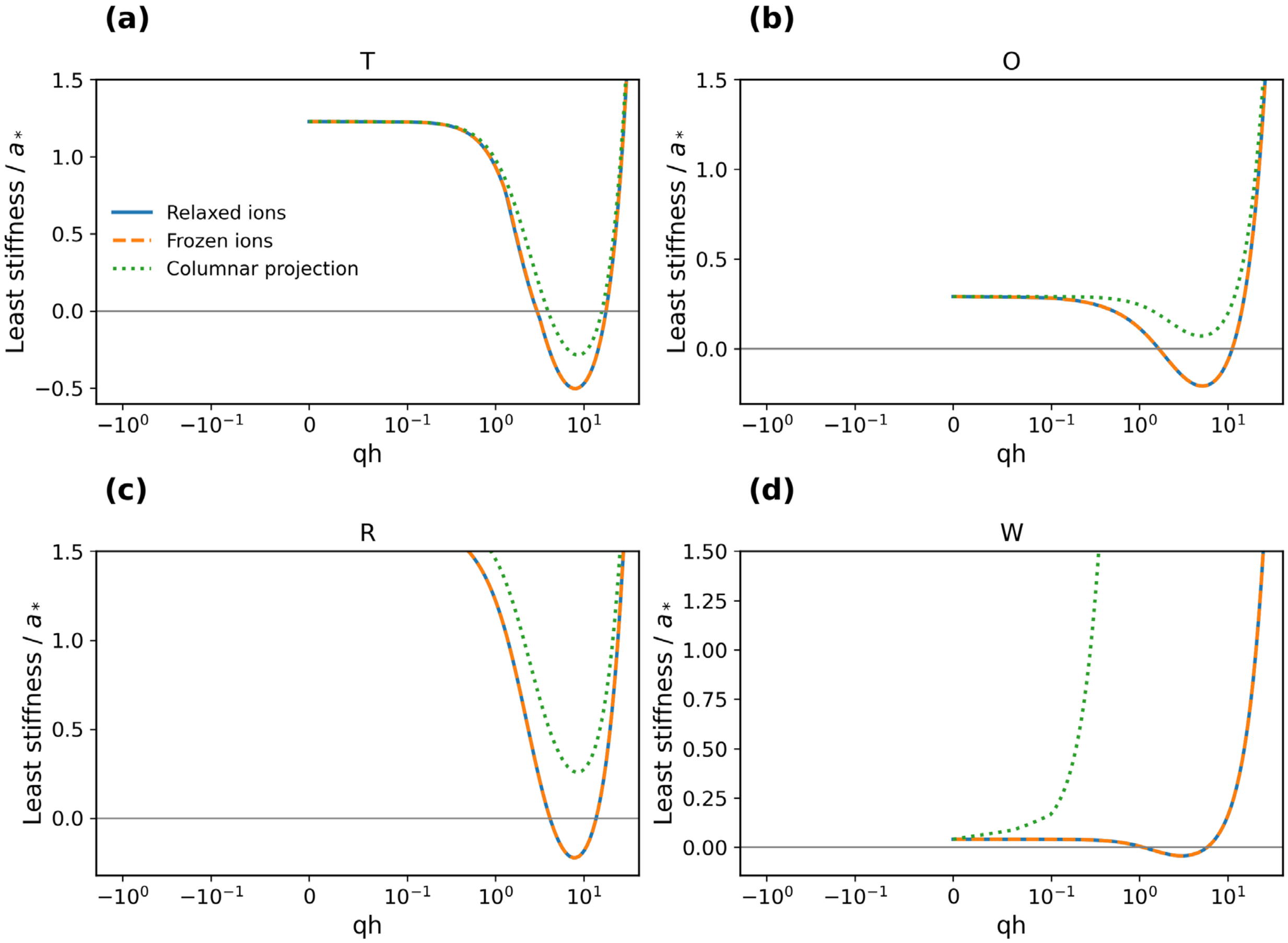

**Figure S12.** Static reference spectra for T, O, R, and W. Relaxed-ion and frozen-ion curves share the same equilibrium reference. The columnar projection restricts polarization to be constant through depth. A negative eigenvalue establishes local instability in the tested space; a positive projected curve alone does not establish full-depth stability.

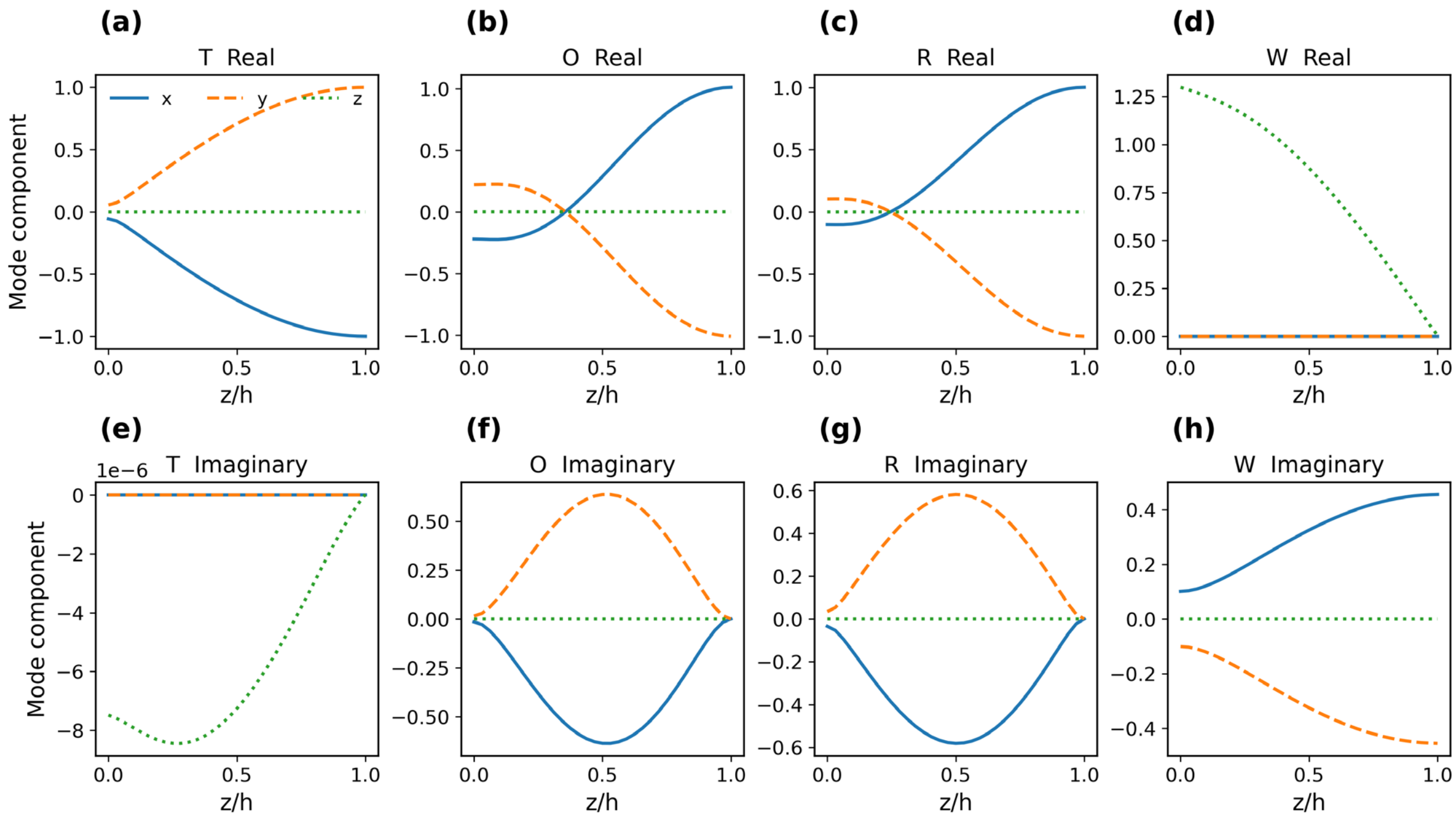

**Figure S13.** Depth profiles and spatial phases of the least-stiff reference modes. (a–d) Real parts for T, O, R, and W. (e–h) Imaginary parts. A common phase makes the largest component real. Imaginary components represent spatial quadrature in a Fourier mode, rather than an oscillation in time.

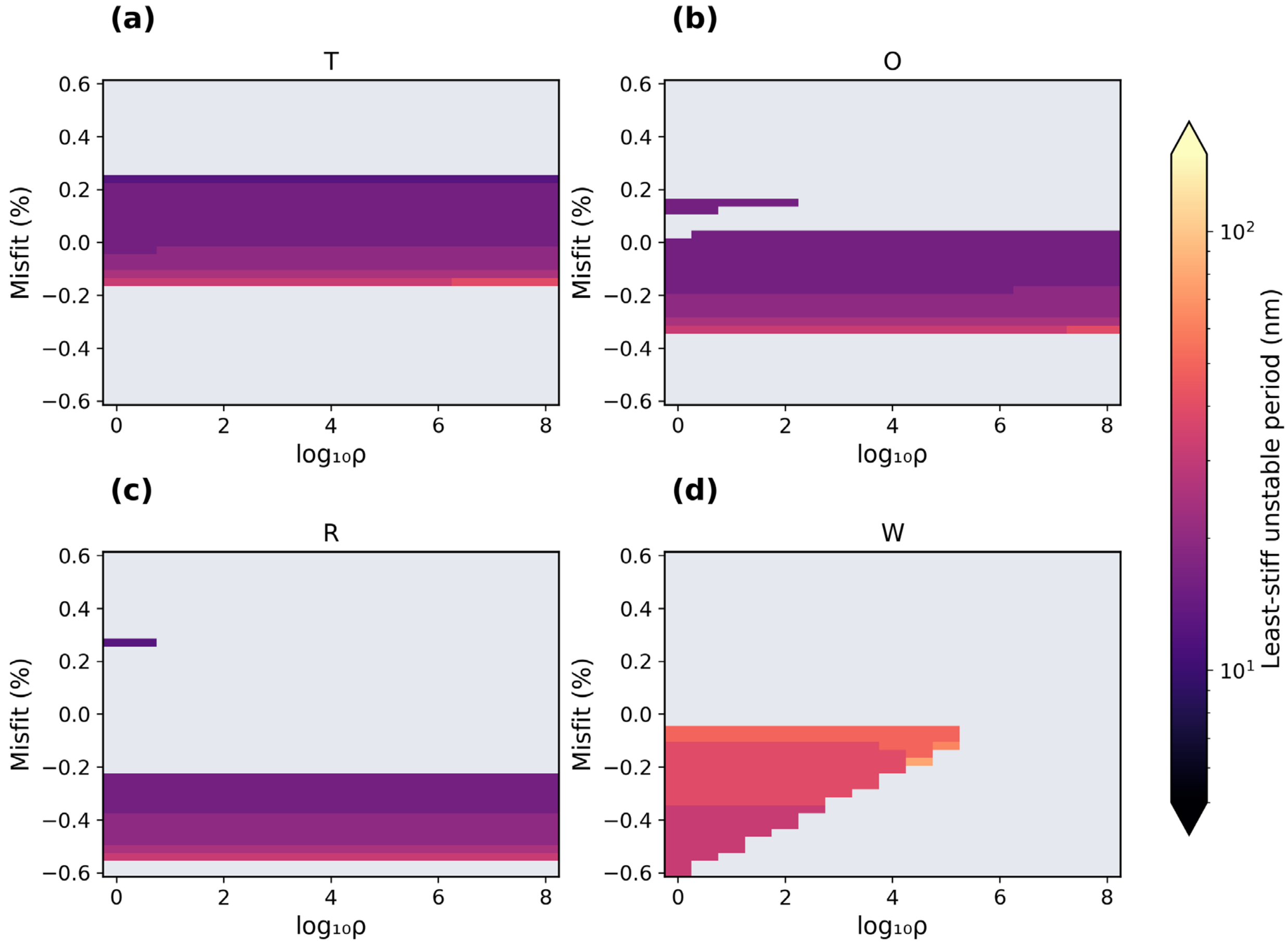

**Figure S14.** Least-stiff unstable periods on the strain–activity grids for (a–d) T, O, R, and W. Gray indicates no selected negative finite-wavevector mode. These maps supplement the component and eigenvalue maps in Figure 7 and do not specify a developed domain repeat.

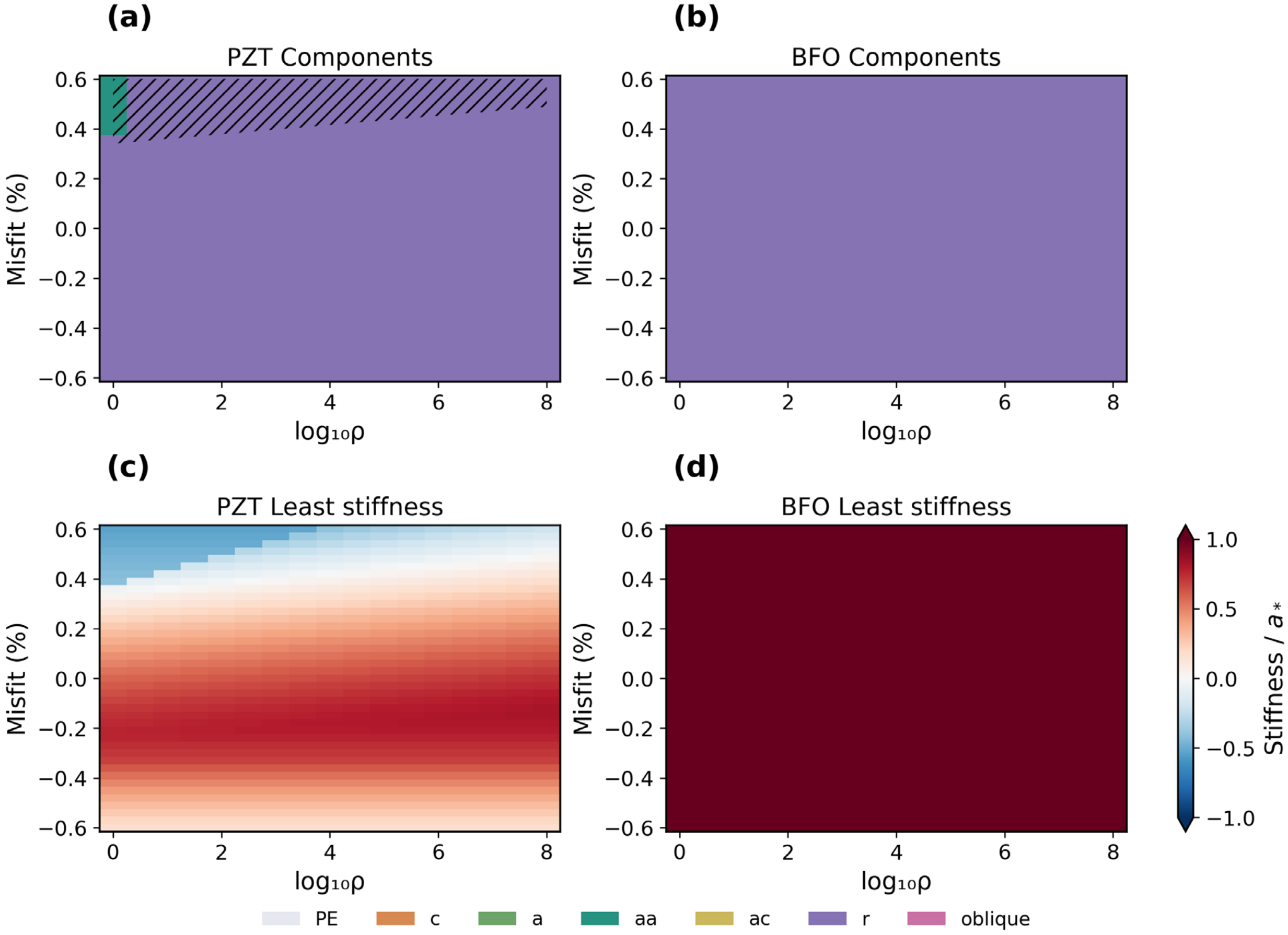

**Figure S15.** Additional strain–activity comparisons at 300 K. (a,b) PZT 50/50 and $BiFeO_3$ component maps with instability hatching. (c,d) Their least sampled spatial stiffnesses. The $BiFeO_3$ comparison retains its stated ionic-capacity and constitutive limitations.

At an ideal rhombohedral bulk minimum the two local rotation curvatures are degenerate. At an ideal orthorhombic minimum they are generally distinct. Clamping, the surface normal, the reference tilt, and wavevector orientation break or redistribute these symmetries in the film. The calculations therefore diagonalize the complete film tangent, rather than attaching a fixed amplitude/rotation classification to every point of a component map. An angular scan at a selected magnitude provides a useful sensitivity test but does not replace joint optimization over magnitude and angle. Both calculations are retained in the code.

## S13 Developed domains, period selection, and numerical verification

### S13.1 One nonlinear energy and its exact derivatives

The nonlinear film solver uses a lateral Fourier representation and a piecewise-linear depth representation of all three polarization components. Electrical and displacement fields are solved on an enriched depth grid with $2n_z - 1$ nodes. Polarization is interpolated to that grid, and the local energy and its derivatives use the same quadrature there. Eliminating the auxiliary fields produces the nonlinear energy, gradient, and Hessian action used by both optimization and Bloch analysis. The mass matrix is the projected quadrature matrix $R^T W R$, not an unrelated diagonal approximation. This construction is essential for agreement between the Hessian of a nonlinear cell and the independently assembled linear operator.

The surface potential and common-site occupations are solved together. Log-sum-exp evaluation avoids overflow of the Langmuir weights, while the chemical tangent uses the full two-species covariance. No mean-charge subtraction is applied. The energy reference is the homogeneous state evaluated on the same mesh and in the same reservoir. Nonlinear gradient residuals are reported after inversion of the polarization mass matrix, so refining the number of nodes cannot artificially improve a residual solely by reducing each quadrature weight.

The initial stripe search tests seven lengths equal to 0.65, 0.80, 1.00, 1.25, 1.60, 2.00, and 2.60 times the least-stiff reference period. It uses 33 lateral points and 17 polarization nodes, with critical-mode, variant-like, and continued seeds. The best state is relaxed on 65 lateral points and 25 polarization nodes. A second scan samples seven lengths around the resulting primitive repeat, with at least 49 lateral points, 25 depth nodes, and an adaptive lateral spacing target of 1.25 nm. Each refined minimum in Figure S17 lies inside its scanned interval. This establishes a sampled minimum in the stripe family, not a continuum-accurate optimum or a global minimum over arbitrary textures.

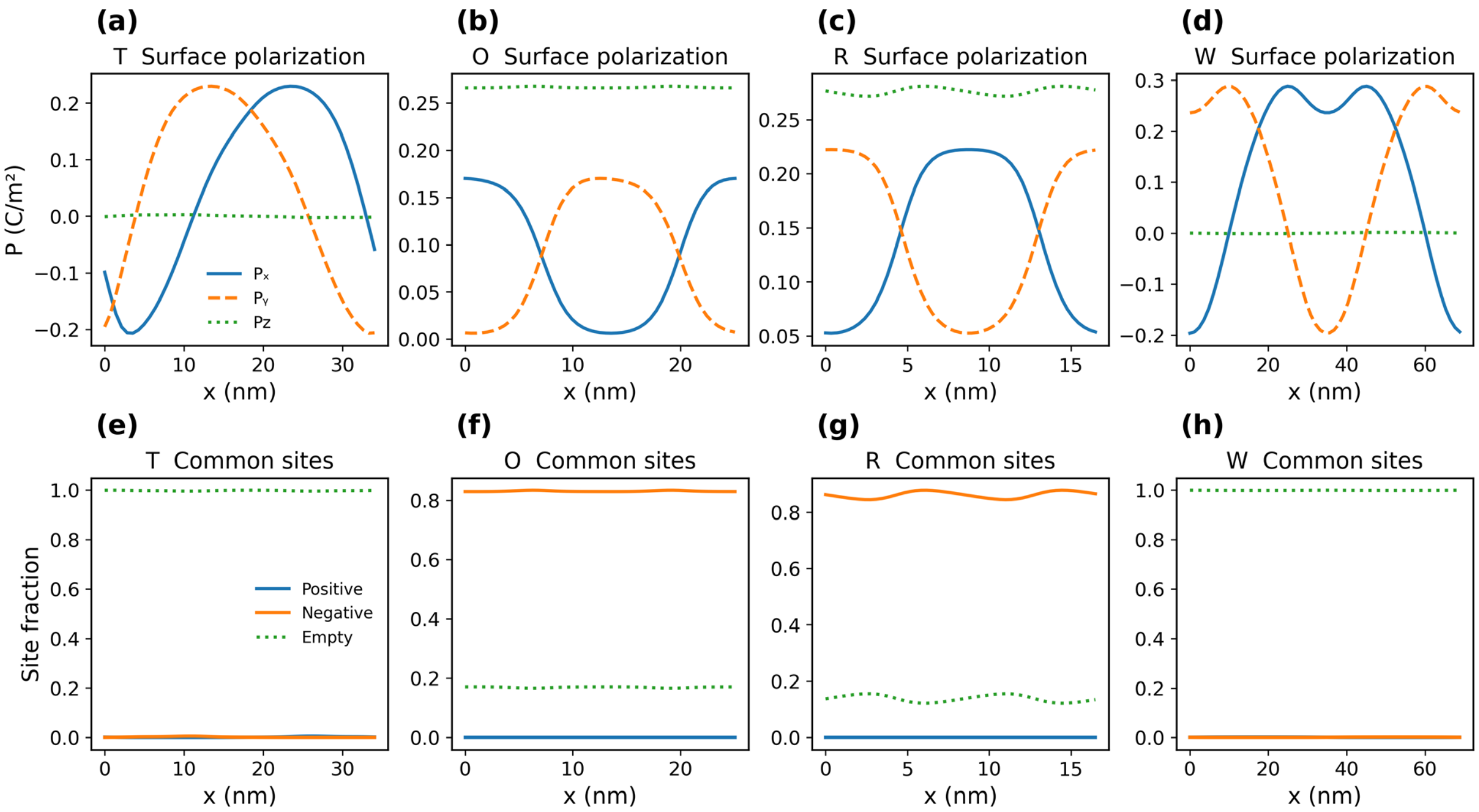

**Figure S16.** Surface profiles of the optimized film candidates. (a–d) Cartesian polarization components for T, O, R, and W. (e–h) Positive, negative, and empty site fractions. The three fractions sum to one locally. The empty fraction is essential to the common-site chemical entropy and its response.

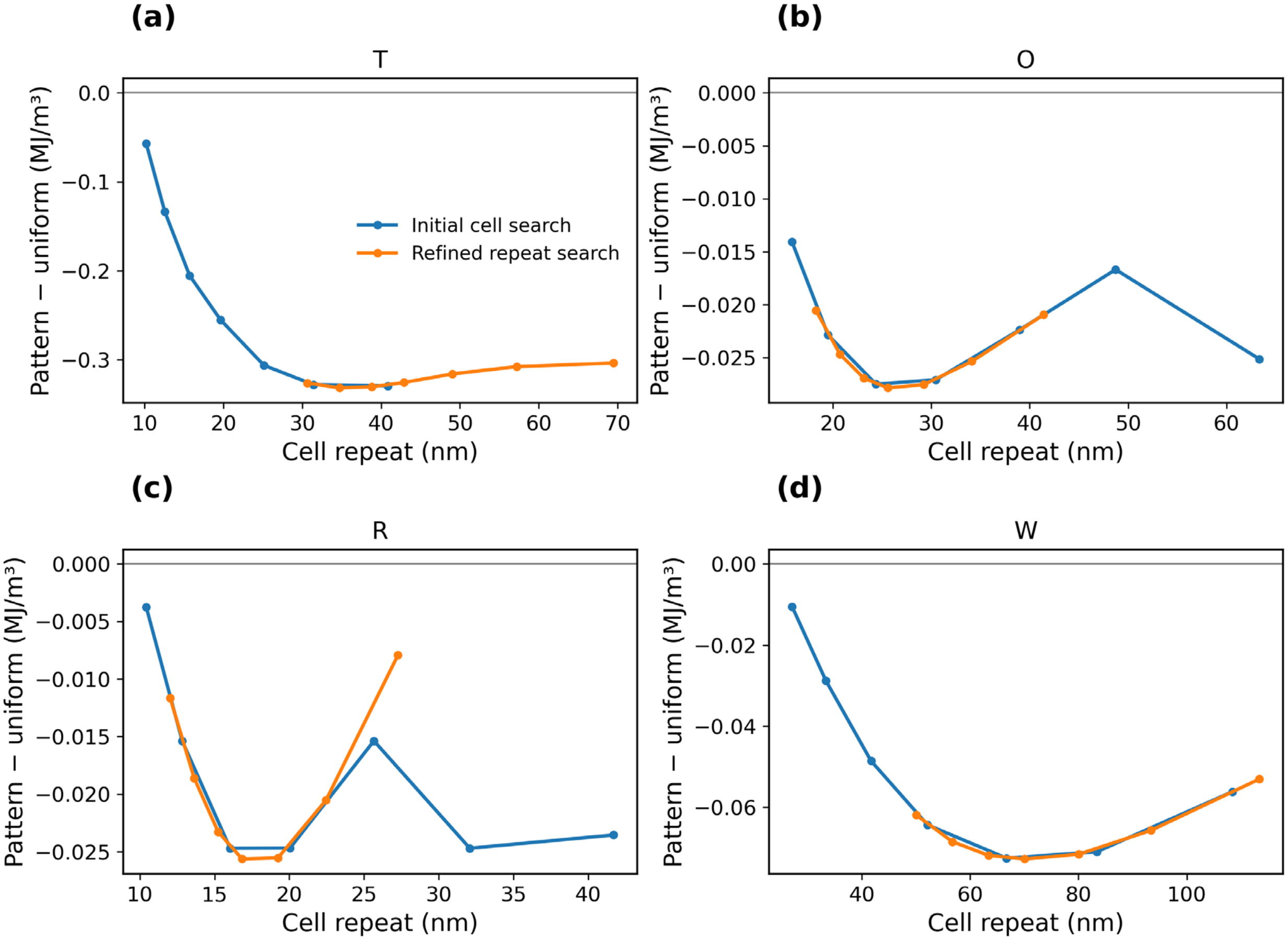

**Figure S17.** Initial and refined period-energy searches. (a–d) T, O, R, and W. The initial search uses cells based on the reference spectral period; the refined search resolves the neighborhood of the primitive-repeat minimum. Energies are relative to a homogeneous field on the same mesh. Changes in box harmonic number are retained in the numerical records.

### S13.2 Primitive repeat and period optimization

The strongest Fourier harmonic is not necessarily the primitive repeat of a vector texture. A normal component can repeat after half a cell while an in-plane component changes sign. The check therefore considers the significant Fourier harmonics of the complete vector field and their common divisor, with direct inspection of the resulting profiles. A repeated cell can then be reduced before period optimization. In the recalculated R example, a larger cell contains two repeats. The refined best states displayed in Figure 9 each have a resolved first harmonic of their full primitive cell.

The nonlinear candidate maps in Figure 10 use nine misfits and five chemical activities for each class, for 180 control points total. Each point tests three lengths, 0.8, 1.2, and 1.8 times the marginal period, with two distinct seeds. This gives 540 cell conditions and 1080 initial relaxations. Selected cells with a force residual above $5 \times 10^{-5}$ are then continued with a larger iteration allowance before their energies are plotted. The homogeneous state and its least spatial stiffness are independently recalculated at the same controls. A candidate is lower in energy only when its reduction exceeds the reporting tolerance. No lower-energy stripe is resolved on the stable side in these sampled film windows. This statement does not exclude a remote minimum, a different repeat outside the search, or a two-dimensional pattern not represented by these cells.

Table S4 compares the spectral periods with the nonlinear repeats.

**Table S4.** Reference spectral periods, lowest-energy sampled primitive repeats, and nonlinear residuals. Energies are relative to matched-grid homogeneous fields.

| Case | $L_{min}$ (nm) | $L_{stripe}$ (nm) | $\Delta f$ (MJ/m$^3$) | Force residual |
|---|---|---|---|---|
| T | 15.7 | 34.7 | -0.332 | 3.19e-06 |
| O | 24.4 | 25.6 | -0.0279 | 1.11e-05 |
| R | 16 | 16.8 | -0.0256 | 9.91e-06 |
| W | 41.7 | 70.1 | -0.0727 | 1.75e-06 |

### S13.3 Bloch stability of an already developed stripe

For a stripe of repeat $L$ and wavevector $q = 2\pi/L$, a perturbation is $e^{i(k_x x + k_y y)} \sum_m u_m(z) e^{imqx}$. The local Landau, prestress, eigenstrain, and chemical tangents are evaluated on the nonlinear periodic field. The compatible electrical and mechanical solves use the shifted wavevectors $(mq + k_x, k_y)$. This is the Hessian of the domain, not the Hessian of its original homogeneous reference. Surface capacitance is spatially varying and follows the local occupations. The full mass matrix is used to whiten the generalized eigenproblem before the lowest eigenvalues are calculated.

The reported tests use offsets $0$, $0.025q$, $0.10q$, $0.25q$, and $0.50q$ along the longitudinal and transverse directions. Each field is relaxed anew on both 33-by-17 and 49-by-25 lateral/depth grids. Three low eigenvalues are retained with eigensolver tolerance $2 \times 10^{-6}$. Translation gives a neutral zero-offset mode in the continuum; lattice pinning and incomplete resolution can shift it. In particular, the T translation error decreases markedly on refinement, and a positive value on a coarse mesh is not assigned a physical restoring force for translation.

Table S5 separates the translation diagnostic from the sampled nonzero-offset stability test.

**Table S5.** Refined optimal-repeat Bloch checks on 49 lateral points and 25 polarization nodes. The zero-offset translation values are reported separately from nonzero-offset minima.

| Case | Zero-offset minimum | Nonzero-offset minimum | Force residual |
|---|---|---|---|
| T | 0.000384 | 0.000417 | 3.19e-06 |
| O | -5.41e-07 | 9.5e-05 | 2.97e-06 |

| Case | Zero-offset minimum | Nonzero-offset minimum | Force residual |
|---|---|---|---|
| R | 1.12e-07 | 0.000219 | 5.18e-06 |
| W | 1.19e-08 | -0.000249 | 3.92e-06 |

The T, O, and R results have no resolved negative nonzero-offset eigenvalue. The W transverse branch is negative at several offsets and remains negative on both spatial grids. Its negative scale exceeds the reported stationarity residual, while the translation value approaches zero under refinement.

The sampled offsets do not constitute an exhaustive scan of the two-dimensional Bloch zone. A negative resolved eigenvalue proves instability in its tested sector; nonnegative samples establish only sampled stability. A stable period band can contain periods with different energies, so a pattern can persist locally without selecting the globally lowest stripe energy. Conversely, a low stripe energy does not protect it against a transverse rotation or bending mode. Figure 11 puts period energetics and domain stability side by side to avoid conflating these criteria.

### S13.4 Second lateral direction and parameter sensitivities

The earlier confined W descendant is tested in a cell with lateral dimensions $L$ and $4L$, using 17-by-49 lateral points and nine polarization depth nodes. The initial stripe is first relaxed on that exact grid. A transverse displacement of the stripe phase plus a small perturbation is then relaxed with all vector components free. The energy difference is measured against the matched-grid stripe rather than against a higher-resolution one-dimensional energy. This test demonstrates an accessible descending route in a finite cell. Its morphology and energy are not claimed as a mesh-converged global equilibrium. The larger square-domain replacement for the main Figure 12(a–c) is documented separately in Section S17.

Figure S18 separates the effect of wavevector orientation from electrostrictive strength. The former scans 73 azimuths at the reference magnitude; the latter changes $Q/Q_{BTO}$ over 25 positive values in the W model at zero misfit, re-solving its homogeneous state and spatial minimum. Figure S20 changes the ratio of the positive depth gradient coefficient to the lateral coefficient from 0.2 to 4 at the reference wavevector. This tests one specified anisotropy and does not purport to span all cubic gradient tensors. No negative gradient coefficient is introduced.

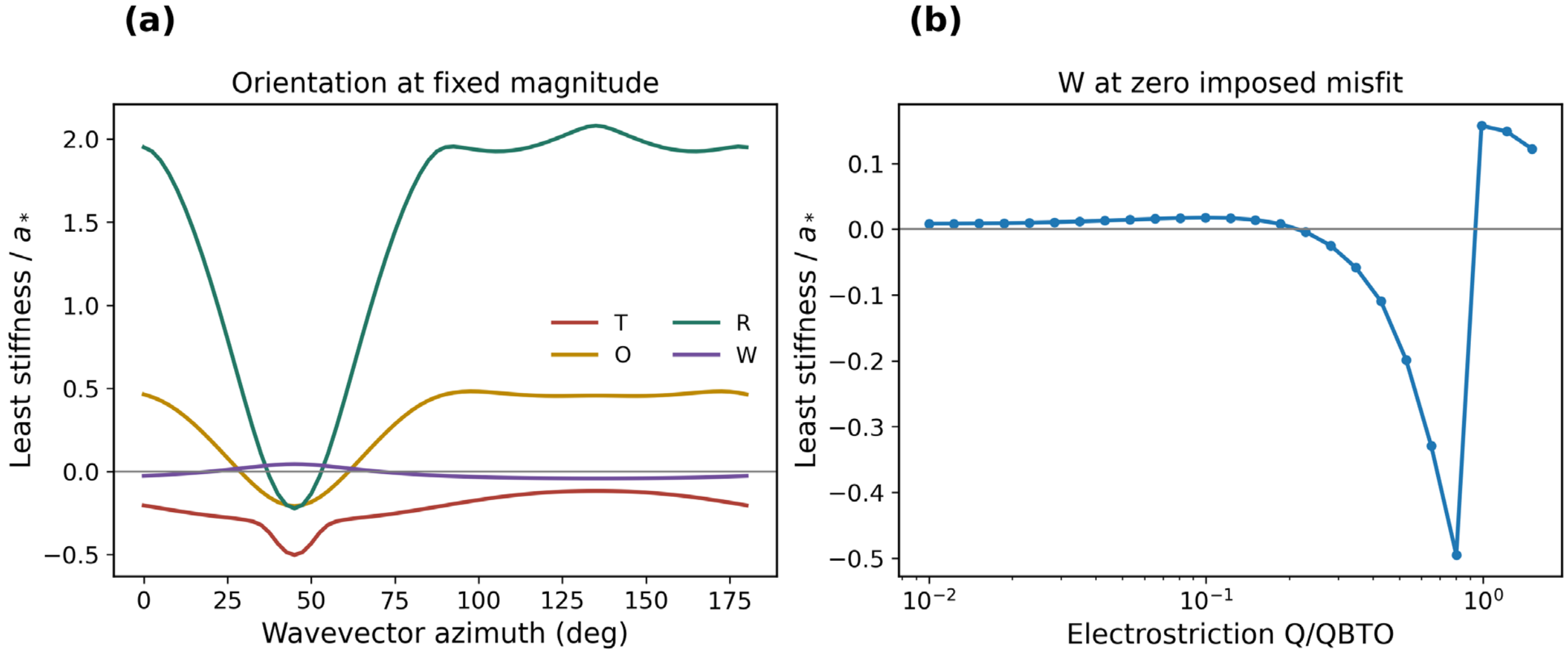

**Figure S18.** Orientation and electrostriction tests. (a) Least stiffness versus wavevector azimuth at each reference magnitude for all four film classes. (b) W spatial minimum as electrostriction is varied at zero misfit, with the homogeneous state re-equilibrated at each value. The two panels vary different physical controls.

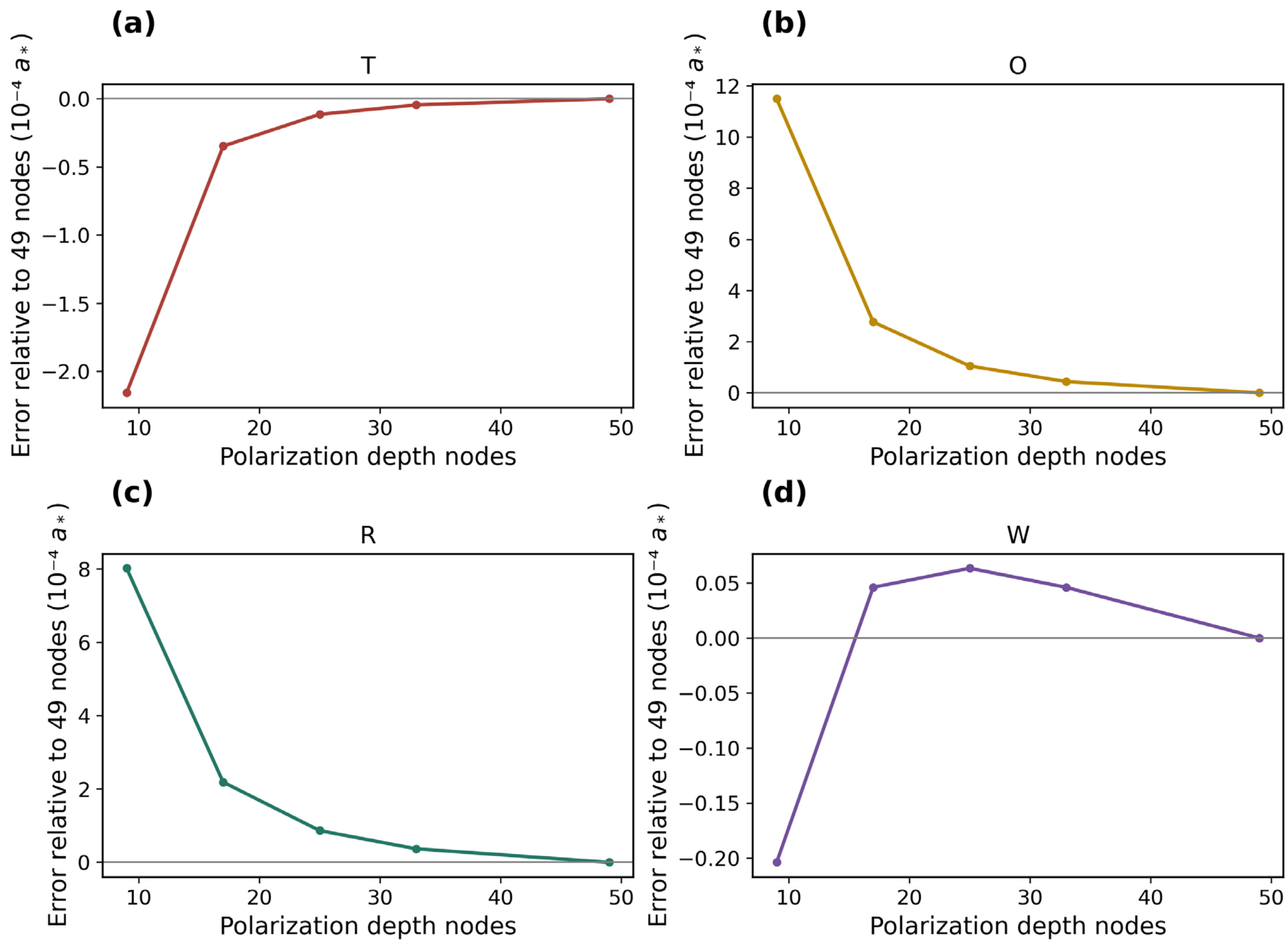

**Figure S19.** Depth convergence of the reference spectra. (a–d) T, O, R, and W. Values show the difference from the 49-node result while magnitude and azimuth are refined at each resolution. Electrical and displacement grids remain enriched. These errors do not quantify uncertainty in the constitutive parameters.

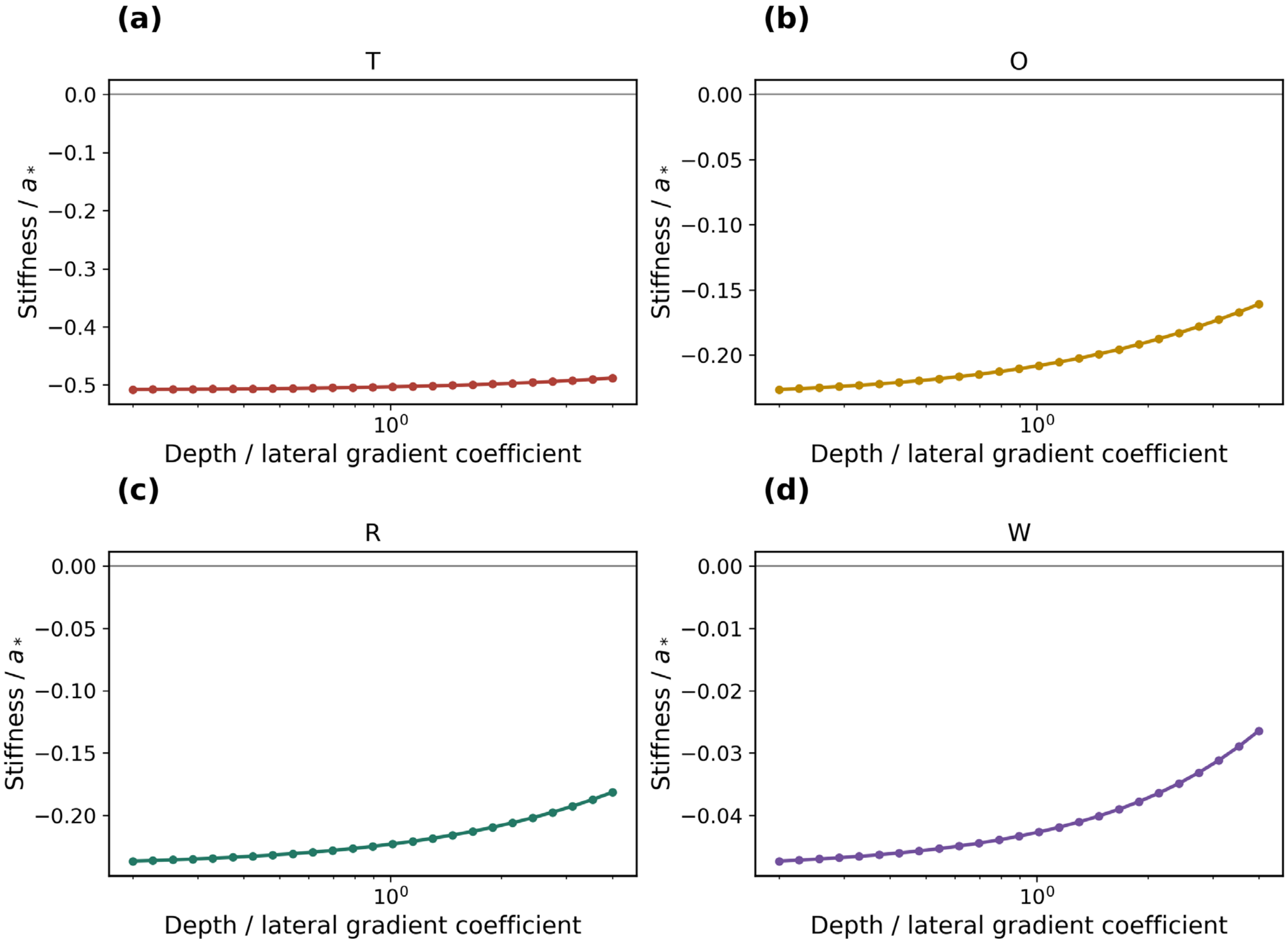

**Figure S20.** Positive gradient-anisotropy sensitivity. (a–d) T, O, R, and W least stiffnesses at the reference wavevector as the depth-to-lateral gradient coefficient ratio varies from 0.2 to 4. The lateral coefficient stays fixed. The calculation does not introduce a negative gradient term or re-optimize the wavevector at every ratio.

The $BiFeO_3$ capacity sweep in Figure S21 varies the common site density over 41 values from $2 \times 10^{17}$ to $5 \times 10^{18}$ m$^{-2}$ at 300 K and $u_m = -0.100\%$. The reference vector, potential, occupation, and differential capacitance are re-equilibrated at each density. The single-sign capacity is $2eN_s$ even though both charge signs use the same sites. Values near saturation can generate substantial unscreened fields and require additional electronic or structural physics for experimental interpretation. Increasing capacity is a sensitivity test, not evidence that the larger site density belongs to the actual surface.

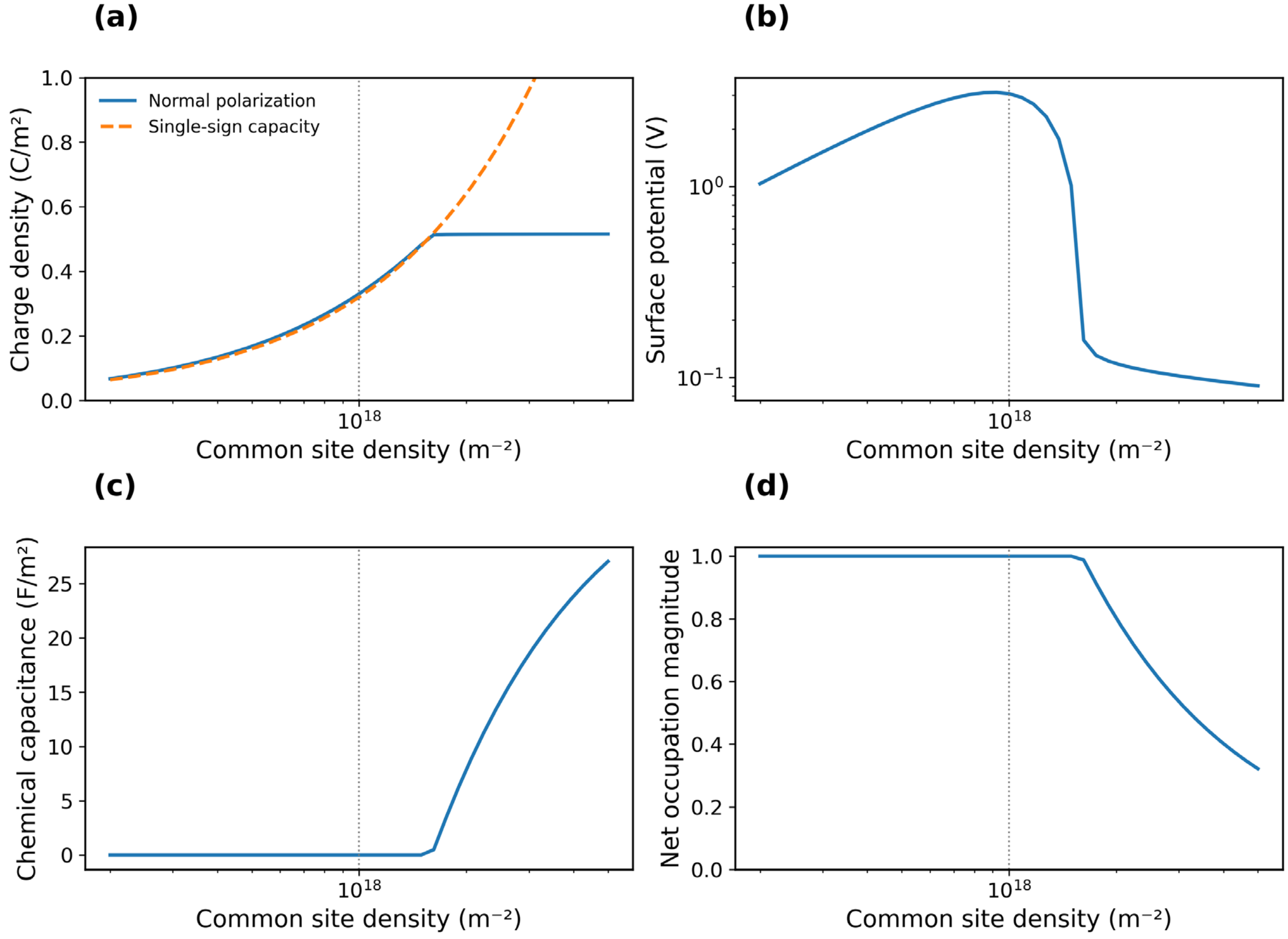

**Figure S21.** Common-site charge-capacity audit for the $BiFeO_3$ reference at 300 K, 20 nm, and $-0.100\%$ misfit. (a) Normal polarization and single-sign ionic capacity. (b) Surface potential. (c) Chemical capacitance. (d) Net occupied-charge fraction. The vertical line marks the baseline density $10^{18}$ $m^{-2}$. Every point is re-equilibrated.

### S13.5 Verification and reproducibility

Table S6 records the independent numerical checks.

**Table S6.** Implementation checks. Unless a physical unit or relative normalization is specified, differences use the dimensionless conventions of the corresponding calculation.

| Independent comparison | Numerical difference |
|---|---|
| Site-fraction excess | 0 |
| Gauss balance (C/m²) | 1.11e-16 |
| Capacitance derivative, relative | 4.69e-10 |
| Matched scalar zero-charge capacitance, relative | 3.33e-16 |

| Independent comparison | Numerical difference |
|---|---|
| Projected open kernel, largest relative difference | 5.31e-05 |
| Uniform polar Hessian | 4.97e-14 |
| Nonlinear energy directional derivative | 1.35e-12 |
| Nonlinear Hessian action, relative | 2.47e-10 |
| Independent linear/nonlinear second variation | 6.14e-09 |
| Batched versus individual eigensolution | 5.11e-15 |
| Surface fixed-charge chemical potential (V) | 6.94e-17 |
| Surface chemical Hessian, relative | 2.55e-15 |
| Surface uniform-kernel limit, relative | 5.09e-09 |
| Surface-kernel interpolation, relative | 1.32e-12 |
| Surface energy directional derivative | 7.08e-12 |
| Early kinetic growth-rate error at step 0.002 | 0.00876 |

The verification compares independently constructed quantities: the derivative of the exact isotherm with the charge covariance; the scalar homogeneous and columnar electrical reductions with the vector implementation; finite differences of the nonlinear energy with its analytic gradient; finite differences of that gradient with the Hessian action; and a real sinusoidal nonlinear-cell second variation with the independently assembled Fourier eigenvalue. The analytical open-surface kernel is compared with a finely resolved variational field solution. These tests are sensitive to an incorrect electrostatic sign, a missing empty-site term, an inconsistent mass matrix, or double-counted elasticity.

The numerical verification is distinct from material validation. A small derivative residual establishes consistency of the stated discretized functional, not accuracy of an uncertain adsorption energy or gradient coefficient. The finite map grids, constrained candidate families, and missing material order parameters remain physical and numerical limits after the implementation checks pass. All such controls are exposed in the notebook. Generated result files are intermediate products of a run; none is required before the notebook starts.

## S14 Surface ordering, shared chemistry, and relaxation

### S14.1 Constitutive limit and its scope

The surface-ordering calculation uses a stable Gaussian vector bulk, $f_b = a|\mathbf{P}|^2/2 + G|\nabla\mathbf{P}|^2/2$, in a film of thickness 300 nm. Its upper surface adjoins the same unbounded dielectric exterior and its bottom electrode is grounded. The values $a = 10^7$ m F$^{-1}$, $G = 10^{-8}$ J m$^3$ C$^{-2}$, and $\epsilon_b = 7.35$ give a soft but locally stable polarization. The surface values $\mathbf{P}_s$ have a nonlinear local energy given explicitly in the main paper. The bottom polarization condition is natural, $\partial_z \mathbf{P}(0) = 0$. The top polarization derivative is determined by variation of the surface energy after bulk elimination.

This is a limit of small bulk angular anisotropy and neglected explicit electrostriction in the Gaussian constitutive sector. Electrostriction is quadratic in polarization and, about a stress-free unpolarized reference, first contributes beyond the linear Gaussian polar operator. A specified background stress would renormalize its quadratic coefficients; compatible electrostriction at finite amplitude would add a nonlocal quartic contribution. Those terms are not implicitly claimed to be fitted by the three surface completions. The separate W film calculations retain explicit electrostriction and show its consequences. The surface calculation instead isolates how normal surface preference and nonlinear completion modify ordering with a stable, soft bulk. Treating it as a quantitative material-specific relaxor model would require those additional coefficients and a disorder model.

The chemical parameters are unchanged: common site density $N_s = 10^{18}\ \mathrm{m}^{-2}$, charges $\pm 2e$, standard formation energy 0.200 eV, and opposite stoichiometries $\nu_\pm = \mp 1/2$. The scale $P_s^* = 0.0500\ \mathrm{C\ m}^{-2}$ is a unit choice, not a site capacity. The charge capacity remains $Q_s = 2eN_s = 0.320\ \mathrm{C\ m}^{-2}$. The energy scale is $E_s^* = 0.00250\ \mathrm{J\ m}^{-2}$, so the quadratic kernel unit is $K_s^* = 1\ \mathrm{m}^2\ \mathrm{F}^{-1}$. The same site constraint and net-charge exchange channel are used in equilibrium, the linear spectrum, full-period minimization, and kinetics.

### S14.2 Exact elimination of the finite Gaussian bulk

Choose a nonzero lateral wavevector along the longitudinal direction and write $P_l = iX$, $P_z = Z$, with Fourier convention $e^{iqx}$. The factor of $i$ makes the sagittal coefficient matrix real. The bulk equations are

$$[a - G(\partial_z^2 - q^2)]X = -q\phi, \qquad [a - G(\partial_z^2 - q^2)]Z = -\phi', \qquad \epsilon_f(\partial_z^2 - q^2)\phi = -qX + Z'. \tag{S46}$$

The sign and the gradient operator in the first two equations follow directly from $(a - G\nabla^2)\mathbf{P} = -\nabla\phi$. The gradient term cannot be dropped when comparing a depth-dependent mode with a surface coefficient. Define

$$k_E = q, \qquad k_L = \sqrt{q^2 + \left(a + 1/\epsilon_f\right)/G}, \qquad k_T = \sqrt{q^2 + a/G}. \tag{S47}$$

For each signed exponent $\kappa$, independent sagittal solutions have vectors

$$\begin{aligned} \kappa = \pm k_E&: \quad (X, Z, \phi) = (\kappa/q, 1, -a/\kappa)e^{\kappa z}, \\ \kappa = \pm k_L&: \quad (X, Z, \phi) = \left(q/\kappa, 1, 1/(\epsilon_f \kappa)\right) e^{\kappa z}, \\ \kappa = \pm k_T&: \quad (X, Z, \phi) = (\kappa/q, 1, 0)e^{\kappa z}. \end{aligned} \tag{S48}$$

Six coefficients are fixed by $X'(0) = 0$, $Z'(0) = 0$, $\phi(0) = 0$, prescribed $X(h) = X_s$ and $Z(h) = Z_s$, and $\epsilon_f \phi'(h) + \epsilon_e q\phi(h) - Z_s = \sigma$. The last condition is the open-surface displacement jump. Positive and negative exponentials are anchored at the nearest boundary in the numerical implementation, so their magnitudes do not overflow for large $k_L h$. Row scaling of the six-by-six system prevents a derivative row measured in inverse length from overwhelming a potential row.

The minimized bulk-plus-electrical energy is quadratic in $(X_s, Z_s, \sigma)$, and its surface conjugate forces are $(GX'(h), GZ'(h), \phi(h))$. Solving the boundary problem for each unit surface variable gives the three-by-three sagittal kernel. The uncoupled transverse-polarization stiffness is $Gk_T\tanh(k_T h)$. Rotating back to $(P_{s,x}, P_{s,y}, P_{s,z}, \sigma)$ restores the imaginary electrical couplings. This procedure gives the full Hermitian four-by-four kernel $K(\mathbf{q}; h)$ used in all surface calculations. It eliminates the Gaussian bulk exactly at finite surface amplitude; nonlinearity resides only in the specified surface and chemical terms.

### S14.3 The uniform charge channel

At $q = 0$ the exterior field is zero, while the grounded electrode permits uniform charge exchange. Define $\chi = (1 + a\epsilon_f)^{-1}$, $k_n = \sqrt{(a + 1/\epsilon_f)/G}$, $k_t = \sqrt{a/G}$, and $K_n = Gk_n\tanh(k_n h)$. The nonzero uniform entries are

$$K_{xx} = K_{yy} = Gk_t\tanh(k_t h), \quad K_{zz} = K_n, \quad K_{z\sigma} = K_{\sigma z} = \chi K_n, \quad K_{\sigma\sigma} = ha\chi + \chi^2 K_n. \tag{S49}$$

One way to obtain these expressions is to use constant displacement $D_z = -\sigma$, giving $E_z = (-\sigma - P_z)/\epsilon_f$. Then $(a + 1/\epsilon_f)P_z - GP_z'' = -\sigma/\epsilon_f$. The natural bottom condition and prescribed top polarization give $P_z(z) = -\chi\sigma + (P_{s,z} + \chi\sigma)\cosh(k_n z)/\cosh(k_n h)$. Its top gradient supplies the normal force. Integrating the field gives the charge-conjugate potential. The tangential solution is the corresponding uncharged cosh profile. This derivation confirms that the uniform kernel is not obtained by deleting the charge row or by setting the exterior potential independently to zero.

The finite-wavevector kernel approaches this uniform kernel continuously as $q \to 0$. The numerical audit checks this limit as well as Hermiticity and energy derivatives. Keeping the uniform sector is necessary to compare periodic candidates with uniformly polarized, ionically compensated states. A zero-mean-charge constraint would suppress that competitor and could incorrectly enlarge the apparent periodic region.

### S14.4 Exact chemical reduction at fixed local charge

For static surface calculations it is efficient to retain $\sigma$ and eliminate the total ionic population locally. Let $m = \sigma/Q_s = \theta_+ - \theta_-$, $x = \Delta g^0/(k_B T)$, $c = e^x/2$, and $\delta = 1/2\ln\rho$. At fixed $m$, the chemical bias contributes the linear term $N_s k_B T\delta m$ and does not change the optimum total population. A Lagrange multiplier $\eta$ for charge gives

$$\eta(m) = \operatorname{artanh} m + \operatorname{arsinh}\left(\frac{cm}{\sqrt{1-m^2}}\right), \qquad Z_m = 1 + 2e^{-x}\cosh\eta, \tag{S50}$$

$$\theta_0 = Z_m^{-1}, \qquad \theta_\pm = e^{-x\pm\eta}/Z_m. \tag{S51}$$

These fractions satisfy the common-site constraint and the prescribed difference exactly for $|m| < 1$. The reduced chemical energy is $g_c = N_s k_B T[\sum_a \theta_a \ln\theta_a + x(\theta_+ + \theta_-) + \delta m]$, up to a reservoir-dependent constant. Its derivatives are

$$\frac{dg_c}{dm} = N_s k_B T(\eta + \delta), \qquad \frac{d^2 g_c}{dm^2} = \frac{N_s k_B T}{1-m^2}\left[1 + \frac{c}{\sqrt{1+(c^2-1)m^2}}\right]. \tag{S52}$$

Thus $dg_c/d\sigma = k_B T(\eta + \delta)/(2e)$ and $d^2 g_c/d\sigma^2 = 1/C_{chem}$ at electrochemical equilibrium. Minimizing $g_c + \sigma\psi$ gives $\eta = -\delta - 2e\psi/(k_B T)$ and exactly recovers the Langmuir weights in Section S2. The reduction is therefore an algebraic change of variables within the same chemical model. It is not a neutral-pair ensemble and imposes no mean-charge condition.

The solver bounds $|m|$ below unity. A smooth convex continuation immediately outside the internal evaluation cutoff is used only to make a line search well defined; accepted physical candidates lie inside the actual occupation domain. Maximum charge fraction and optimizer residual are stored for every surface candidate. The phase classifier excludes a candidate that reaches the numerical occupation limit. The fixed-charge derivative and curvature are independently compared with the potential-based film isotherm at three temperatures, three activities, and both signs of potential.

### S14.5 Normal preference and the common spinodal

The total surface energy is

$$\Omega_s/A = \frac{1}{2}\sum_{\mathbf{q}} \mathbf{u}_s^\dagger K(\mathbf{q})\mathbf{u}_s + \langle f_s(\mathbf{P}_s) + g_c(\sigma)\rangle, \qquad \mathbf{u}_s = (P_{s,x}, P_{s,y}, P_{s,z}, \sigma). \tag{S53}$$

Its three polar completions are those of Eq. (16b). They share $a_{s,n}(T) = -3.18 + 0.0200(T - 300\,\mathrm{K})/\mathrm{K} + \Delta a_s$ in $\mathrm{m^2\ F^{-1}}$. They also share the same chemical energy and Gaussian kernel. At balanced activity, the unpolarized state has equal occupations of the two ionic signs. Eliminating the positive tangential/charge block of its Hessian gives $\Lambda_n(q) = a_{s,n} + \Gamma(q,T)$. The common spinodal is $\min_q \Lambda_n = 0$, with the uniform sector included in the minimization. A negative shift of $a_{s,n}$ changes the onset condition without changing the wavevector minimizing $\Gamma$ at that temperature.

Figure S22 gives the threshold normal coefficient, its marginal period, the balanced chemical capacitance, and the resulting least normal stiffness for several shifts. The original normal law at zero shift remains linearly stable in the tested temperature interval. The additional preference needed for the surface instability is therefore a quantitative requirement of the revised model. A favorable period is not reported without the associated stiffness sign. At 400 K the marginal period is of order 272 nm, but the corresponding threshold is about $-5.52\ \mathrm{m^2\ F^{-1}}$; the unshifted surface coefficient at that temperature is only $-1.18\ \mathrm{m^2\ F^{-1}}$.

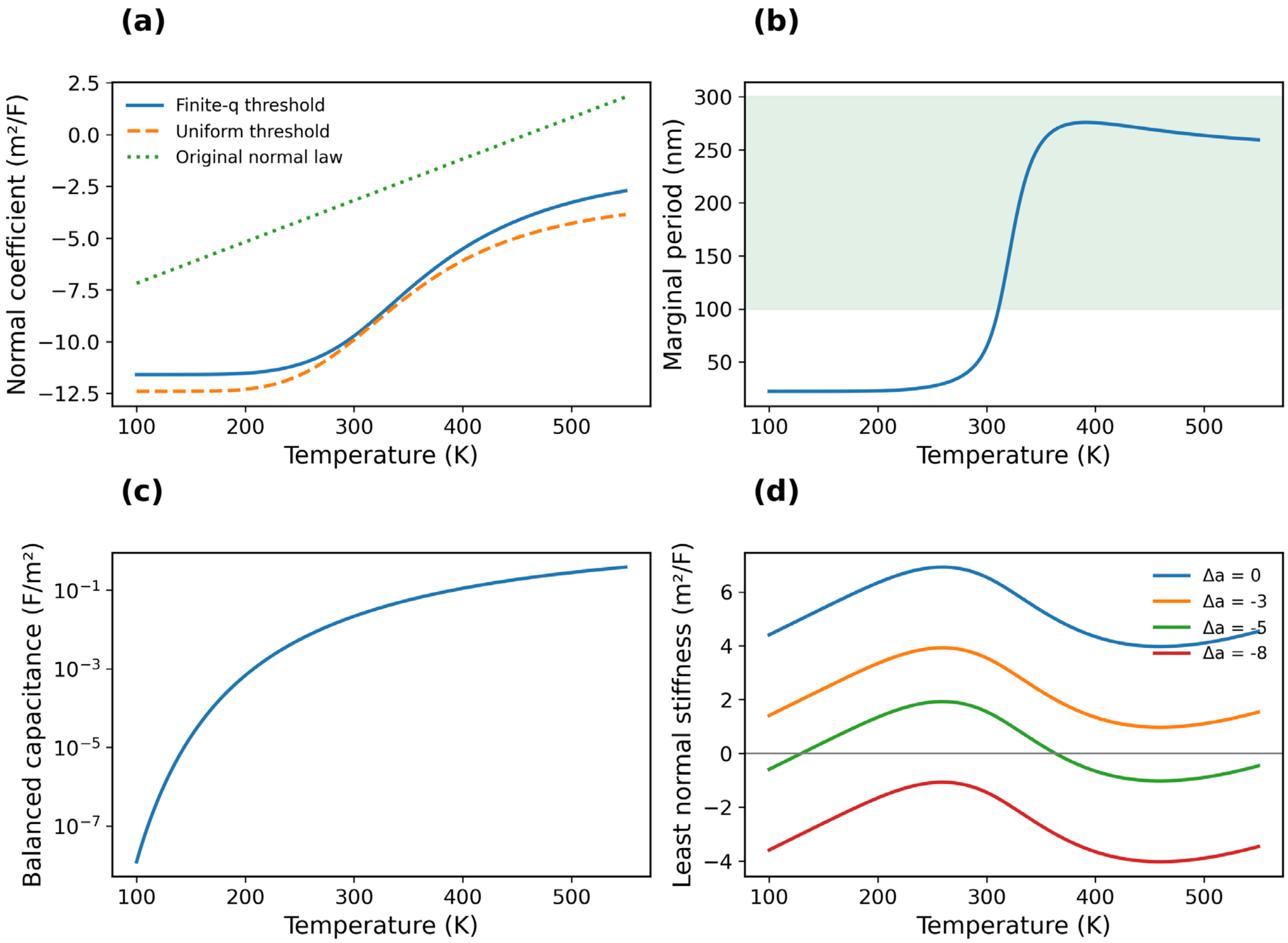

**Figure S22.** Surface-ordering threshold with unchanged common chemistry. (a) Critical normal coefficient at the finite-wavevector minimum and in the uniform sector, compared with the unshifted normal law. (b) Marginal period; shading marks 100–300 nm. (c) Balanced chemical capacitance. (d) Least normal stiffness for several independent normal shifts. The unshifted law remains above the instability threshold in the tested interval.

The allowed microscopic origins of a normal quadratic preference include surface bonding and reconstruction, a modified structural order parameter, a localized stress acting through electrostriction, and a consistently reduced flexoelectric interaction. Surface symmetry permits the coefficient but does not fix its sign or magnitude. The present calculation does not derive a material-specific value from the Langmuir entropy. A direct adsorption–polarization bonding term could make the coefficient occupation dependent, but that would be a new interaction requiring its own parameterization and derivatives. The shift shown in the maps is instead an explicit polar control, while the chemistry is held common.

### S14.6 Candidate maps and nonlinear period searches

The first-harmonic ansatz is $\mathbf{u}_s(x) = \bar{\mathbf{u}} + \mathbf{a}\cos(qx) + \mathbf{b}\sin(qx)$, with four-component vectors and an optimized $\ln q$. It has 13 real variables. The mean is unconstrained; it can develop normal polarization and ionic charge. Local nonlinear energies are averaged on 65 quadrature points. A kernel spline on 1001 logarithmic wavevectors accelerates the period derivative; its knots are calculated from the exact boundary problem at runtime, and the audit compares interpolated and direct kernels. The full-field calculations use the direct kernel. Periods are searched over 8–3000 nm for this restricted ansatz.

The main candidate atlas contains 31 temperatures from 250 to 500 K and 25 normal shifts from $-10$ to zero $\mathrm{m}^2\ \mathrm{F}^{-1}$ for each completion. Uniform normal candidates are minimized from several amplitudes and both charge signs. Completion III additionally includes several in-plane initializations. Multiple first-harmonic and continued seeds are compared with these uniform minima. The candidate relation map uses both the energy ordering and the common unpolarized spinodal. At nonzero chemical activity, the unpolarized state is generally no longer stationary, so the zero-bias common spinodal is not drawn as a physical instability curve on an activity sweep.

Figure S23 supplies activity cuts from $\log_{10}\rho = -8$ to eight at 300, 400, and 450 K, holding $\Delta a_s = -5\ \mathrm{m}^2\ \mathrm{F}^{-1}$. It compares the periodic candidate with the best uniform state and shows its amplitude. A lower energy in this restricted family is a valid variational improvement over the uniform candidate. Failure to find such an improvement is not proof that all higher-harmonic patterns are unfavorable.

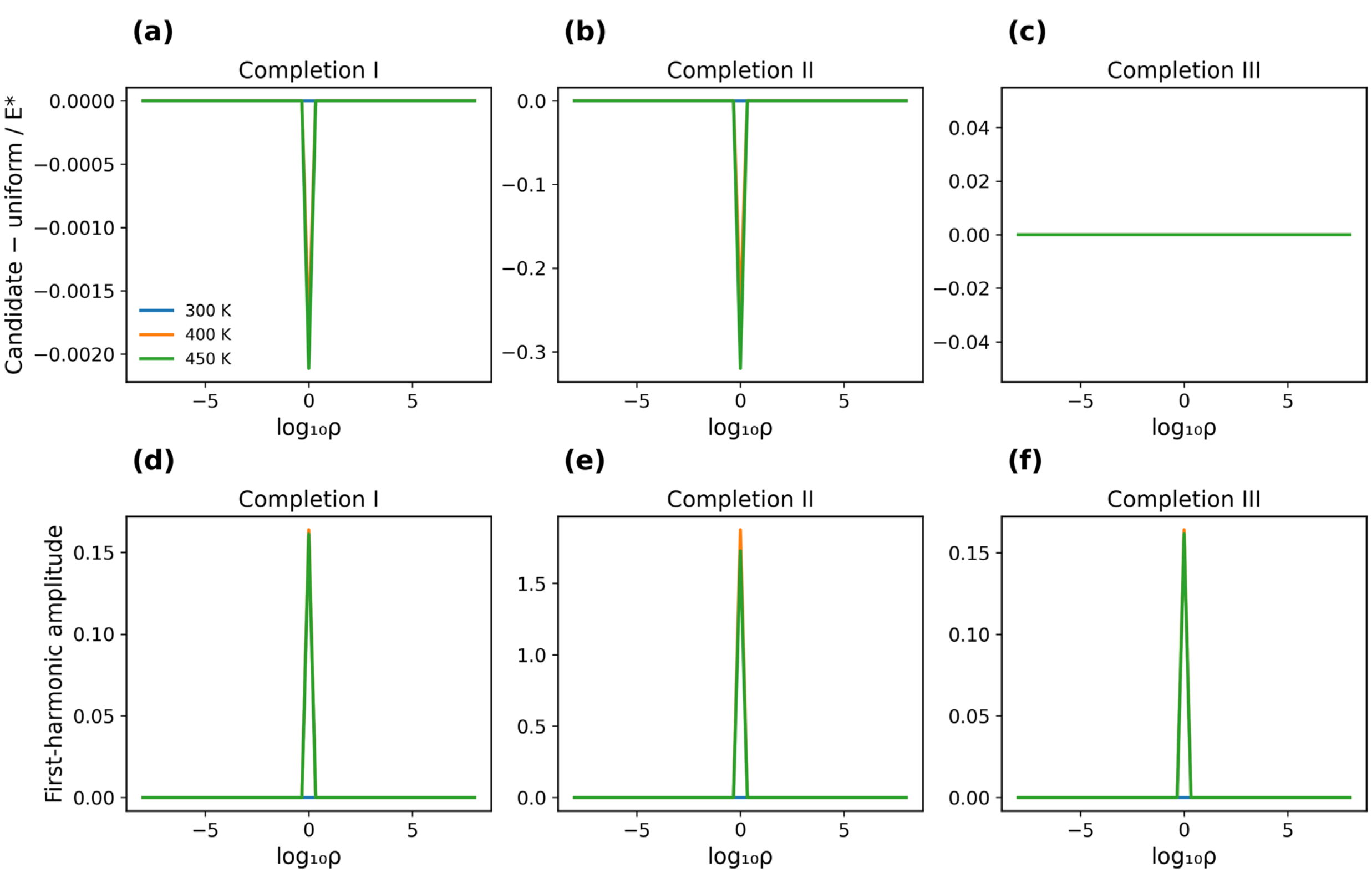

**Figure S23.** Surface candidate comparison under chemical bias at $\Delta a_s = -5\ \mathrm{m}^2\ \mathrm{F}^{-1}$. Columns correspond to completions I, II, and III. (a–c) First-harmonic candidate energy relative to the best uniform state, with positive differences clipped to zero for readability. (d–f) First-harmonic amplitudes. Curves compare 300, 400, and 450 K. A nonzero amplitude is not by itself a thermodynamic preference.

Full nonlinear stripes use 257 lateral points for completions I and III and 1025 for the sharper completion-II walls and retain every resolved harmonic of the four fields. At 350, 400, and 450 K and $\Delta a_s = -5\ \mathrm{m^2\ F^{-1}}$, six periods spanning the neighborhood of the first-harmonic optimum are relaxed from small-mode, larger-amplitude, and continued seeds. The best state is compared with all uniform candidates, including the charged normal state. Figure S24 gives the 400 K profiles, and Figure S25 gives representative period-energy and Bloch comparisons. Completion III becomes uniform in-plane in the representative full-field searches; its arbitrary simulation box size is not a domain wavelength.

Table S7 summarizes the full-field surface candidates.

**Table S7.** Full-field surface candidates at balanced activity and normal shift $-5\ \mathrm{m^2\ F^{-1}}$. Energies are divided by $E_s^*$. A uniform solution has no selected repeat.

| Completion | T (K) | Repeat (nm) | $\Omega_{cand}/E_s^*$ | $\Omega_{unif}/E_s^*$ | Gradient |
|---|---|---|---|---|---|
| I | 350 | 265 | -0.000157 | -2.15e-16 | 1.43e-06 |
| I | 400 | 312 | -0.00245 | -5.28e-05 | 1.14e-07 |
| I | 450 | 308 | -0.00284 | -0.000123 | 7.09e-08 |
| II | 350 | 370 | -2.21 | -1.37 | 1.99e-06 |
| II | 400 | 369 | -1.5 | -0.778 | 1.2e-06 |
| II | 450 | 295 | -0.912 | -0.318 | 1.11e-06 |
| III | 350 | Uniform | -0.364 | -0.364 | 7.03e-07 |
| III | 400 | Uniform | -0.364 | -0.364 | 1.11e-06 |
| III | 450 | Uniform | -0.364 | -0.364 | 4.27e-07 |

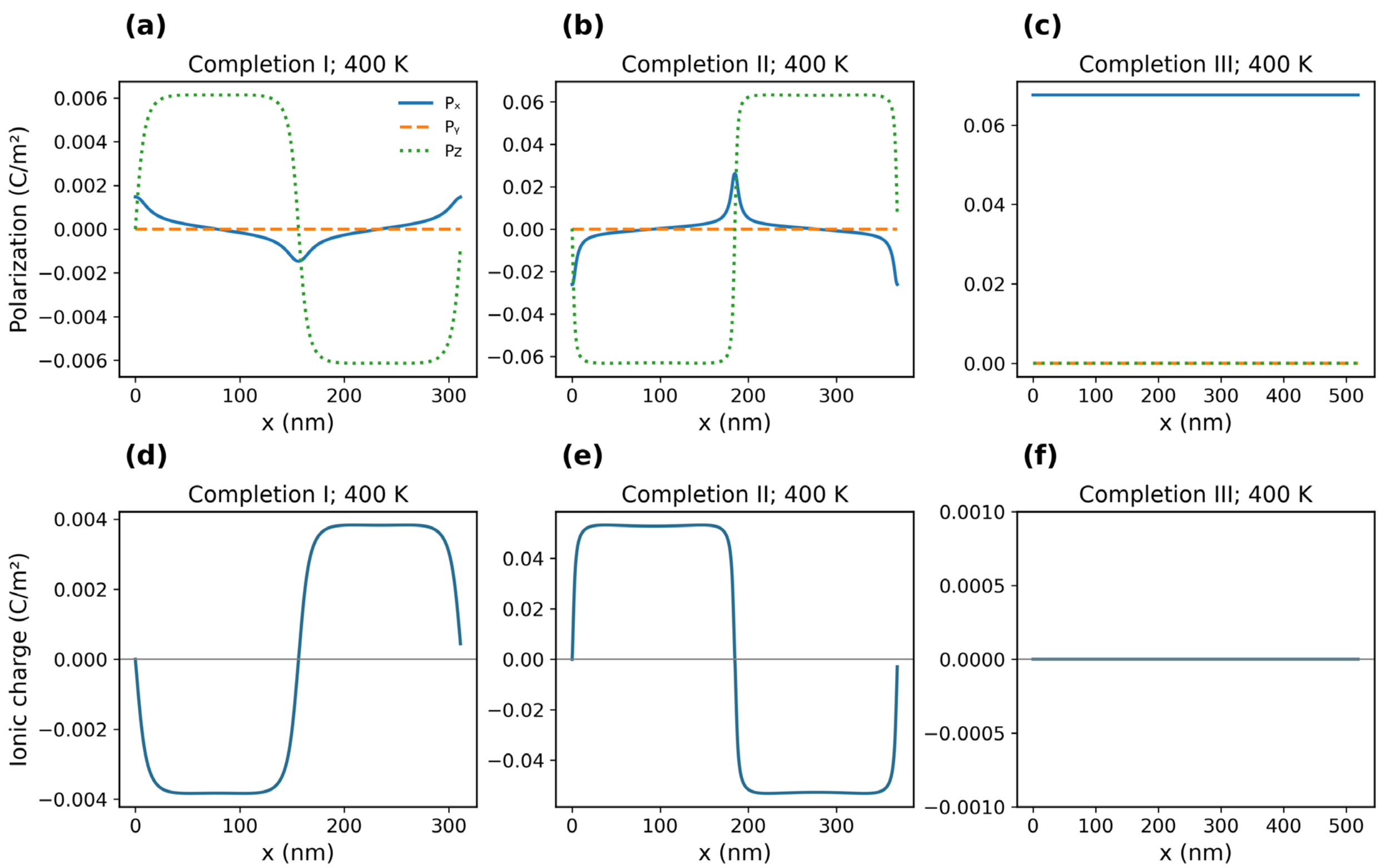

**Figure S24.** Full-field surface profiles at 400 K and $\Delta a_s = -5\ \mathrm{m}^2\ \mathrm{F}^{-1}$. (a–c) Cartesian polarization components for completions I, II, and III. (d–f) Ionic charge. Completions I and II retain nonuniform states; III relaxes to uniform in-plane order, so its cell length is arbitrary. The mean charge was allowed to relax in every case.

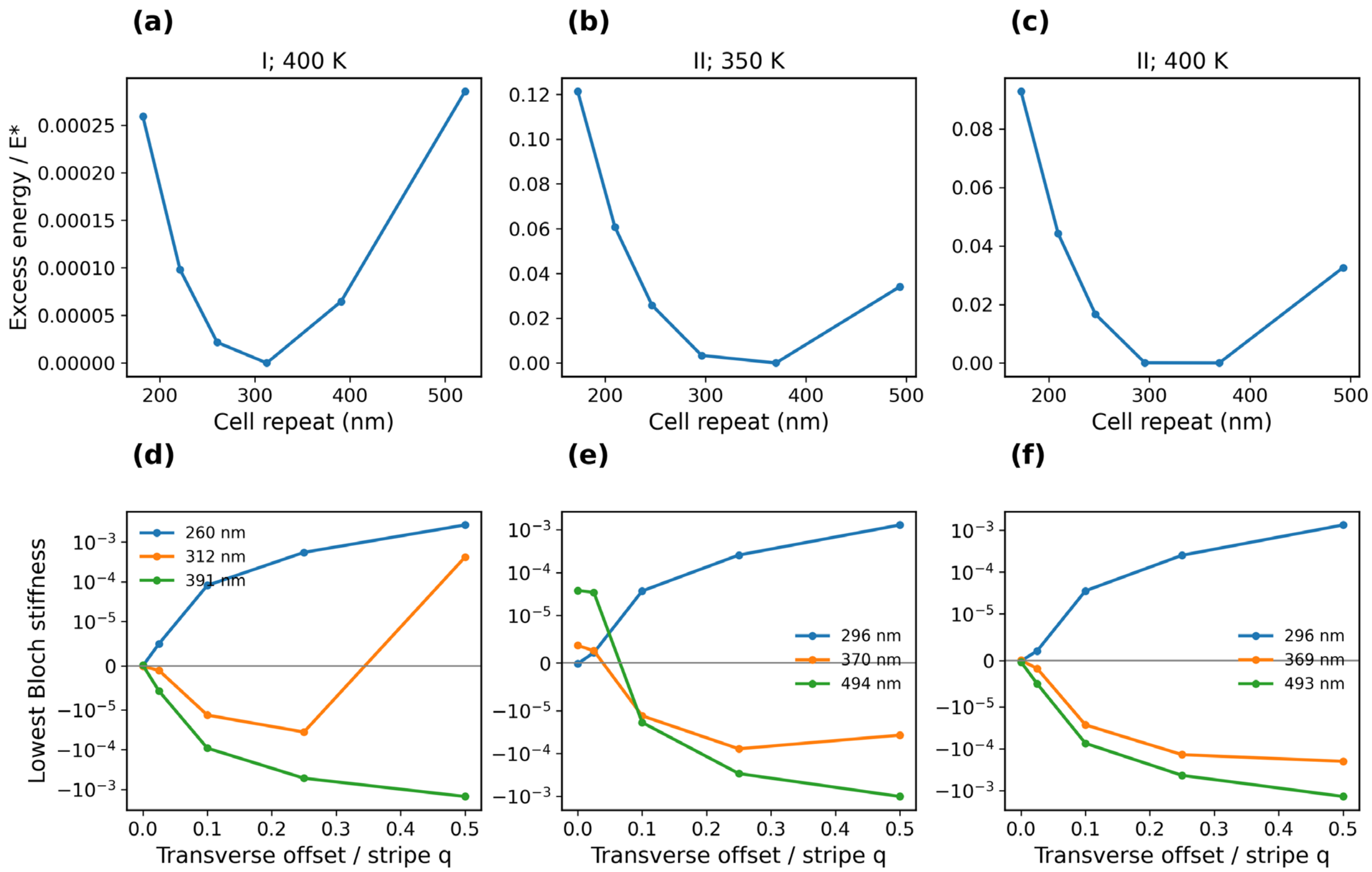

**Figure S25.** Surface period selection and transverse stability on representative slices. Columns show completion I at 400 K, II at 350 K, and II at 400 K. (a–c) Full-field period-energy scans relative to their lowest sampled energies. (d–f) Lowest transverse Bloch stiffnesses at the best and neighboring sampled periods. A low energy does not guarantee stability against every transverse offset. Energies and stiffnesses use the surface normalization.

The full-field calculation reveals a finite-amplitude completion-I stripe at 350 K that is not selected by the corresponding first-harmonic candidate. This provides a direct estimate of a truncation failure: the higher-harmonic search can improve the candidate diagram before any claim of global equilibrium is made. Completion II also supports periodic states where infinitesimal perturbations of the unpolarized state decay, and its best uniform normal competitor is included in the energy difference. These findings establish why a common linear spectrum cannot supply the nonlinear phase diagram. They also mean that the solid boundary in Figure 13 is a boundary of the stated first-harmonic candidate comparison, not an exact coexistence line for the full functional.

Surface Bloch perturbations use the exact shifted bulk kernel and the local Hessian of both the surface polynomial and reduced common-site chemical energy on the periodic solution. Longitudinal and transverse offsets are sampled at 0, $0.025q$, $0.10q$, $0.25q$, and $0.50q$ for the best period and its adjacent sampled periods. The surface kernel is already exact through depth, so the remaining spatial approximation is lateral Fourier resolution. Small zero-offset translation values are treated as numerical limits. Lateral refinement is particularly important for completion II: at 400 K and the best sampled repeat, the apparent zero-offset stiffness decreases from about 0.15 on 257 points to a value of order $10^{-7}$ or smaller on 1025 points, while its energy changes by less than $3 \times 10^{-5} E_s^*$. An apparently positive coarse-grid spectrum was therefore not accepted as evidence of continuum stability. The finer calculation resolves transverse negative modes on representative energy-preferred surface stripes. A pattern that is lower than all tested uniform states can still have a negative Bloch stiffness; the energy and stability results are stored separately.

Table S8 quantifies the lateral-resolution test and its translation diagnostic.

**Table S8.** Lateral-resolution check of completion II at 400 K and its best sampled repeat.

| **Lateral points** | $[\Omega(n) - \Omega(2049)] / E_s^*$ | **Zero-offset stiffness** | **Gradient** |
|---|---|---|---|
| 65 | -0.000932 | 0.789 | 1.87e-07 |
| 129 | -0.000478 | 0.78 | 8.16e-07 |
| 257 | -2.86e-05 | 0.154 | 8.63e-07 |
| 513 | -2.88e-08 | 0.00162 | 4.89e-07 |
| 1025 | 1.72e-12 | 7.61e-09 | 1.09e-06 |
| 2049 | 0 | 8.48e-07 | 9.97e-06 |

### S14.7 Reversible reactions and polarization relaxation

The kinetics retain both ionic occupations rather than their instantaneous fixed-charge minimum. At balanced activity the reaction affinities are $\Delta_i = \Delta g^0 + q_i \psi$. A bounded forward/reverse pair is

$$k_i^+ = \frac{\Gamma}{1+\exp[\Delta_i/(k_B T)]}, \qquad k_i^- = \frac{\Gamma}{1+\exp[-\Delta_i/(k_B T)]}, \qquad \dot{\theta}_i = k_i^+ \theta_0 - k_i^- \theta_i. \tag{S54}$$

Their ratio is $k_i^+/k_i^- = \exp[-\Delta_i/(k_B T)]$, so the stationary fractions are precisely the competitive Langmuir fractions. Each positive or negative adsorption event uses an empty site. There is no reaction term allowing both species to occupy the same site. For a nonunit activity, replacing the standard formation energy by $\Delta g_i^0 - \nu_i k_B T \ln\rho$ gives the same generalized law; the explicit kinetic examples here use $\rho = 1$.

Polarization evolves by $\dot{\mathbf{p}}_s = -\delta(\Omega_s/E_s^*)/\delta \mathbf{p}_s$ in reduced time. The potential entering the reaction rates is the charge-conjugate derivative of the same exact bulk kernel. The total energy uses the unreduced three-state entropy. Detailed balance makes the reaction contribution nonpositive: each channel has a flux proportional to the difference of forward and reverse populations, multiplied by the logarithm of their ratio with the opposite sign in the energy derivative. Coupled with positive polar mobility, the passive evolution decreases the grand potential. An imposed zero mean of $\theta_+ - \theta_-$ would break this ensemble and is not applied.

At the balanced unpolarized state, the occupation block of the Hessian is $\beta[\mathrm{diag}(1/\boldsymbol{\theta}) + \mathbf{1}\mathbf{1}^T/\theta_0]$, with $\beta = N_s k_B T/E_s^*$. The charge map is $\sigma/P_s^* = (Q_s/P_s^*)(\theta_+ - \theta_-)$. Applying this map to the four-field bulk kernel gives the five-variable polar/occupation Hessian. The reaction mobility is diagonal at equilibrium, with entries $\Gamma\theta_i/\left[\beta\left(1 + e^{-\Delta g^0/k_B T}\right)\right]$. The eigenvalues of its symmetrically weighted Hessian give the real growth rates. This derivation uses the actual two-occupation entropy, not a scalar capacitance assigned to an arbitrary kinetic timescale.

**S14.8 Numerical relaxation and reproducibility**

The kinetic examples use 129 points in one lateral dimension and 65-by-65 points in two, except that the sharp finite-amplitude completion-II seed uses 4097 points over four repeats to retain the stationary-wall resolution. The cell initially contains four periods of the analytically calculated fastest mode, except that a finite-amplitude seed is initialized from four repeats of its corresponding stationary stripe. Polarization uses a semi-implicit update of the positive bulk kernel. Occupations use a local backward-Euler step for the three-state Markov generator at the current potential. The two-by-two linear solve preserves nonnegative fractions and their sum below one. A backtracking check rejects an increase of the complete grand potential.

Early noise growth on the unstable side uses a reduced step no larger than 0.002 until the polarization variance exceeds $10^{-4}$; the later maximum is 0.0200. At the representative 400 K unstable wavevector, the discrete early growth rate differs from the continuous five-variable rate by less than one percent. A further tenfold step reduction decreases that error by approximately tenfold. The complete step-convergence numbers are retained in the surface verification record. This check matters because a strongly implicit polar update can otherwise make an apparent ionic delay mostly a time-discretization effect.

The five one-dimensional examples are completion I from noise at 400 K, completion II from noise and from a finite-amplitude stripe at 350 K, and completion III from noise and from an in-plane seed at 400 K. The two-dimensional example is completion I from noise at 400 K. All use $\Delta a_s = -5\ \mathrm{m}^2\ \mathrm{F}^{-1}$, balanced activity, and the common chemical parameters. Figure S26 reports energy, polarization variance, mean charge, and the dependence of the least-decaying or fastest period on the reaction-rate ratio. The 350 K unpolarized state is linearly stable, so its reported linear spectral maximum is a decay mode, not an unstable wavelength.

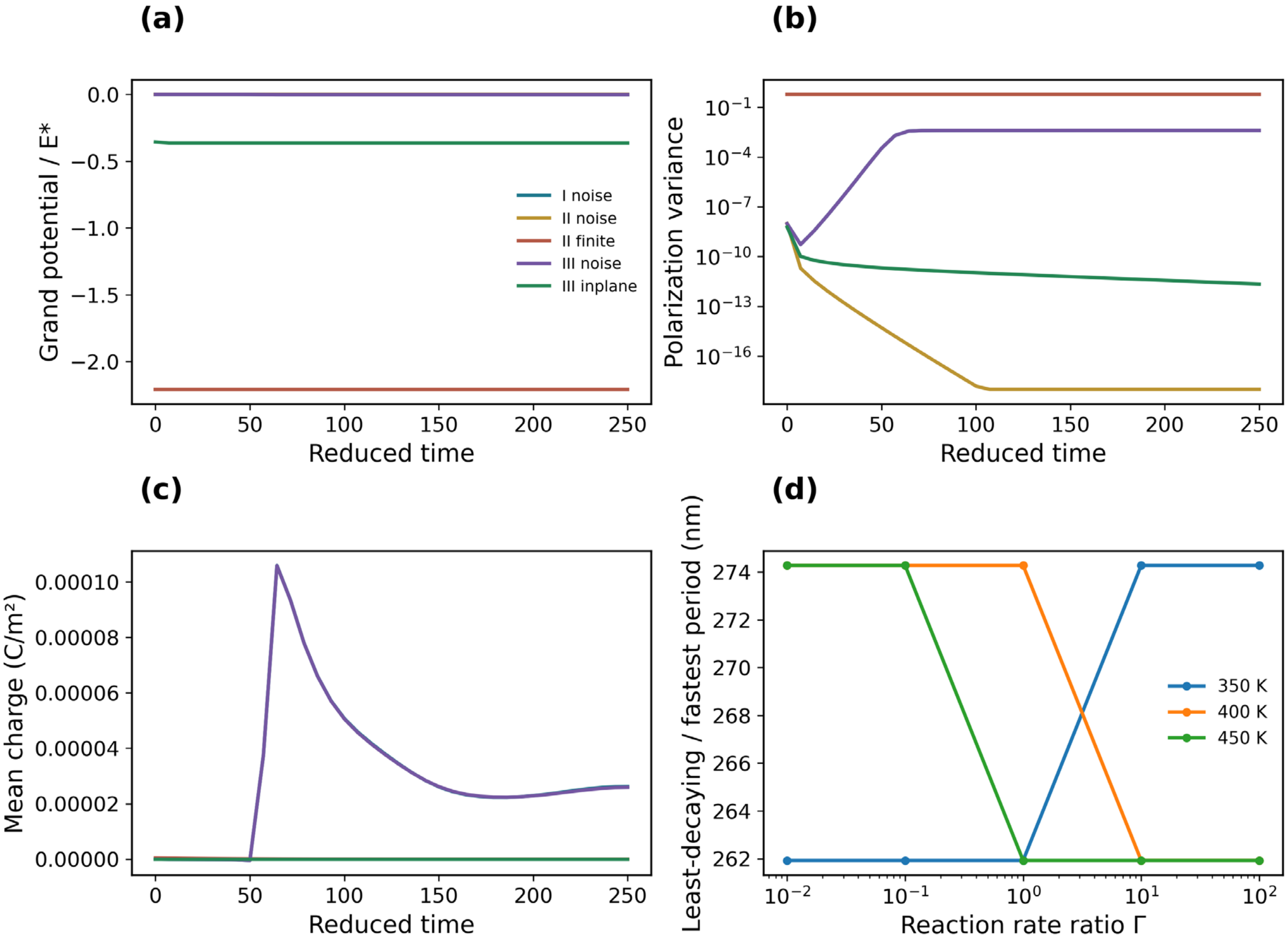

**Figure S26.** Relaxation with explicit common-site reaction kinetics. (a) Grand potential. (b) Polarization variance. (c) Mean ionic charge, which is not projected to zero. Curves compare noise, finite-amplitude, and in-plane preparations as specified in Section S14. (d) Fastest growing period at 400 and 450 K, and least-decaying period at the linearly stable 350 K reference, versus reaction-rate ratio. Time is reduced and has no fitted conversion to seconds.

No conversion of reduced time to seconds is implied. Such a conversion requires independently measured polarization and reaction mobilities. Likewise, a noise-generated finite-cell image is evidence of an allowed instability pathway, not proof of a unique final morphology. The final energy and the occurrence of a Fourier peak are interpreted alongside the stationary candidate and Bloch calculations. The main paper uses this combination to compare onset, thermodynamics, and persistence without identifying them as one calculation.

The notebook begins with dependency installation and visible source-code cells, followed by verification. Its first calculation part rebuilds the main figures from model inputs; its second part generates the expanded supplementary atlas and audits. No output cells, serialized internal package, external source archive, or previously calculated numerical array is needed to start. Solver-created files are ordinary intermediate products within the runtime. Changing a grid control and rerunning the dependent stage produces new results, while changing the chemistry or boundary conditions requires rerunning the complete downstream chain. The archival source and results provide a record of this particular publication calculation, rather than a hidden input to it.

## S15 Stability at the initial and neighboring film repeats

The optimal-repeat tests do not determine whether a pattern formed at the initial spectral period can survive. Figure S27 therefore supplements the energy scans with re-relaxed stripes at the least-stiff reference period and at the two refined periods neighboring the energy minimum. Each uses 33 lateral points, 17 polarization nodes, and enriched auxiliary fields. The tested Bloch offsets are $0$, $0.10q$, $0.25q$, and $0.50q$ in both longitudinal and transverse directions. The plot compares nonzero offsets; all zero-offset values and optimization residuals remain in the result records.

The tetragonal initial-period candidate at about 15.7 nm has a negative longitudinal stiffness, whereas the longer neighbors of the optimal repeat have no negative mode in the sampled nonzero sectors. The orthorhombic and rhombohedral initial periods are closer to their nonlinear energy minima and pass these sampled tests. The weak-anisotropy outcome is nonmonotone with repeat: the initial-period candidate and the longer sampled neighbor have no resolved negative nonzero-offset mode, but the shorter neighbor and the energy-preferred stripe have transverse instabilities. Thus an initially formed repeat can remain locally stable even if a stripe of another repeat has a lower energy. Conversely, minimizing the stripe energy alone does not remove a secondary instability.

These are finite-resolution, finite-offset tests. In particular, numerical translation pinning is appreciable in the sharper tetragonal pattern on the coarse stability mesh, while the more resolved optimal-repeat calculation reduces that error. The sampled points are not interpolated into a claimed complete stability band. Their purpose is to make the comparison between the initial period, nonlinear energy preference, and persistence explicit.

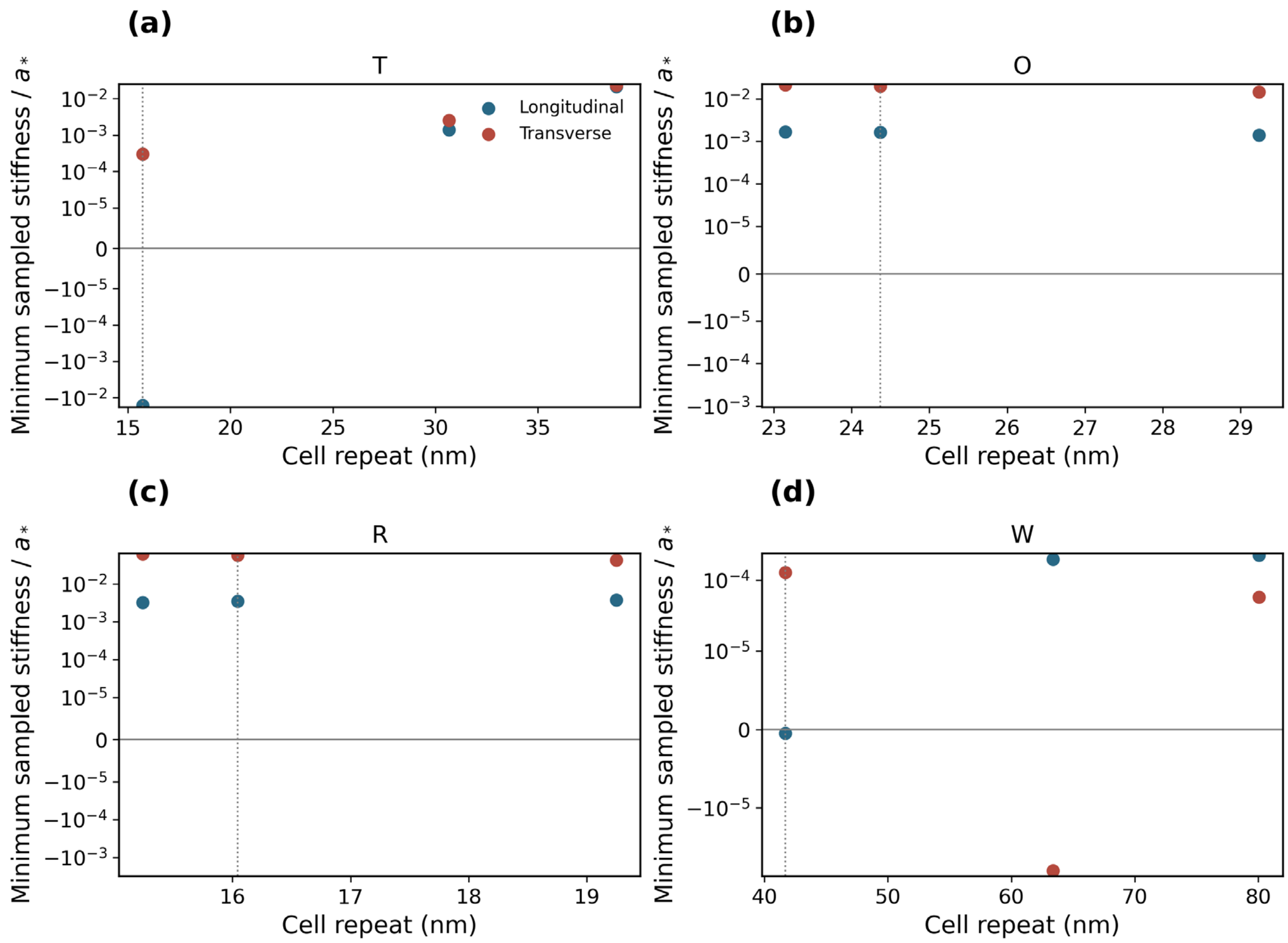

**Figure S27.** Sampled stability of developed stripes at initial and detuned repeats. (a–d) T, O, R, and W. Each group compares longitudinal and transverse minimum stiffness over the tested nonzero Bloch offsets at the initial least-stiff period, a shorter neighboring repeat, and a longer neighboring repeat. The calculation re-relaxes each field on a 33-by-17 grid. The refined optimal-period tests are given separately in Figure 11. Negative points demonstrate a secondary instability; nonnegative samples do not establish a complete stability band.

## S16 Electrical boundary conditions and the sensitivity of domain formation

### S16.1 Exposed surfaces and a dielectric layer beneath a top electrode

An exposed, chemically compensating surface and a surface separated from a metal electrode by a dielectric layer describe different experimental geometries. The first is appropriate to a film with a grounded bottom electrode and an unbounded exterior. The second can represent a top contact separated from the ferroelectric by an insulating interfacial layer, an adsorbate layer, or a water-containing gap. Surface water and adsorption can participate in compensation [6,37,41,43], but identifying a layer as water does not specify its dielectric response, conductivity, or reaction law. In particular, the relative permittivity of unity used in the earlier gap calculation is an idealized dielectric value, not a measured permittivity of interfacial water. If the layer conducts ions throughout its thickness, a dielectric-layer model alone is insufficient.

For a Fourier component with lateral wave number $q > 0$, the exposed exterior has potential $\phi_e = \psi\exp[-q(z-h)]$. Its electrical contribution at the surface is an admittance $Y_{open}(q) = \epsilon_e q$, where the permittivity is absolute. In the alternative geometry, let a dielectric layer of thickness $d$ and permittivity $\epsilon_g$ lie between the surface and a top electrode held at zero perturbation potential. Solving Laplace's equation in that layer gives

$$Y_{open}(q) = \epsilon_e q, \qquad Y_{gap}(q) = \epsilon_g q\coth(qd). \tag{S55}$$

The respective limits at zero wave number are $Y_{open}(0) = 0$ and $Y_{gap}(0) = \epsilon_g/d$. The latter is the top-gap capacitance per area. Thus, for a columnar normal-polarization perturbation, the electrostatic stiffness after chemical equilibration is

$$d_n(q) = \left\{h\left[\epsilon_f q\coth(qh) + Y(q) + C_{chem}\right]\right\}^{-1}. \tag{S56}$$

At zero wave number, the film contributes $C_f = \epsilon_f/h$. Replacing the exposed boundary by a grounded top electrode therefore changes the normal stiffness from $\left[h\left(C_f + C_{chem}\right)\right]^{-1}$ to $\left[h\left(C_f + \epsilon_g/d + C_{chem}\right)\right]^{-1}$ at the same reference state and chemical capacitance. A finite imposed top voltage also adds a bias and must be included in the reference-state calculation. The fixed-reference comparison does not include that additional effect.

A dielectric or adsorbate coating without a top electrode is a third geometry. If its outer side adjoins an unbounded dielectric of permittivity $\epsilon_e$, its effective admittance is

$$Y_{coat}(q) = \epsilon_g q\,\frac{\epsilon_e + \epsilon_g\tanh(qd)}{\epsilon_g + \epsilon_e\tanh(qd)}. \tag{S57}$$

Its zero-wave-number admittance vanishes. A water layer on an exposed surface consequently cannot be treated as a finite gap capacitor merely because it has a finite thickness. The remote electrical termination matters. For equal exterior and gap permittivities, the grounded-gap admittance exceeds the open admittance and reduces the fixed-reference electrical penalty. If both geometry and permittivity change, their effects must be evaluated together.

### S16.2 Common-site chemistry and independent adsorption families

Changing the electrical geometry does not require changing the adsorption law. The present calculations use one population of empty, positive, and negative sites in every geometry considered analytically. Both ionic species compete through the denominator $1 + w_+ + w_-$. Their differential capacitance is the variance of the charge on one site, as derived in Section S2. An alternative with two independent binary site families has separate denominators $1 + w_i$ and a different entropy. These assumptions can be compared, but their nonlinear energies and fluctuations cannot be mixed within a single calculation.

At balanced activity, equal formation energies, zero potential, and equal charge magnitudes $q_{ion}$, define $w = \exp[-\Delta g^0/(k_B T)]$. If each independent family has density $N_s$, matching the single-sign saturation charge of the common-site model, the capacitances are

$$C_{common} = \frac{2N_s q_{ion}^2}{k_B T}\frac{w}{1+2w}, \qquad C_{independent} = \frac{2N_s q_{ion}^2}{k_B T}\frac{w}{(1+w)^2}. \tag{S58}$$

Their ratio is $(1 + w)^2/(1 + 2w)$, and both recover $2N_s q_{ion}^2 w/(k_B T)$ in the dilute limit. The occupancy correction can therefore be small near a dilute balanced reference, although site competition remains essential for consistency and can matter at appreciable occupation or under bias. This statement does not establish equivalence of the two nonlinear thermodynamic models.

### S16.3 Why the selected film modes respond weakly

The ionic response couples to the potential created at the surface by a polarization fluctuation. A nearly charge-neutral rotation can release elastic energy while generating very little surface potential. Section S4.4 expresses the change from frozen to equilibrated ionic fluctuations as

$$H_{rel} = H_{fr} - \frac{C_{chem}}{1+C_{chem}R_q} w_q w_q^{\dagger}. \tag{S59}$$

For a normalized, nondegenerate mode $v$, at fixed reference polarization and stress,

$$\frac{\partial \lambda}{\partial C_{chem}} = -\frac{\left|w_q^{\dagger} v\right|^2}{\left(1+C_{chem}R_q\right)^2}. \tag{S60}$$

Weak coupling to the surface-potential response explains weak incremental sensitivity even when the equilibrium screening charge is substantial. Chemical conditions can still select a different homogeneous orientation or prestress, thereby changing the state about which the instability is evaluated [57]. The derivative above holds that reference fixed and must not be interpreted as the total derivative of a chemically selected equilibrium branch.

Table S9 quantifies this distinction using the stored reference spectra. At each case's sampled minimum, the calculation is repeated with the ionic fluctuation set to zero, retaining the same stationary polarization, stress, and surface charge. This is a controlled test of incremental ionic relaxation in the exposed geometry, not a comparison of two independently equilibrated samples.

**Table S9.** Incremental ionic sensitivity at the sampled reference-spectrum minimum. The last column is the absolute change divided by the magnitude of the relaxed-ion stiffness.

| Case | Relaxed stiffness $\lambda/a_*$ | Frozen stiffness $\lambda/a_*$ | Relative change |
|---|---|---|---|
| T | -0.503 | -0.503 | Below $10^{-6}$% |
| O | -0.208 | -0.208 | Below $10^{-6}$% |
| R | -0.223 | -0.223 | Below $10^{-6}$% |
| W | -0.0427 | -0.0427 | 0.072% |

The T, O, and R differences are negligible on the scale of their negative stiffnesses. The W difference is approximately 0.072%. This result concerns these selected modes and reference conditions; it does not imply that every branch of the multiaxial phase maps is insensitive to surface chemistry. In the earlier-versus-revised benchmark comparison, the marginal periods remained approximately 13.8, 14.5, 15.1, and 49.4–49.5 nm. The weak-anisotropy ferroelectric film is therefore also a case of small onset sensitivity.

Finite-amplitude domains need not retain the electrical character of their initial mode. Normal polarization, higher harmonics, spatially varying occupation, and local saturation can become appreciable during nonlinear evolution. The developed tetragonal candidate illustrates this distinction: the earlier primitive vector repeat was 63.3 nm, with its strongest Fourier period at 31.7 nm; the revised candidate has both quantities near 34.7 nm. The full repeat changed substantially even though the dominant modulation scale changed by about 10%. Geometry, branch search, and resolution also differed between those calculations, so this comparison does not isolate an occupancy-law effect.

**S16.4 Why the surface-ordering threshold changes more strongly**

The surface-ordering limit of Section S14 has a positive Gaussian bulk energy. Its destabilizing tendency comes from a negative normal surface coefficient, while bulk polarization, gradients, and screened electrostatics supply restoring costs. The relaxed normal stiffness is $\Lambda_n(q) = a_{s,n} + \Gamma(q)$. A normal surface fluctuation is electrically active, so its ordering threshold depends strongly on the screening available to reduce the depolarizing penalty.

The earlier illustrative surface example used approximately 0.132 eV for the formation energy, two independent adsorption families, a semi-infinite bulk, and a zero-mean ionic-charge constraint. The revised example uses 0.200 eV, common sites, a 300 nm film above a grounded bottom electrode, and exchange of mean charge with the reservoir. These are several distinct changes. The earlier chemical parameters gave approximately 0.300 F m$^{-2}$ at 300 K, whereas the common film parameter set gives 0.0216 F m$^{-2}$. The dominant scale of this difference is explained by the dilute formation-energy dependence,

$$\frac{C_{new}}{C_{old}} \simeq \exp\left[-\frac{0.068\ \text{eV}}{k_B T}\right] \simeq 0.072 \quad (T = 300\ \text{K}). \tag{S61}$$

Equivalently, the numerator in the exponential is an energy difference of 0.068 eV and the denominator is the thermal energy in the same units. This approximately fourteenfold reduction must not be attributed to common-site exclusion alone. The ordinary film benchmarks already used 0.200 eV and did not undergo this change in formation energy.

At 300 K the earlier surface calculation had $\min_q \Lambda_n \simeq -0.00471$ m$^2$ F$^{-1}$ with $a_{s,n} = -3.18$ m$^2$ F$^{-1}$. Positive and negative contributions therefore nearly cancelled: the instability margin was only about 0.15% of the normal surface term. In the revised model, $\min_q \Gamma \simeq 9.73$ m$^2$ F$^{-1}$, giving $\min_q \Lambda_n \simeq 6.55$ m$^2$ F$^{-1}$ at the unchanged surface coefficient. The original parameter-specific instability is removed. Additional normal preference can restore it, as quantified by the threshold maps, but remains an independent constitutive requirement.

The large revision is specific to this electrically active surface-ordering construction. It is not evidence that relaxors universally respond more strongly than conventional ferroelectrics to an occupancy-law correction. The equilibrium model also preserves several general results across the electrical geometries: positive chemical capacitance relaxes an electrical penalty;

compatible strain and vector polarization remain necessary; onset and developed-domain stability are different tests; and reciprocal passive kinetics has real linear relaxation rates. Absolute phase boundaries, charged-mode thresholds, and nonlinear periods require recalculation for the actual geometry. Quantitative attribution of the earlier-versus-revised differences to individual assumptions would require a sequence of otherwise identical calculations, which is not claimed here.

## S17 Square-domain calculation and activity-sweep movies

### S17.1 A common lateral field of view for the film texture

Figure 12(a–c) is recalculated on a $1000 \times 1000$ nm periodic lateral domain for the W film at 300 K, $h = 20$ nm, and $u_m = -0.150\%$. The film-axis azimuth is the same as in the selected W stripe. The displayed result uses 462 points across the stripes and 132 along them, 9 polarization depth nodes, and 17 displacement/potential depth nodes. A reference containing fourteen primitive vector repeats has a commensurate period of 71.4 nm, close to the independently optimized 70.1 nm stripe. The descendant lowers the matched-grid energy by 1.4 kJ $\mathrm{m}^{-3}$. Its maximum mass-normalized force residual is $1.12 \times 10^{-5}$ in the dimensionless film convention. The optimizer reports 21 iterations. Refining the nonlinear seed from 17 to 33 points along its 250 nm transverse period changes its energy lowering from 1.45 to 1.4 kJ per cubic meter, a 3.7% change relative to the finer result. The matched primitive-stripe calculation independently verifies a negative transverse eigenvalue at the resolved stripe repeat. The full two-dimensional morphology remains conditional on the stated lateral and depth discretization; no continuum-converged global domain minimum or final-state Bloch stability is claimed.

The constitutive energy, film thickness, clamped substrate, free upper mechanical boundary, open electrical boundary, and common-site chemical model are unchanged. The enlarged calculation uses a Fourier representation in both lateral directions and the same piecewise-linear depth representation as the earlier nonlinear functional. It does not stretch or tile the earlier relaxed two-dimensional image. A straight stripe reference is first relaxed on the same lateral and depth resolution used for the square calculation; a perturbed field is then independently minimized with all three polarization components and all lateral Fourier modes free.

For the large lateral mesh, the surface Newton equation is solved without assembling a dense point-to-point capacitance matrix. The electrical convolution is applied by Fourier multiplication, and the Newton correction is obtained by preconditioned conjugate gradients with the local chemical tangent retained. The original and matrix-free implementations agree in energy and analytic gradient on a small two-dimensional cell; the recorded energy difference is below $10^{-9} f_*$ and the largest gradient difference is below $10^{-8}$ in the original discrete normalization. This is a change in numerical representation of the same functional.

The straight reference is generated from a primitive stripe containing 33 lateral points and is repeated fourteen times across the square, giving 462 points in that direction. It is relaxed at the commensurate repeat of 71.4 nm. The transverse seed is the negative eigenmode of this same discretized stripe at $k_y = 8\pi/(1000\ \mathrm{nm})$, calculated from its developed-state Hessian. Its small finite amplitude is checked to lower the full nonlinear energy. A new nonlinear periodic solution is first calculated in the 71.4-by-250 nm symmetry cell, with 17 and then 33 points in the longer direction. That solution is used only as an initial condition for minimization on the full 1000-by-1000 nm square, after adding a Gaussian perturbation of standard deviation $10^{-6}P_*$ with seed 501. All represented square-domain variables are then free; its calculation does not impose the smaller cell as a continuing constraint. The periodic embedding is verified by comparing the primitive and full-square energies before perturbation. The final square mesh has 462 points across the stripes and 132 along them. The physical domain remains square despite these different sampling densities. The 250 nm seed modulation is an allowed unstable period selected for this continuation example, not an independently optimized intrinsic second period.

An under-resolved grid spanning many stripes can pin their walls and change their energy by more than the secondary modulation energy. We therefore construct the reference, its unstable perturbation, and its descendant on the same commensurate discretization. The primitive-stripe mode is checked with 17, 21, 25, and 33 points per repeat. At the selected transverse wavevector, the least stiffnesses are approximately $-5.35 \times 10^{-4}$, $-2.89 \times 10^{-4}$, $-2.97 \times 10^{-4}$, and $-3.10 \times 10^{-4}$ in units of $a_*$. The negative direction persists with refinement; the 21–33-point values differ by less than ten percent. This check concerns the initial secondary instability, and does not by itself establish convergence of a fully developed two-dimensional morphology.

To improve conditioning of the polarization minimization, each Fourier amplitude is written as $\widehat{\mathbf{P}}_{\mathbf{q}} = B_{\mathbf{q}}^{-1/2}\hat{\mathbf{y}}_{\mathbf{q}}$. The fixed positive Hermitian matrix $B_{\mathbf{q}}$ combines the gradient kernel, an equilibrated-ion electrostatic kernel, and a positive regularization of the local polar curvature. Its construction and floors are explicit in the source. The energy and gradient are transformed consistently, and L-BFGS minimization is performed in $\mathbf{y}$. This is an invertible change of numerical coordinates; the regularization is not added to the physical free energy. Forces quoted here are evaluated after transforming back to polarization and applying the inverse finite-element mass matrix. The surface Newton residual, polarization force residual, energy history, and termination message are stored separately.

The square has periodic lateral boundaries and therefore admits discrete Fourier wavevectors $q_x = 2\pi m/L_x$ and $q_y = 2\pi n/L_y$. Its finite area and initialization can influence morphology and repeat number. An observed energy decrease establishes a descending pathway within this domain, while a small force residual establishes stationarity only at the stated discretization. Neither condition proves a unique large-area equilibrium texture. The earlier approximately 70.1-by-280 nm calculation is retained in Section S13 as a separate confinement example; the square calculation supplies the revised Figure 12(a–c).

### S17.2 Equilibrium activity sweeps

Supplementary Movies 1–4 show the homogeneous component maps and corresponding polarization tilt for $BaTiO_3$, $BiFeO_3$, W with $Q = 0.1Q_{BTO}$, and W with $Q = Q_{BTO}$, respectively. Every movie contains 49 activity values from $\log_{10}\rho = -12$ to $+12$ at increments of 0.5. Two frames per decade interval give a visible, uniform progression through the reservoir control. Playback is two frames per second, so one sweep lasts 24.5 seconds. Playback time has no physical kinetic meaning.

Each frame uses 41 temperatures and 33 logarithmically spaced thicknesses between 1 and 300 nm. $BaTiO_3$ and W span 100–500 K; the $BiFeO_3$ constitutive example spans 300–1200 K. The misfit, material parameters, and electrostriction choices are those of Figures 3–5. The homogeneous calculation enumerates and refines the normal-polarization minima after algebraic minimization over the in-plane components, initially sampling 601 normal-polarization values. Integer component labels are plotted directly without interpolation into fictitious phases. The color ranges and axes remain fixed throughout each movie.

Equal standard formation energies, opposite ionic charges, and opposite reaction stoichiometries give an exact activity-inversion symmetry. Under $\rho \mapsto 1/\rho$, the equilibrium normal polarization, surface potential, and ionic charge reverse sign, while the component magnitudes and tilt remain unchanged. The numerical calculation therefore evaluates the 25 nonnegative logarithmic activity values and reflects the sign-dependent quantities to form the full 49-frame sweep. The movie identifies the selected sign; at balanced activity the two signs are degenerate when normal order is present. This symmetry belongs to the stated reservoir model and need not hold for an asymmetric experimental surface reaction.

The plotted activity is a dimensionless reservoir variable. Values above unity should not be read as oxygen pressures in atmospheres. Their thermodynamic effect follows from $\Delta g_i = \Delta g_i^0 - \nu_i k_B T \ln\rho$. A strongly oxidizing environment may explore a large chemical-potential bias, but changing chemical species can also change site chemistry, formation energies, and stoichiometry. The extreme frames are therefore controlled model extrapolations. In particular, the $BiFeO_3$ movie retains the charge-capacity limitation discussed in the main paper and does not add electronic compensation at large surface potentials.